\documentclass[floatfix,prc,aps,twocolumn]{revtex4}
\usepackage{amssymb}
\usepackage{graphicx,psfrag}

\begin{document}
\title{$b \overline b$ Kinematic Correlations in Cold Nuclear Matter}

\author{R. Vogt}

\affiliation{
  Nuclear and Chemical Sciences Division,
  Lawrence Livermore National Laboratory, Livermore, CA 94551, USA\break
  Physics Department, University of California, Davis, CA 95616, USA
  }

\begin{abstract}
  \begin{description}
  \item[Background]
  The LHCb Collaboration has studied a number of kinematic correlations between
  $B$-hadron pairs through their subsequent decays to $J/\psi$ pairs in $p+p$
  collisions at 7 and
  8~TeV for four minimum values of the $J/\psi$ $p_T$.
  \item[Purpose]
  In this work, these measurements are compared to calculations of
  $b \overline b$ pairs and their hadronization and inclusive decays to
  $J/\psi J/\psi$ are compared to the same observables.  Potential cold matter
  effects on the $b \overline b$ pair observables are discussed to determine
  which are most likely to provide insights about the system and why.
  \item[Methods]
  The calculations, employing the exclusive HVQMNR code, assume the same
  intrinsic $k_T$-broadening and fragmentation as in [R. Vogt,
  Phys. Rev. C {\bf 98} (2018) 034907].  The
  pair distributions presented by LHCb are calculated in this approach, both
  for the parent $b \overline b$ and the $J/\psi J/\psi$ pairs produced in their
  decays.  The sensitivity of the results to the intrinsic $k_T$ broadening
  is shown.  The theoretical uncertainties due to the $b$ quark mass and scale
  variations on both the initial $b \overline b$ pairs and the resulting
  $J/\psi$ pairs are also shown, as is the dependence of the results on the
  rapidity range of the measurement.
  Possible effects due to the presence of the
  nucleus are studied by increasing the size of the $k_T$ broadening and
  modifying the fragmentation function.
  \item[Results]  Good agreement with the LHCb data is found for all
  observables.  The parent $b \overline b$ distributions are more sensitive to
  the $k_T$ broadening than are the final-state $J/\psi$ pairs.
  \item[Conclusions] Next-to-leading order calculations with $k_T$ broadening,
  as in [R. Vogt,
  Phys. Rev. C {\bf 98} (2018) 034907], can describe all correlated observables.
  Multiple measurements of correlated observables are sensitive to different
  nuclear effects which can help distinguish between them.
  \end{description}
\end{abstract}
\maketitle                 
\section{Introduction}
\label{intro}

Heavy flavor pair production has long been of interest in elementary $p+p$
and $p + \overline p$ collisions as a way to test perturbative QCD.
Measurements of heavy flavor correlations provide information about how heavy
quark pairs are produced in perturbative QCD, indeed much more information that
can be gained from single inclusive heavy flavor production alone. 
In the case of $b \overline b$ production, measurements of pair observables
can improve measurements of $B^0 - \overline B^0$ mixing \cite{UA1}.
Finally, a good understanding of multiple correlated observables provides a
better baseline for their production and modification in heavy-ion collisions.

Correlated $b \overline b$ studies have been carried out at hadron colliders.
The first measurements were in $p + \overline p$ collisions and carried out
at the CERN S$p \overline p$S, 
UA1 ($\sqrt{s} = 0.63$~TeV) \cite{UA1} and Fermilab Tevatron, D0
($\sqrt{s} = 1.8$~TeV) \cite{D0}, and CDF
($\sqrt{s} = 1.8$~TeV \cite{CDF180}
and $\sqrt{s} = 1.96$~TeV \cite{CDF196}).  
These measurements were primarily through studies of lepton pairs.  The
backgrounds for these measurements include $c \overline c$ decays, Drell-Yan
dileptons and leptons from light meson decays.  The light hadron decay leptons
can be removed by like-sign subtraction.  The Drell-Yan rate is much lower than
the heavy flavor production rate because Drell-Yan is an electroweak
process.  In addition,
if relatively high $p_T$ leptons are selected, the charm rate will be
suppressed.
CDF \cite{CDF180} and, more recently, at the LHC,
ATLAS \cite{ATLAS} studied $B$ hadron
pair production through $J/\psi +$lepton final states in $p + \overline p$
collisions at $\sqrt{s} = 1.8$~TeV and $p+p$ collisions at $\sqrt{s} = 8$~TeV
respectively.

More recent measurements of heavy flavor contributions to low mass
dilepton production in $p+p$ collisions have been reported by PHENIX at the
BNL relativistic heavy-ion collider ($\sqrt{s} = 0.2$~TeV) \cite{PHENIX} and
ALICE at the LHC ($\sqrt{s} = 7$~TeV \cite{ALICE7} and 13~TeV \cite{ALICE13}).
In these measurements, the low mass kinematic region of interest makes
the $c \overline c$ contributions larger so that both contributions have
to be taken into account in the analysis.

Previous dilepton analyses \cite{UA1,D0,CDF180,ATLAS} have generally focused
on tests of NLO calculations, usually in conjunction with diagrams of
different topologies in a leading order event generator such as ISAJET
\cite{ISAJET}, HVQJET \cite{HVQJET}, HERWIG \cite{HERWIG} or
PYTHIA \cite{PYTHIA}.  For example, UA1 favorably
compared calculations at next-to-leading order made with the HVQMNR code
\cite{MNR} to ISAJET as long as the ISAJET ``higher'' order contributions
(flavor excitation and gluon splitting) were included.  They tried to separate
different topologies but the results were rather inconclusive due to momentum
requirements and low statistics.  They did, however, determine that the NLO
contribution was at least 40\% of the measured cross section, depending on
the muon $p_T$ \cite{UA1}.  The D0 Collaboration compared their data to
calculations with HVQJET.  They showed that their measured azimuthal separation,
$|\Delta \phi |$,
between the decay leptons was compatible with HVQJET with higher order
corrections and not with the leading order contributions alone \cite{D0}.
The CDF measurement at $\sqrt{s} = 1.8$~TeV \cite{CDF180} compared their
final-state $J/\psi +$lepton data
to HVQMNR calculations as well as to PYTHIA and HERWIG simulations.  They
found that $\approx 25$\% of the $b \overline b$ production as a function of
azimuthal separation was found at $|\Delta \phi | < 90^\circ$, suggesting the
importance of higher-order corrections.  CDF also studied the dependence of the
$|\Delta \phi|$ distribution to the bottom quark mass, factorization and
renormalization scales, and the intrinsic $k_T$.  They found that mass and
scale variations could alter the magnitude of the cross section but not its
shape: changing the shape of the distribution was only possible by modifying
the $k_T$ \cite{CDF180}.  These findings are in accord with the hadron-level
studies of $b \overline b$ correlations in Ref.~\cite{QQazi}.  ATLAS
compared their final-state $J/\psi +$lepton results with several event
generators, finding good agreement between the simulations and the data
\cite{ATLAS}.  The low mass dilepton measurements of PHENIX attempted to
separate the dilepton data into $c \overline c$ and $b \overline b$ components,
as well as separating the heavy flavor cross sections into their topological
components, as if they were independent production mechanisms \cite{PHENIX}.

Most of the measurements mentioned so far have focused on the central rapidity
region.
The LHCb Collaboration \cite{LHCb} is the first to study $b \overline b$
correlations through $J/\psi J/\psi$ final states as well as
at forward rapidity.
While such a measurement is less direct than reconstructed $D$ or $B$ mesons, 
as discussed in Ref.~\cite{QQazi}, along with comparison to $D \overline D$
pairs measured by CDF at $\sqrt{s} = 1.96$~TeV \cite{CDFDDpairs} and by LHCb
at $\sqrt{s} = 7$~TeV \cite{,LHCbDDpairs} and $B$ hadron-$b$ jet pairs
measured by CMS at $\sqrt{s} = 7$~TeV \cite{CMSbbpairs},
it allows a more straightforward comparison
to calculations than the dilepton decay channel.

In this work, the model of $Q \overline Q$ production developed in
Ref.~\cite{QQazi}, with modified fragmentation function parameters and
$k_T$ broadening, is employed to study
$b \overline b \rightarrow J/\psi J/\psi$ pair production.  The measurement is
discussed in more detail in Sec.~\ref{sec:exp}.  In
Sec.~\ref{sec:model}, the model employed for $b \overline b$ production
is briefly discussed.  The pair observables are 
calculated both for the initial $b \overline b$ production and the final
$J/\psi J/\psi$ pairs, assuming the same minimum $p_T$ for both the parent $B$
meson and the decay $J/\psi$ in Sec.~\ref{sec:calcs}.  
The results are compared to the LHCb observables
and their dependence on the experimental
$p_T$ cut is shown.  Section~\ref{sec:kTcut} shows
how neglecting $k_T$ broadening affects the calculated observables.  
The mass and scale dependence of the results and how they change with $p_T$
cut is discussed in Sec.~\ref{sec:unc} while the dependence of the results on
the rapidity range of the measurement is shown in Sec.~\ref{sec:rap}.
Section~\ref{sec:cnm} describes
possible nuclear effects on the correlations.
The work is then summarized in
Sec.~\ref{sec:summary}.

\section{LHCb measurement of $b \overline b \rightarrow J/\psi J/\psi X$}
\label{sec:exp}

LHCb reconstructed two $J/\psi$s from their decays to dimuons in the forward
rapidity region, $2 < y < 4.5$.  The two $J/\psi$s must be associated with the
same primary vertex to ensure that they came from the same collision.
The $J/\psi$s were also required to be displaced from their primary vertex to
be $b$-decay candidates.  This
requirement essentially eliminated prompt $J/\psi$s from different collisions
as well as events with two prompt $J/\psi$s and associated $J/\psi$
and $b$ quark production.

They chose different minimum $J/\psi$ transverse momenta, $p_T$, to study the
effect of an increasing minimum $p_T$ on the pair correlations.  The data from
proton-proton collisions at $\sqrt{s} = 7$ and 8~TeV were combined for greater 
statistics.  Because the shapes of the distributions at this energy are
independent of $\sqrt{s}$ even if the integrated
production cross sections differ, the
results were presented as $(1/\sigma) d\sigma/dX$ where $X$
refers to the pair observable.  This way of displaying the data makes it
easier to compare the shapes of the
distributions with different minimum $p_T$.

LHCb presented results for six pair observables, $| \Delta \phi^* |$, the
difference in azimuthal angle between the $b$ and $\overline b$ mesons;
$| \Delta \eta^*|$, the difference in pseudorapidity between the $b$ and
$\overline b$ mesons; $A_T$, the asymmetry between the transverse momenta of
the $J/\psi$s; and the mass, $M$,
transverse momentum, $p_{T_p}$, and rapidity, $y_p$ of the $J/\psi$ pair.  The
first two observables, $|\Delta \phi^*|$ and $|\Delta \eta^*|$, were assumed
to be directly related to the parent $b$ mesons because $\phi^*$ and $\eta^*$
were estimated from the direction of the vector from the primary vertex to the
$J/\psi$ decay vertex \cite{LHCb}.  They also included, in an appendix, the
distributions $|\Delta \phi |$, $|\Delta \eta |$, and $| \Delta y|$,the
differences in azimuthal angle, pseudorapidity,
and rapidity respectively between the $J/\psi$ mesons themselves.  In this work,
$|\Delta y |$ is presented rather than $|\Delta \eta|$ for the parent
$b \overline b$.  All the pair observables studied by LHCb will be calculated
for both the parent $b \overline b$
mesons and the subsequent $J/\psi J/\psi$ decays.

In Ref.~\cite{LHCb}, the LHCb Collaboration compared their data to
PYTHIA \cite{PYTHIA6,PYTHIA8} and POWHEG
\cite{POWHEG} calculations as well as simulations of uncorrelated
$b \overline b$
production \cite{LHCb1,LHCb2} based on the transverse momentum and rapidity
distributions for single $b \rightarrow J/\psi X$ decays measured by LHCb.  
They noted that the pair distributions generated by both
PYTHIA6 \cite{PYTHIA6} and PYTHIA8 \cite{PYTHIA8}
were identical and thus the results from the two simulations
were combined in their comparison to the data.

As in a number of the previous $b \overline b$ measurements analyzed via
dilepton decays \cite{UA1,D0,CDF180}, LHCb looked for evidence of different
topological contributions to heavy flavor production in their data:
gluon splitting, flavor excitation and flavor creation.

As discussed in detail in Ref.~\cite{QQazi}, these artificial designations are
not indicative of different production mechanisms but of distinct diagram
topologies at leading order (flavor creation) and next-lo-leading order
(including gluon
splitting and flavor excitation).  These processes are distinguished by having
two (flavor creation), one (flavor excitation), or no (gluon splitting) heavy
quarks in the hard scattering.  Heavy quarks not involved in the hard scattering
are produced in an initial- or final-state parton shower \cite{NorrbinSjo}.
This separation
is necessary because PYTHIA includes only LO diagrams.  While all three
topological contributions are part of the $gg \rightarrow Q \overline Q X$
production process, when they are treated as
individual components, not all
NLO production diagrams are actually included (such as virtual corrections)
and the interferences between diagrams are not accounted for.
However, none of these contributions to $b \overline b$ production
constitute a new production mechanism.  The implementation of heavy flavor
production in PYTHIA is more completely described in Ref.~\cite{NorrbinSjo}.

There are parameters that can be tuned, depending on the generator employed,
that can match the distributions from the LO generator to those of a NLO
calculation, see for example Ref.~\cite{Bedjidian:2004gd} for more detail.
However, such tuning may change the relative contributions of distinct
topologies from the same initial state in PYTHIA relative to a NLO calculation.
Double counting of these processes is avoided
by requiring that the hard scattering should be of greater virtuality than the
parton shower \cite{Bedjidian:2004gd}.  The parton showers also effectively
provide a leading-log resummation of light emissions while keeping the pair
distributions finite over all phase space.

POWHEG, a NLO generator using PYTHIA for hadronization and decay \cite{POWHEG},
does not separate these topologies in the same way that PYTHIA does, all
diagrams, with their interference terms, are included.  In Ref.~\cite{LHCb},
they conclude that, because POWHEG and PYTHIA both describe
the data, NLO effects on $b \overline b$ production are small.
They also note that, aside from the $| \Delta \phi^*|$ distributions,
the data are consistent with uncorrelated $b \overline b$ production.  They
reach this conclusion by suggesting that gluon splitting is a small
contribution to $b \overline b$ production.  However, it is not feasible to
separate this diagram from all other NLO contributions because it
interferes with the amplitudes of other $gg$ diagrams.

The conclusion the NLO contributions to $b \overline b$ production are small,
reached by the LHCb Collaboration in Ref.~\cite{LHCb}, can be
tested by comparison to a NLO calculation of both the $b \overline b$ and
$J/\psi J/\psi$ final states.  This comparison, in Sec.~\ref{sec:calcs}, is at
the center of this work.

\section{Model Description}
\label{sec:model}

The calculations here, using the HVQMNR code \cite{MNR} designed to calculate
$Q \overline Q$ pair production at NLO, follow those outlined
in Ref.~\cite{QQazi}.
The bottom quark mass, $m_b$, factorization
scale, $\mu_F$,
and renormalization scale $\mu_R$ and their uncertainties were set by comparison
to the $b \overline b$ total cross section data with $m_b = 4.65 \pm 0.09$~GeV,
$\mu_F/m = 1.40^{+0.77}_{-0.49}$, and $\mu_R/m = 1.10^{+0.22}_{-0.20}$ \cite{QQazi}.

Hadronization was accomplished through the use of the Peterson fragmentation
function \cite{Peterson} and $k_T$ broadening.
The value of $\epsilon_P$, the Peterson function
parameter, was set by comparison to the FONLL $B$ meson $p_T$ distribution in
Ref.~\cite{QQazi} while $\langle k_T^2 \rangle$, the average broadening, was
fixed previously by comparing the measured $\Upsilon$ $p_T$ distributions to
calculations of $\Upsilon$ production in the color evaporation
model.  Here, for $b \overline b$ production,
\begin{eqnarray}
\epsilon_p & = & 0.0004 \, \, , \label{eq:epsP} \\
\langle k_T^2 \rangle & = & 1 + \frac{\Delta}{3}\ln \left(
  \frac{\sqrt{s}}{20\, {\rm GeV}} \right) \, \, {\rm GeV}^2 \, \, ,
  \label{eq:kt2}
\end{eqnarray}
where the parameter $\Delta$ was introduced to study the sensitivity of the
azimuthal correlations to the amount of broadening.  The value $\Delta = 1$ is
the default value \cite{QQazi}, resulting in
$\langle k_T^2 \rangle  \approx 3$~GeV$^2$ for $\sqrt{s} = 7$~TeV.
For further details on the determination of
$\epsilon_P$ and the sensitivity of the $Q \overline Q$ results to the magnitude
of $\langle k_T^2 \rangle$, see Ref.~\cite{QQazi}.

Note that it is necessary to use a code such as the exclusive HVQMNR calculation
because the $Q \overline Q$ pair quantities are calculable in such an approach
while only single inclusive quark distributions are so far
available with the FONLL
\cite{FONLL} and generalized mass - variable flavor number approaches 
\cite{GMVFN,Helenius:2018uuf}.

The HVQMNR code \cite{MNR} uses negative weight
events to cancel divergences numerically.  Without a
$k_T$ kick there can be a mismatch in the cancellation, leading to a negative
value of the cross section for pair $p_T$ at $p_{T_p} = 0$ and azimuthal
separation at $\phi = \pi$, as can be seen in some of the $b \overline b$
distributions with $\langle k_T^2 \rangle = 0$.  Smearing the parton
momentum through the introduction of intrinsic transverse momenta, $k_T$,
reduces the importance of the negative weight events at low $p_T$.

HVQMNR
does not include any resummation; the broadening plays this role in the code
at low $p_T$ \cite{CYLO}.  Open charm results
at fixed-target energies required
transverse momentum broadening to obtain agreement with the
data after fragmentation was applied \cite{MLM1}.  Such broadening was also
used as a proxy for resummation in Drell-Yan production, 
see {\it e.g.} Refs.~\cite{CYLO,APP,CGreco}.  

In HVQMNR, the kick is added in the
final state using the Gaussian function $g_p(k_T)$ \cite{MLM1},
\begin{eqnarray}
g_p(k_T) = \frac{1}{\pi \langle k_T^2 \rangle} \exp(-k_T^2/\langle k_T^2
\rangle) \, \, ,
\label{intkt}
\end{eqnarray}
which multiplies the parton
distribution functions, 
assuming the $x$ and $k_T$ dependencies
factorize.  As explained in Ref.~\cite{MLM1},
it does not matter whether the $k_T$ dependence is added in the initial or
final state as long as the kick is not too large.

The application of the $k_T$ is now described.
The $Q \overline Q$ system is first boosted to the rest frame
from its longitudinal center-of-mass frame.  The intrinsic transverse
momenta of the incoming partons, $\vec k_{T 1}$ and $\vec k_{T 2}$, are chosen
at random with $k_{T 1}^2$ and $k_{T 2}^2$ distributed according to
Eq.~(\ref{intkt}).   The quarks are then boosted out of the pair rest frame,
changing the initial transverse momentum of the hard scattering
from $\vec p_T$ to $\vec p_T + \vec k_{T 1} + \vec k_{T 2}$.  While the
$k_T$ is here applied to the $Q \overline Q$ pair, it could have alternatively
been given to the entire final-state system, including the light parton in
$2 \rightarrow 3$ processes, as if it were applied directly to the initial
state.  The two methods of introducing $k_T$ are equivalent if the
calculation is LO but at NLO a light parton in the final state can make
the correspondence inexact.  

The effect of a $k_T$ kick on $p_T$-related distributions ($p_{T_p}$, $M$, $A_T$)
should decrease as $\sqrt{s}$ increases because $\langle p_T \rangle$
also increases with $\sqrt{s}$.  Because $\langle k_T^2 \rangle$ is assumed to
increase with $\sqrt{s}$, see Eq.~(\ref{eq:kt2}), the effect is most important
at low $p_T$.  The effect of a $k_T$ kick also decreases with increasing quark
mass, as shown in Ref.~\cite{QQazi}, requiring a larger $\langle k_T^2 \rangle$
for bottom quarks relative to charm quarks
to have a non-negligible effect on bottom production at higher energies.

Although LHCb suggested in Ref.~\cite{LHCb} that the similarity of the PYTHIA
and POWHEG simulations are indicative of a small NLO contribution, it is
important to recall that gluon splitting is an
explicit contribution to $gg \rightarrow Q \overline Q X$ at
$\mathcal{O} (\alpha_s^3$) and thus a real NLO contribution, as is flavor
excitation.  As previously discussed, it is not feasible to separate 
individual diagrams since such a procedure no longer allows for interferences
between diagrams.  The LO flavor creation contributions,
$gg \rightarrow Q \overline Q$ and $q \overline q \rightarrow Q \overline Q$,
only produce back-to-back $Q \overline Q$ pairs, a delta function for
$| \Delta \phi^* | = \pi$, $A_T = 0$ and $p_{T_p} = 0$ without broadening.
The NLO contributions are modeled in PYTHIA by flavor excitation and gluon
splitting.  These contributions have a significantly weaker $\Delta \phi$
dependence in PYTHIA.  Flavor excitation is weakly enhanced at
$|\Delta \phi| \approx \pi$ while, since gluon splitting generally produces
collinear $Q \overline Q$ pairs, it results in a weak enhancement at
$|\Delta \phi| \approx 0$ \cite{NorrbinSjo}.
These contributions and summed together with flavor creation at
$\mathcal{O} (\alpha_s^2$), without interference terms but with parton showers.

The introduction of $k_T$
broadening at NLO in HVQMNR softens and widens the peak at
$|\Delta \phi | = \pi $ for $b \overline b$ with a 
finite tail as $|\Delta \phi| \rightarrow 0 $. It does not produce a significant
enhancement at $|\Delta \phi| \rightarrow 0$ as it does for charm pairs at
similar values of $p_T$ because
$\langle k_T^2 \rangle < m_b^2$ while, for charm,
$\langle k_T^2 \rangle \approx m_c^2$ 
\cite{QQazi}.  The effect of $k_T$ broadening also depends strongly on the quark
momentum, as discussed for the
$|\Delta \phi|$ distributions in Ref.~\cite{QQazi}.

Observables related to the rapidity, either the rapidity difference or the pair
rapidity, should be independent of the broadening.  However, the other pair
observables studied by LHCb should be affected by broadening, at least for the
parent $b$ quarks.  Thus the
calculations here compare results with and without $k_T$ broadening on both
the initial $b \overline b$ pairs and the final state $J/\psi J/\psi$ pairs.

\section{Comparison to the LHCb data}
\label{sec:calcs}

In this section, the pair quantities, $|\Delta \phi|$, $|\Delta y|$, $y_p$,
$A_T$, $p_{T_p}$ and $M$ are calculated and compared to the LHCb data.  The
$b \overline b$ pair distribution include both fragmentation and $k_T$
broadening as described in the previous section.  The $J/\psi J/\psi$
pair distributions are calculated with the
$B \rightarrow J/\psi X$ decay with a
1.094\% branching ratio \cite{BtoJpsi_decay}.

All results are shown for the minimum $p_T$ cuts of
2, 3, 5 and 7~GeV on the $J/\psi$ and the parent $B$ meson.  Note that the
$J/\psi$s from $B$ decay would, of course, generally arise from parent $B$
mesons with $p_T$ larger than those of the final-state $J/\psi$.

The LHCb data are also shown for comparison
on each plot.  All quantities are divided by the total cross section,
$(1/\sigma)(d\sigma/dX) \equiv d \ln \sigma/dX$ where $X$ denotes the
observables on the $y$-axes of the plots,
so that the 7 and 8~TeV LHCb measurements can be combined for improved
statistics.  Note that even though the LHCb data shown here are from the
$\sqrt{s} = 7$ and 8~TeV runs combined, the
$\approx 15$\% difference in $\sqrt{s}$ between the two data sets gives only a
1-2\% change in $\langle k_T^2 \rangle$ based on Eq.~(\ref{eq:kt2}).  Given the
small change in $\langle k_T^2 \rangle$ for $p+p$ collisions and the uniform
shapes of $d\ln \sigma/dX$, the calculations compared to the data in this
section are all done for $\sqrt{s} = 7$~TeV.  It was verified that the
normalized pair distributions calculated here remain unchanged at $\sqrt{s} = 7$
and 8~TeV in $p+p$ collisions.

Note that, in
the calculations, the $J/\psi$s from $B$ decays
have lower statistics than the parent $B$ mesons, especially as the minimum
$p_T$ increases.  Thus red histograms are generally used to represent the
$J/\psi$ pair quantities while black curves are used for the $b \overline b$
pair distributions.  The LHCb results for $J/\psi$ pairs are rendered as red
points while the reported $b \overline b$ quantities are given as black points.

\subsection{$| \Delta \phi|$ and $|\Delta \phi^*|$}

\begin{figure}[htpb]\centering
\begin{tabular}{cc}
  \includegraphics[width=0.5\columnwidth]{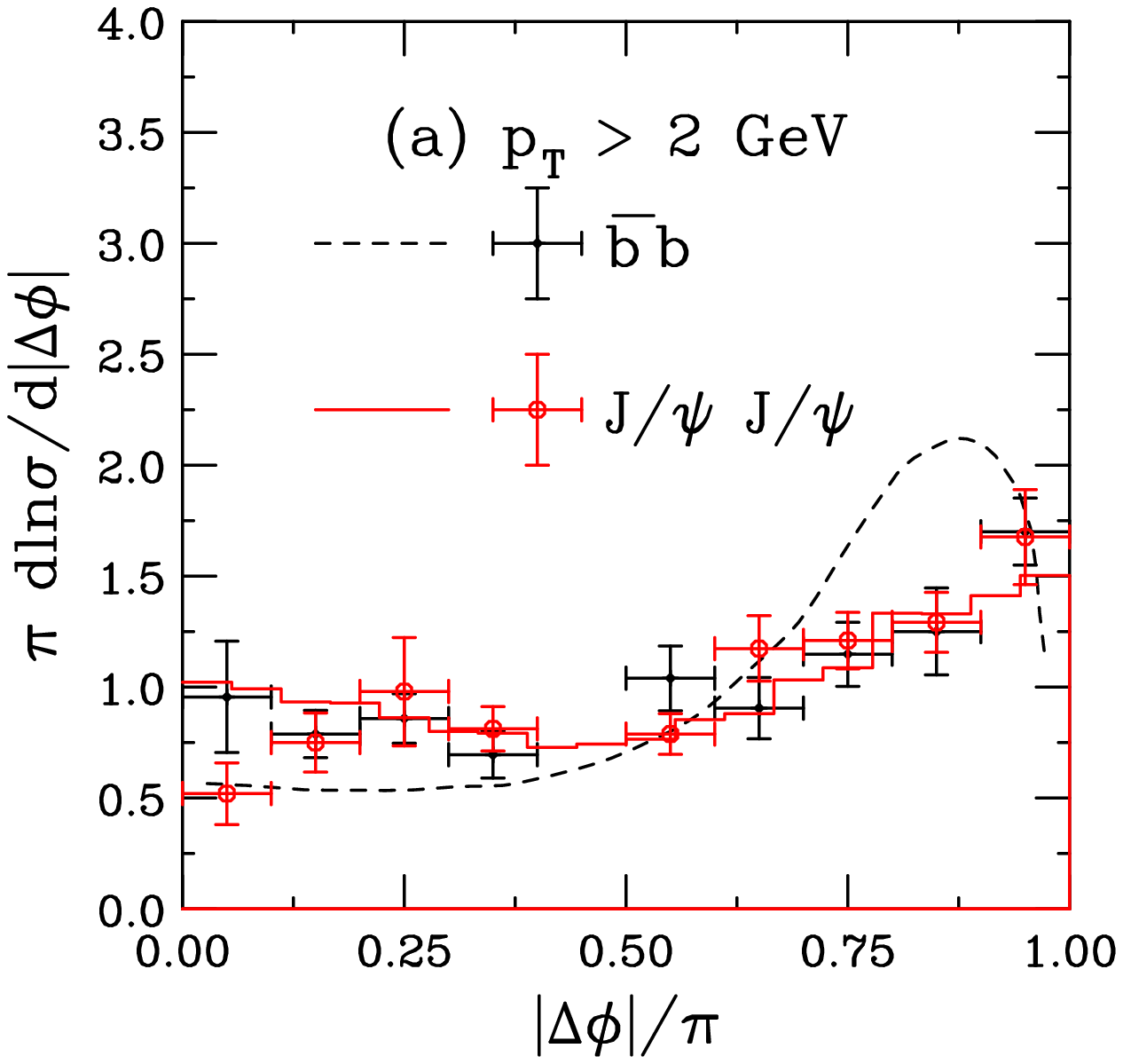} &
  \includegraphics[width=0.5\columnwidth]{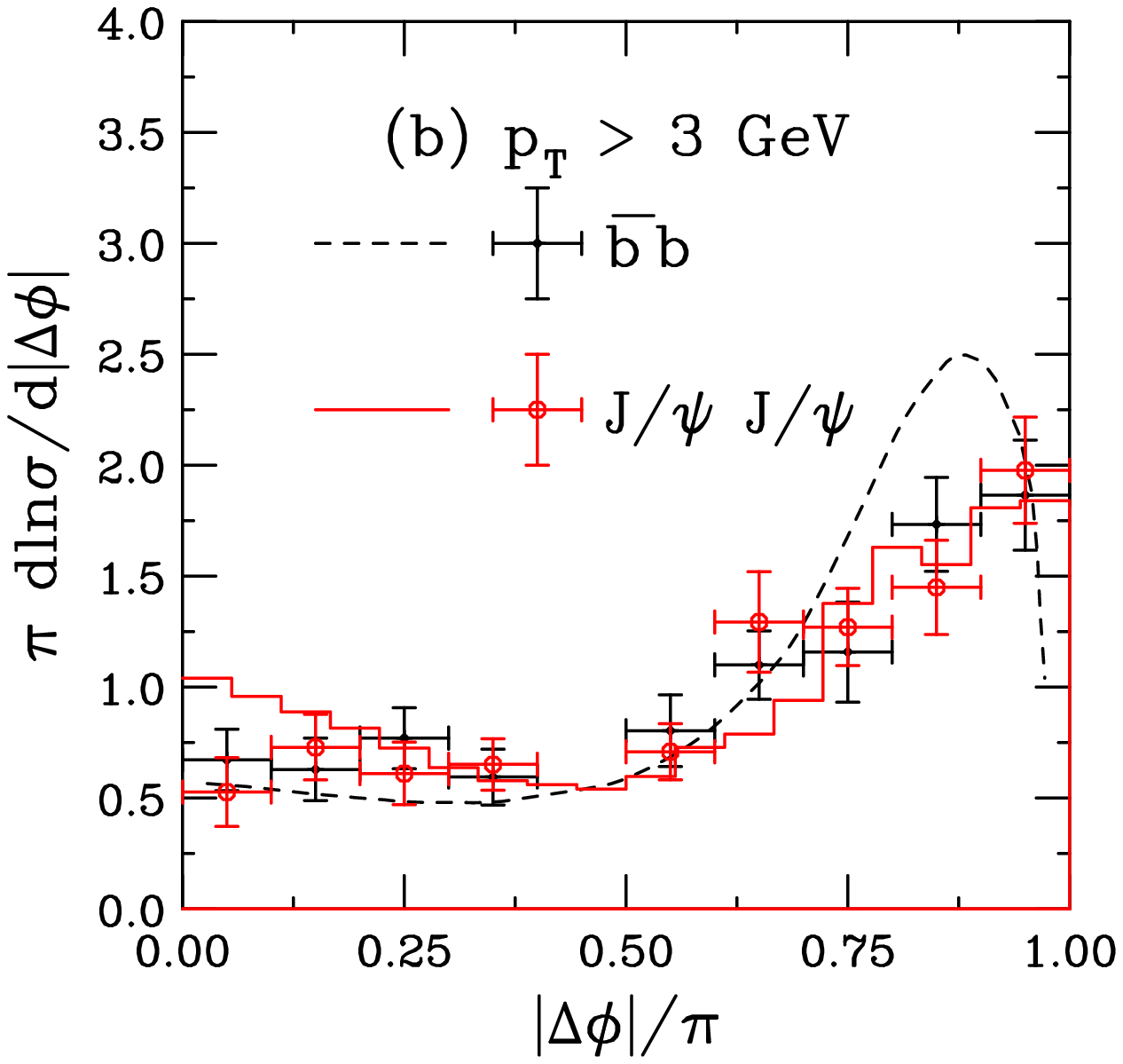} \\
  \includegraphics[width=0.5\columnwidth]{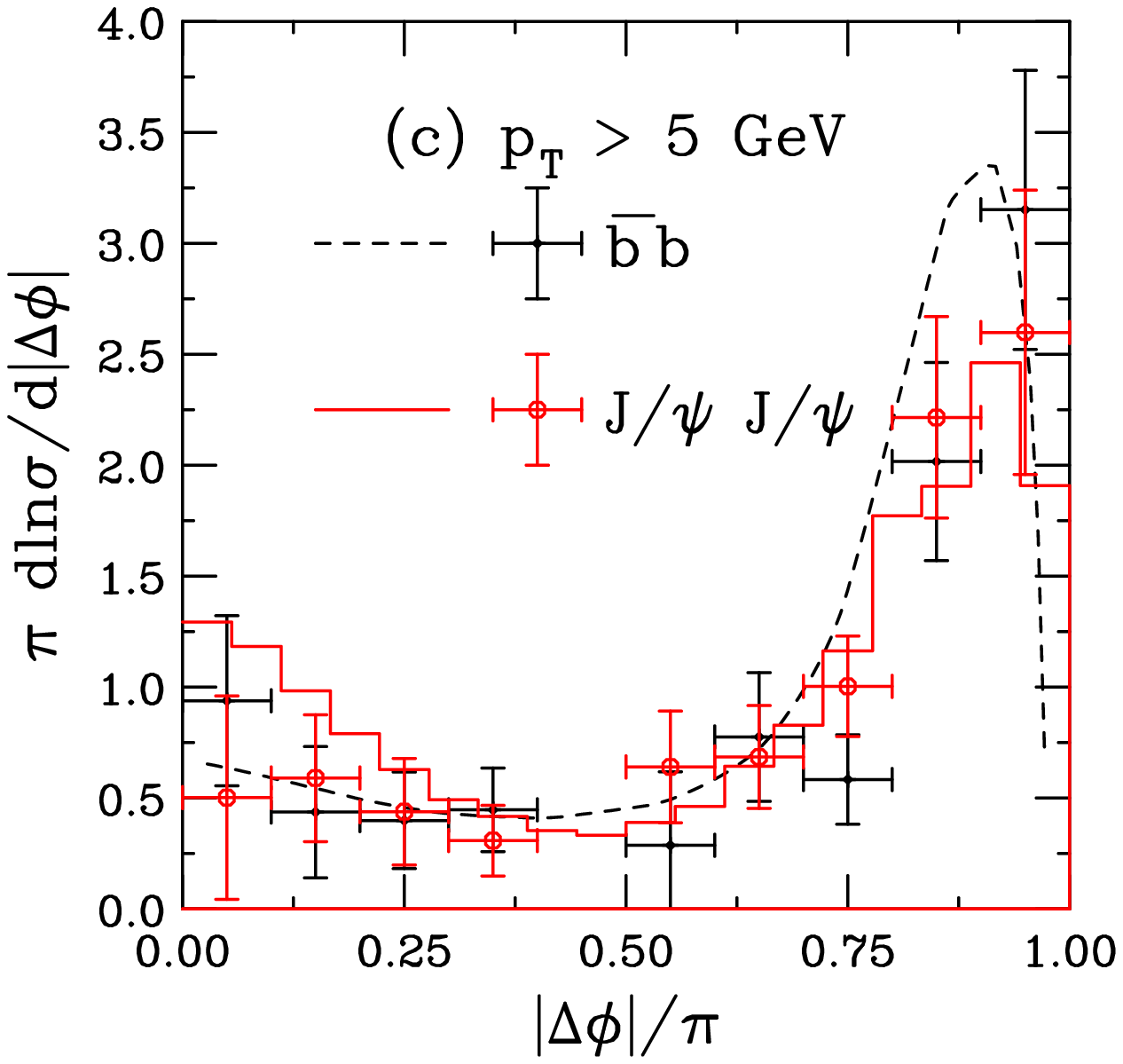} &
  \includegraphics[width=0.5\columnwidth]{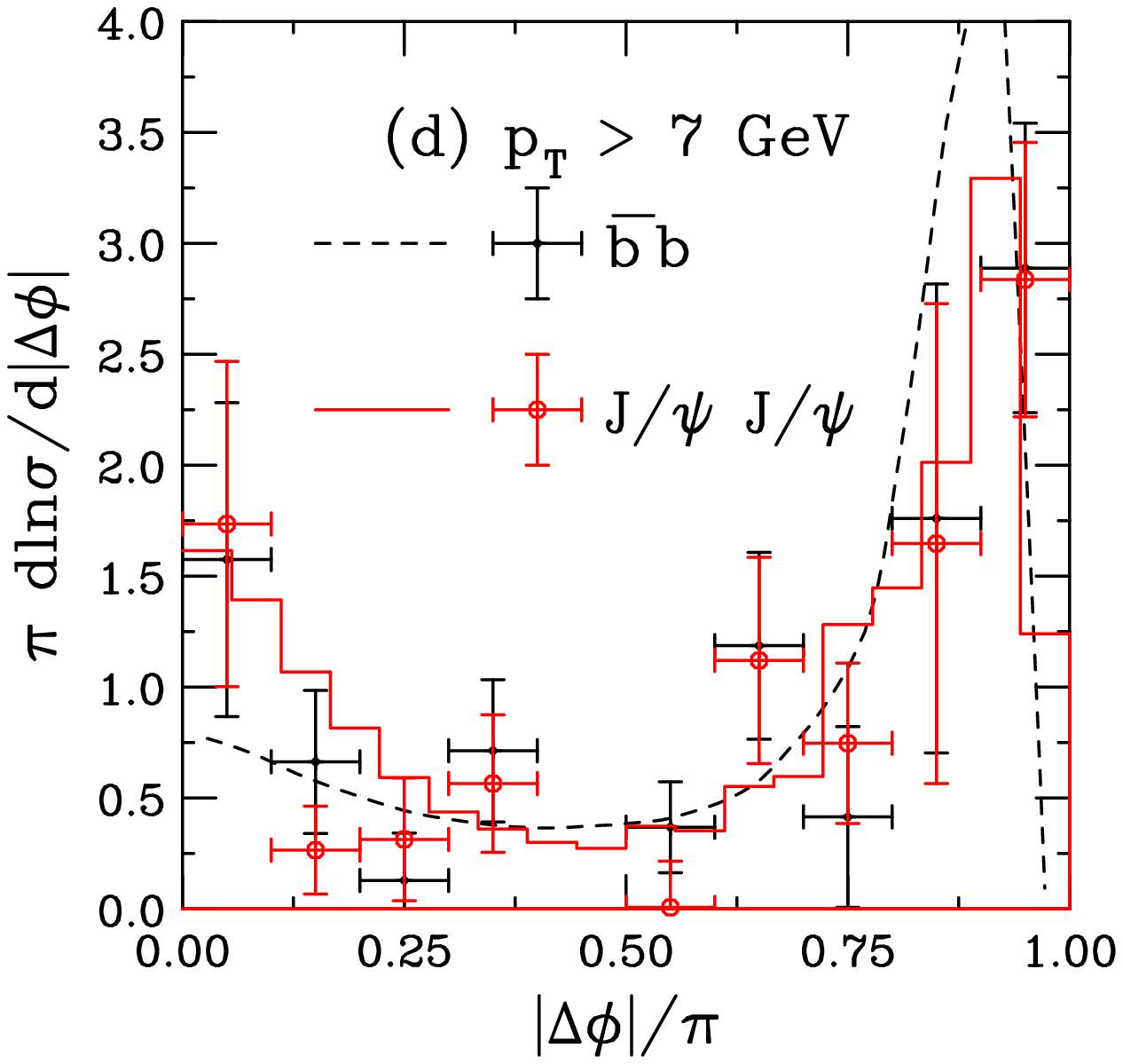} \\
\end{tabular}
  \caption[]{(Color online) The azimuthal angle difference between the $b$ and
    $\overline b$ (black dashed curves)
    and the $J/\psi$'s resulting from the bottom
    quark decays (red histograms) are shown compared to the LHCb data
    \protect\cite{LHCb} (black for $b \overline b$,
    red circles for $J/\psi$ pairs)
    for the $p_T$ cuts
    on the $b$ quarks and the $J/\psi$ of 2 (a), 3 (b), 5 (c) and 7~GeV (d).
  }
  \label{fig_azi}
\end{figure}

LHCb presented $|\Delta \phi |$ distributions for both the initial $B$ meson
pair, reported as $b \overline b$ in Fig.~\ref{fig_azi} and in all the figures
in this section,
and the $J/\psi$ pairs.  Recall that LHCb estimated the azimuthal angle of each
$B$ meson from the direction of the vector from the primary vertex to the
$J/\psi$ decay vertex.  They also determined the azimuthal angles for the
$J/\psi$s individually.  As shown in Fig.~\ref{fig_azi}, the
$|\Delta \phi^*|$ and $|\Delta \phi|$
distributions for $b \overline b$ and $J/\psi J/\psi$ respectively are
compatible with each other within the uncertainties.  

The $b \overline b$ $|\Delta \phi^*|$ distribution has a peak slightly below
$|\Delta \phi^*| \approx \pi$ with a flatter distribution as
$|\Delta \phi^*| \rightarrow 0$
relative to that of the $J/\psi$ pair.  As the minimum $p_T$ grows, the peak
near back-to-back ($|\Delta \phi^*| \approx \pi$) grows higher and becomes narrower
for the $b \overline b$ pairs.  Likewise, the distribution at
$|\Delta \phi^*| \approx 0$
starts to increase from approximately flat at low $|\Delta \phi^*|$ to a slight
enhancement that becomes more pronounced with increasing minimum $p_T$.
This is because that, as the minimum $p_T$ grows from 2 to 7~GeV, the
relative values of $\langle k_T^2 \rangle^{1/2}$ and $m_T = \sqrt{p_T^2 + m_b^2}$
change from $m_T/\langle k_T^2 \rangle^{1/2} \approx 3$ to 
$m_T/\langle k_T^2 \rangle^{1/2} \approx 4.9$.  The larger $m_T$ allows the
development of a double-peaked $\Delta \phi^*$ distribution, more closely
connected to diagrams with a high $p_T$ $b \overline b$ pair balanced
against a hard parton in the opposite direction, such as `gluon splitting'.

This trend in the increase of $(\pi/\sigma) d\sigma/d |\Delta \phi |$ can
especially be seen for the lighter mass $J/\psi$ decay products.  In this
case, because $m_{J/\psi}/m_b \approx 2/3$ and the minimum $J/\psi$ $p_T$ is
generally smaller than that of the parent $B$ meson and the $k_T$ kick is
applied to the parent meson, not the $J/\psi$ decay product,
the enhancement seen in the
$b \overline b$ distributions sets in at lower $p_T$ for $J/\psi$ pairs and
is correspondingly larger.  Here $m_{T_\psi} = \sqrt{p_T^2 + m_\psi^2}$
so that $m_{T_\psi}/\langle k_T^2 \rangle^{1/2} \approx 2.1$ to 
$m_{T_\psi}/\langle k_T^2 \rangle^{1/2} \approx 4.4$, assuming
$p_T^{J/\psi} \equiv p_T^B$.  Because the $k_T$
kick is on the bottom quarks as they hadronize rather than on the $J/\psi$
itself, the $p_T$ selected is larger relative to the primary
$B$ hadron so that the enhancement grows faster with minimum
$p_T$ for $J/\psi$ pairs, as shown in Fig.~\ref{fig_azi}.  If a higher minimum
$p_T$ were chosen for $B$ mesons, to more closely match the average $p_T$ of
the $B$ meson producing the minimum $J/\psi$ $p_T$, the enhancement at
$|\Delta \phi| \rightarrow 0$ would
grow larger, closer to matching the peak at $|\Delta \phi | \approx \pi$, as shown
for $c \overline c$ correlations with $p_T > 10$~GeV
in Ref.~\cite{QQazi}.

\subsection{$| \Delta y|$ and $y_p$}

The difference in rapidity, $\Delta y$, (or, in the case of the LHCb
measurement, $\Delta \eta$), was determined both for the initial
$b \overline b$ pairs and the
final-state $J/\psi$ pairs.  The pair rapidity, $y_p$, was only determined for
the $J/\psi$ pairs.  Given the acceptance of the LHCb spectrometer of
$2 < y < 4.5$, the limits on $\Delta y$ is constrained to be in the range
$0 < \Delta y < 2.5$ while
the pair rapidity reported by LHCb lies in the range $2 < y_p < 4.5$.

\begin{figure}[htpb]\centering
\begin{tabular}{cc}
  \includegraphics[width=0.5\columnwidth]{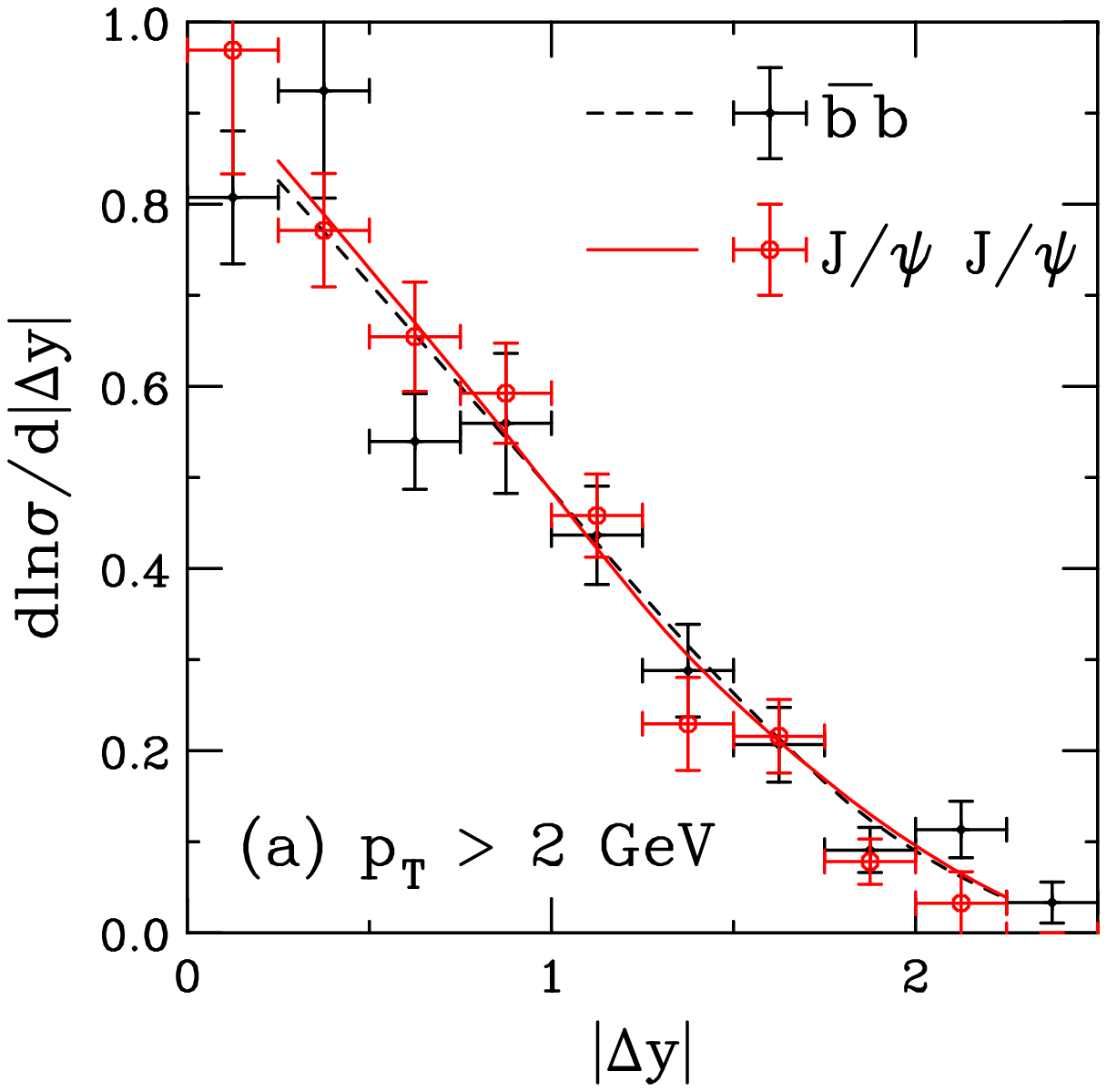} & 
  \includegraphics[width=0.5\columnwidth]{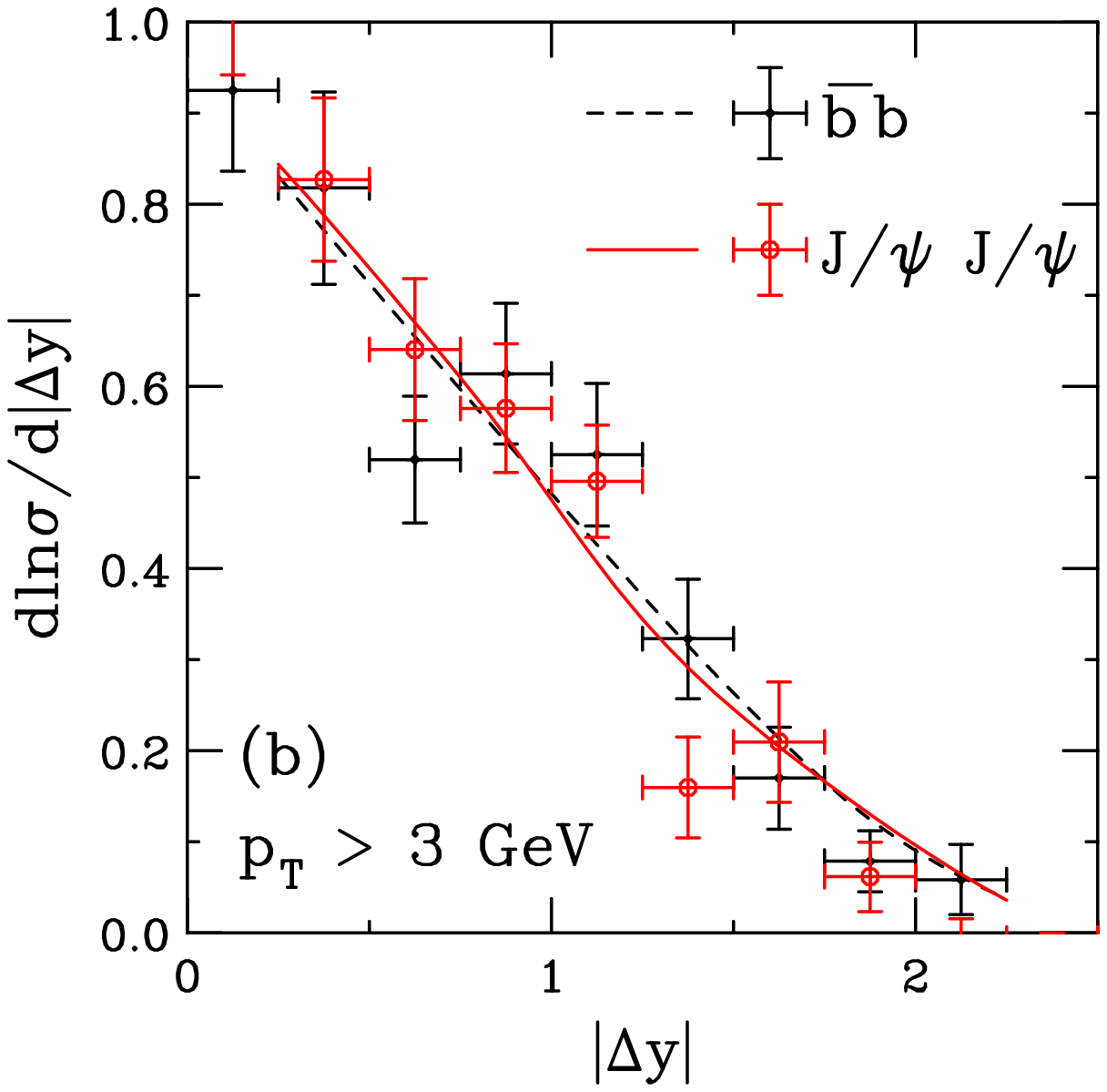} \\
  \includegraphics[width=0.5\columnwidth]{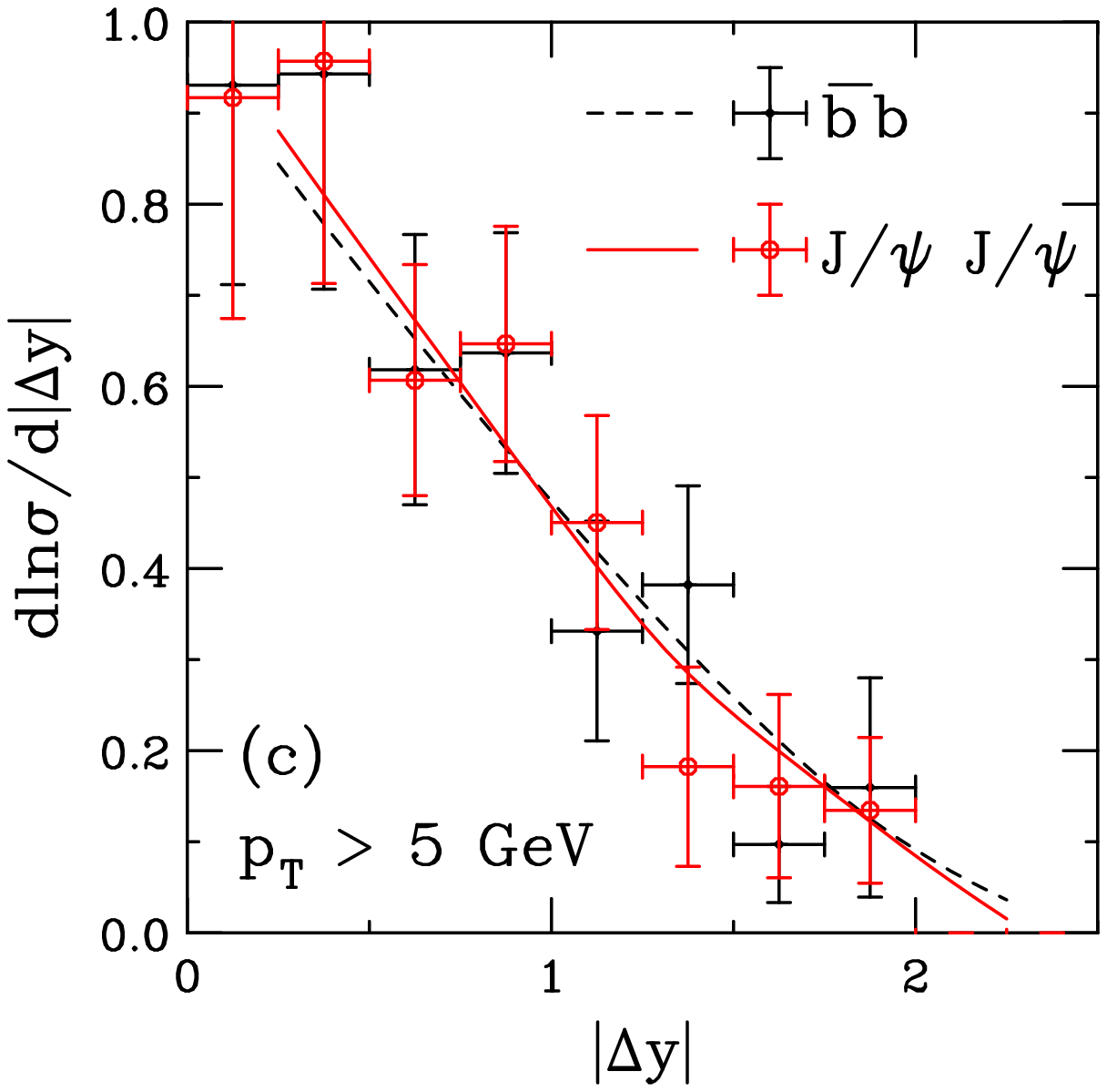} &
  \includegraphics[width=0.5\columnwidth]{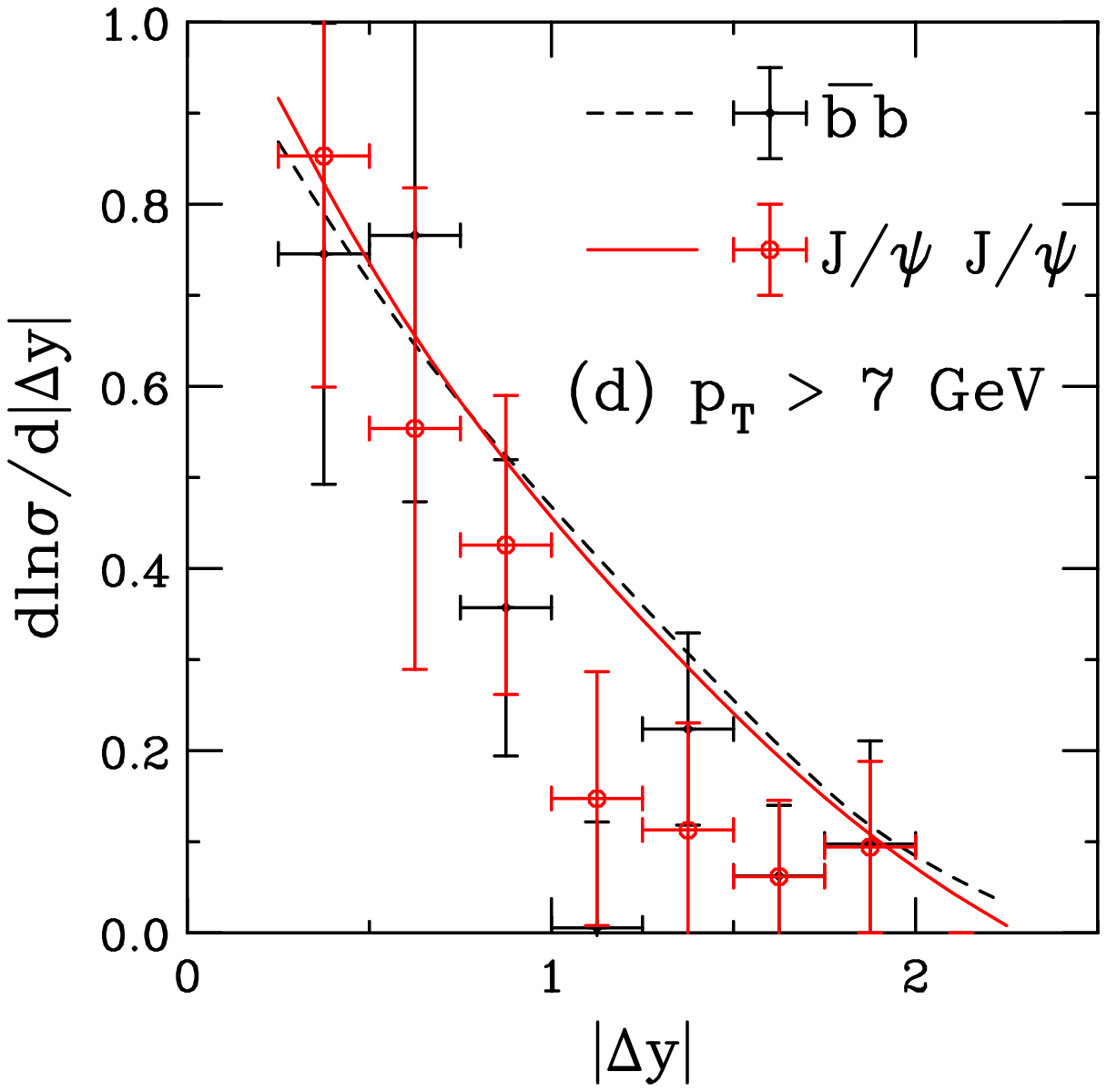} \\
\end{tabular}
\caption[]{(Color online) The rapidity difference $|\Delta y|$
  between the $b$ and
  $\overline b$ (black dashed curve)
  and the $J/\psi$'s resulting from the bottom
  quark decays (red solid curve) are shown compared to the LHCb data
    \protect\cite{LHCb} (black for $b \overline b$,
    red circles for $J/\psi$ pairs) for the $p_T$ cuts
    on the $b$ quarks and the $J/\psi$ of 2 (a), 3 (b), 5 (c) and 7~GeV (d).
  }
  \label{fig_dely}
\end{figure}

As is evident from Fig.~\ref{fig_dely}, the $|\Delta y|$ distribution for
$b \overline b$ and $J/\psi J/\psi$ are in good agreement.  They decrease from
a peak at $|\Delta y | = 0$ to 0 at
$|\Delta y| = 2.5$.  The shape of both distributions is more concave than
linear but is approximately identical for
$b \overline b$ and $J/\psi J/\psi$.  The behavior is also relatively
independent
of the minimum $p_T$.  In the case of the $b \overline b$ pairs, the average
$|\Delta y|$ decreases from 0.75 for $p_T > 2$~GeV to 0.73 for $p_T > 7$~GeV, 
only a 2\% difference.
On the other hand, the average values of $|\Delta y|$ for the $J/\psi$
pairs decreases from 0.74 to 0.69 as the minimum $p_T$ increases from 2 to
7~GeV.  At the highest minimum $p_T$, the average $|\Delta y|$ is reduced by
5\% for $J/\psi$ pairs relative to $b \overline b$ pairs.  The differences,
while not significant, are not zero.

\begin{figure}[htpb]\centering
\begin{tabular}{cc}
  \includegraphics[width=0.5\columnwidth]{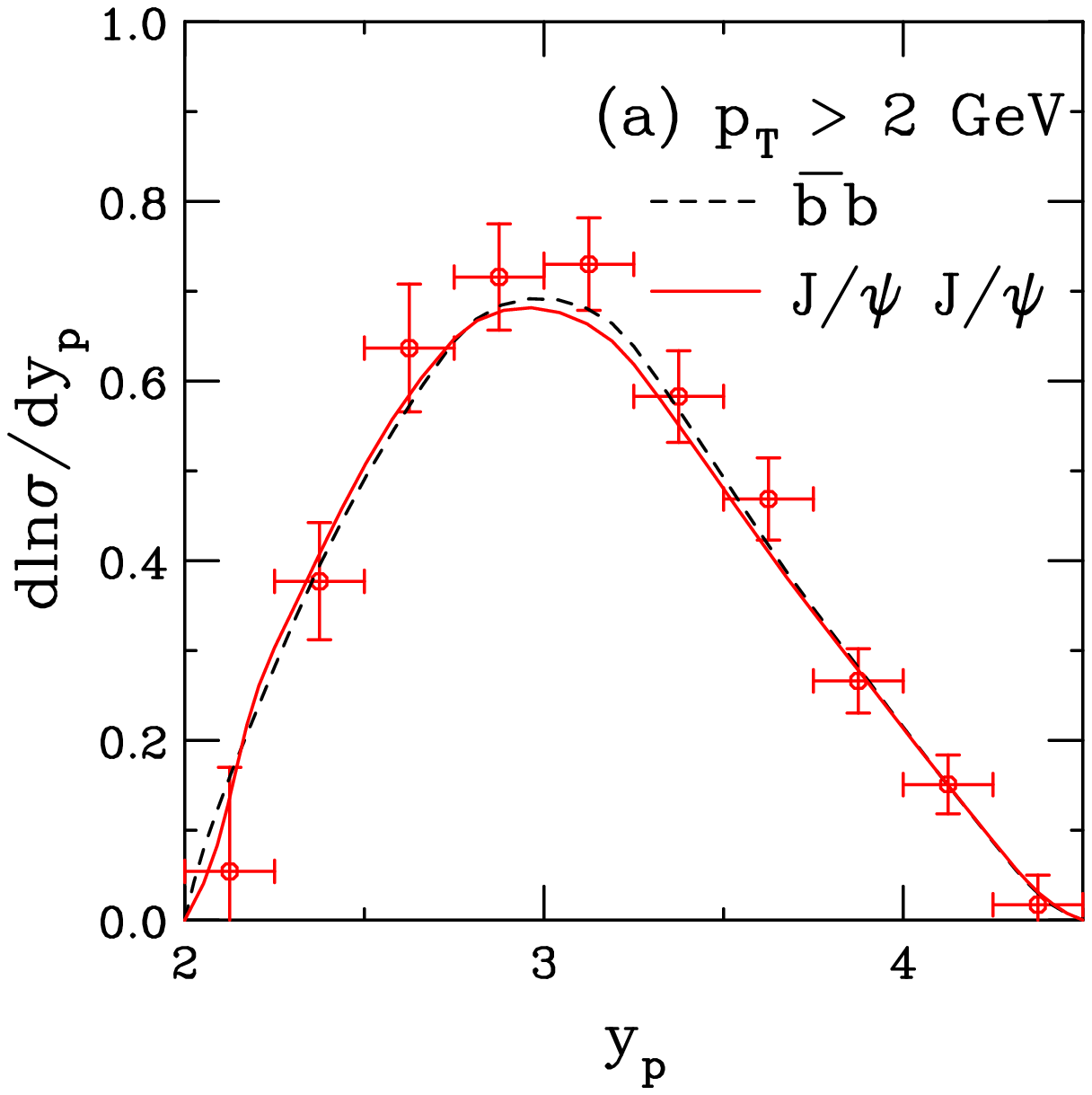} &
  \includegraphics[width=0.5\columnwidth]{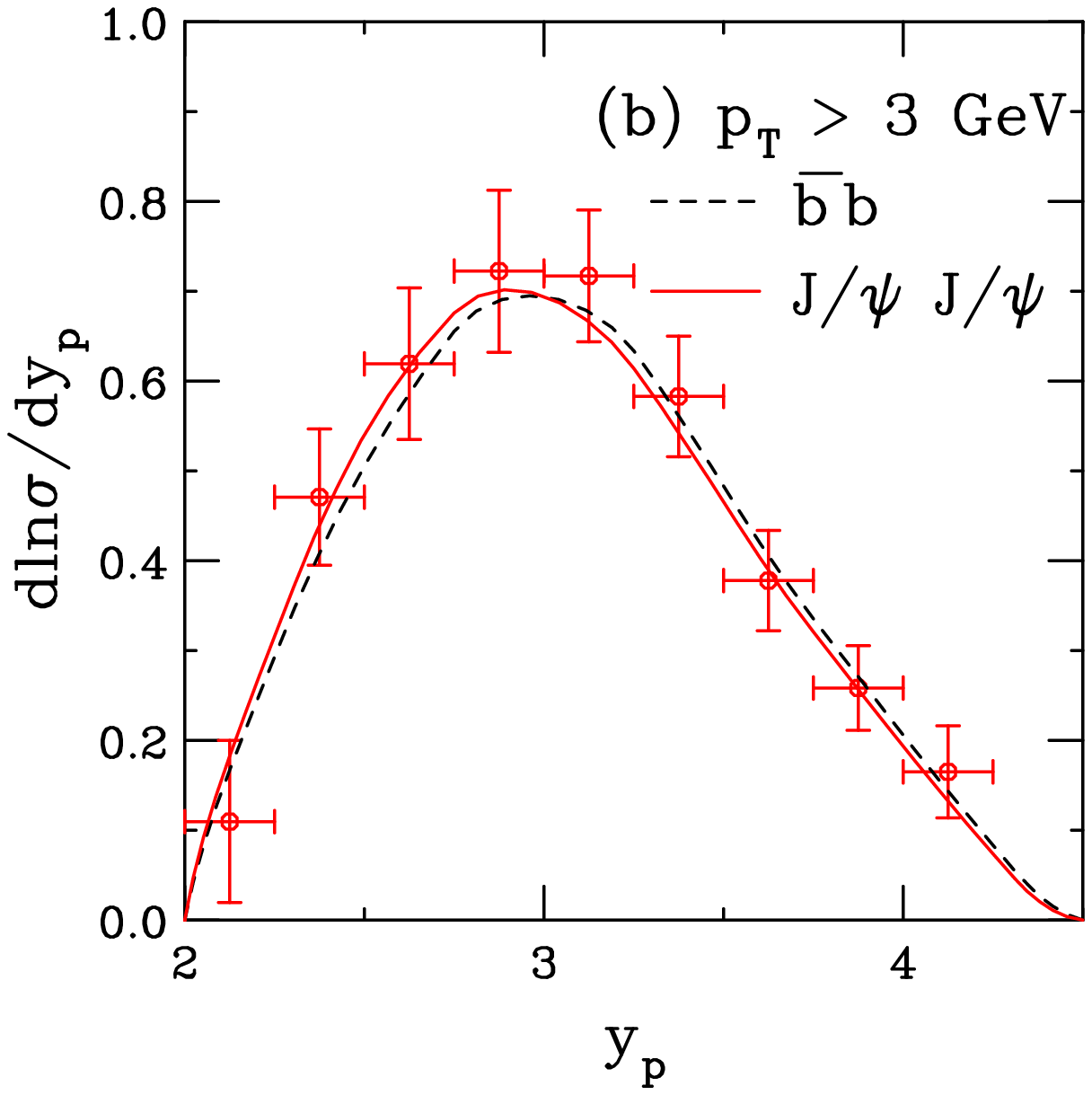} \\
  \includegraphics[width=0.5\columnwidth]{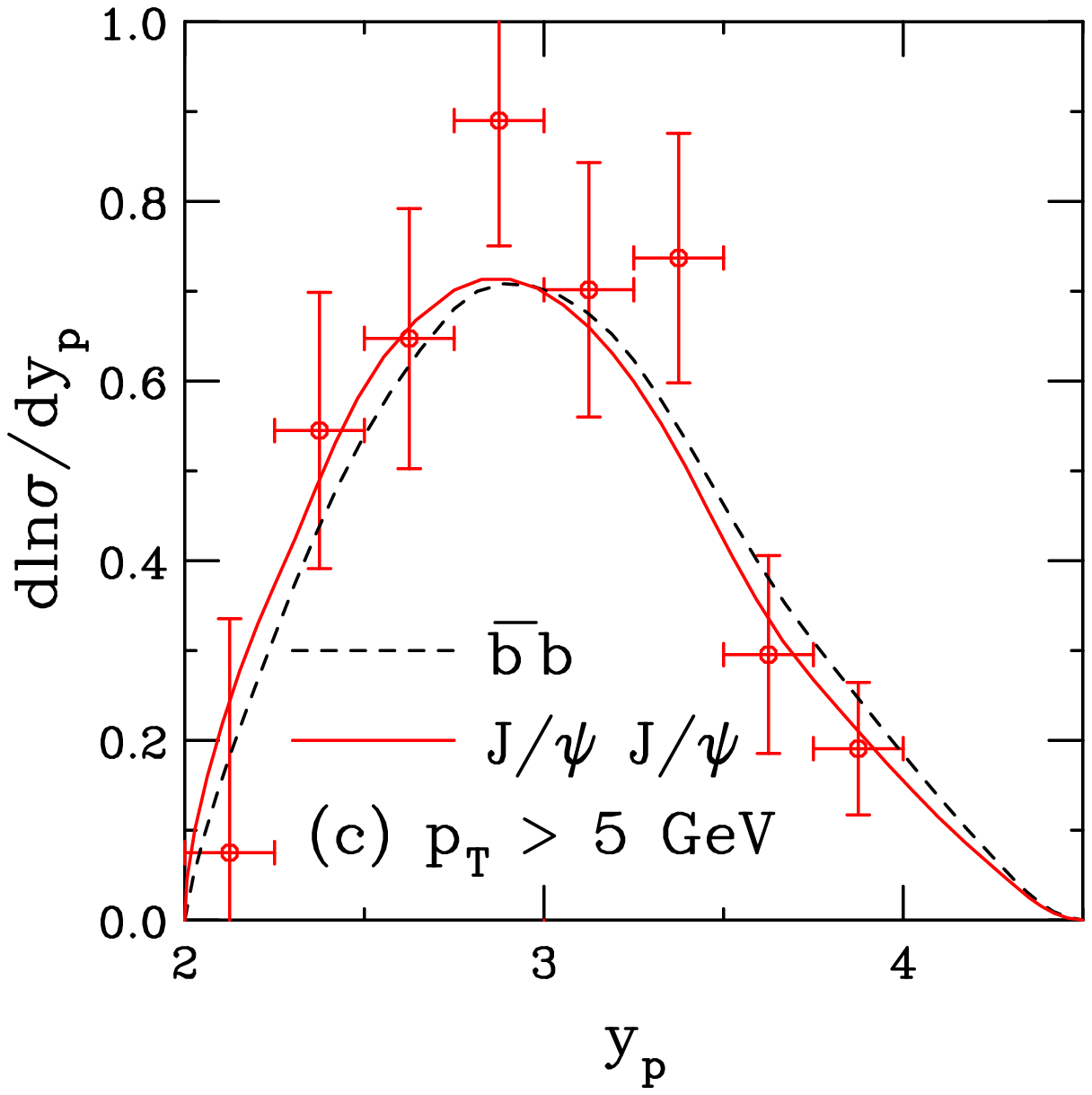} &
  \includegraphics[width=0.5\columnwidth]{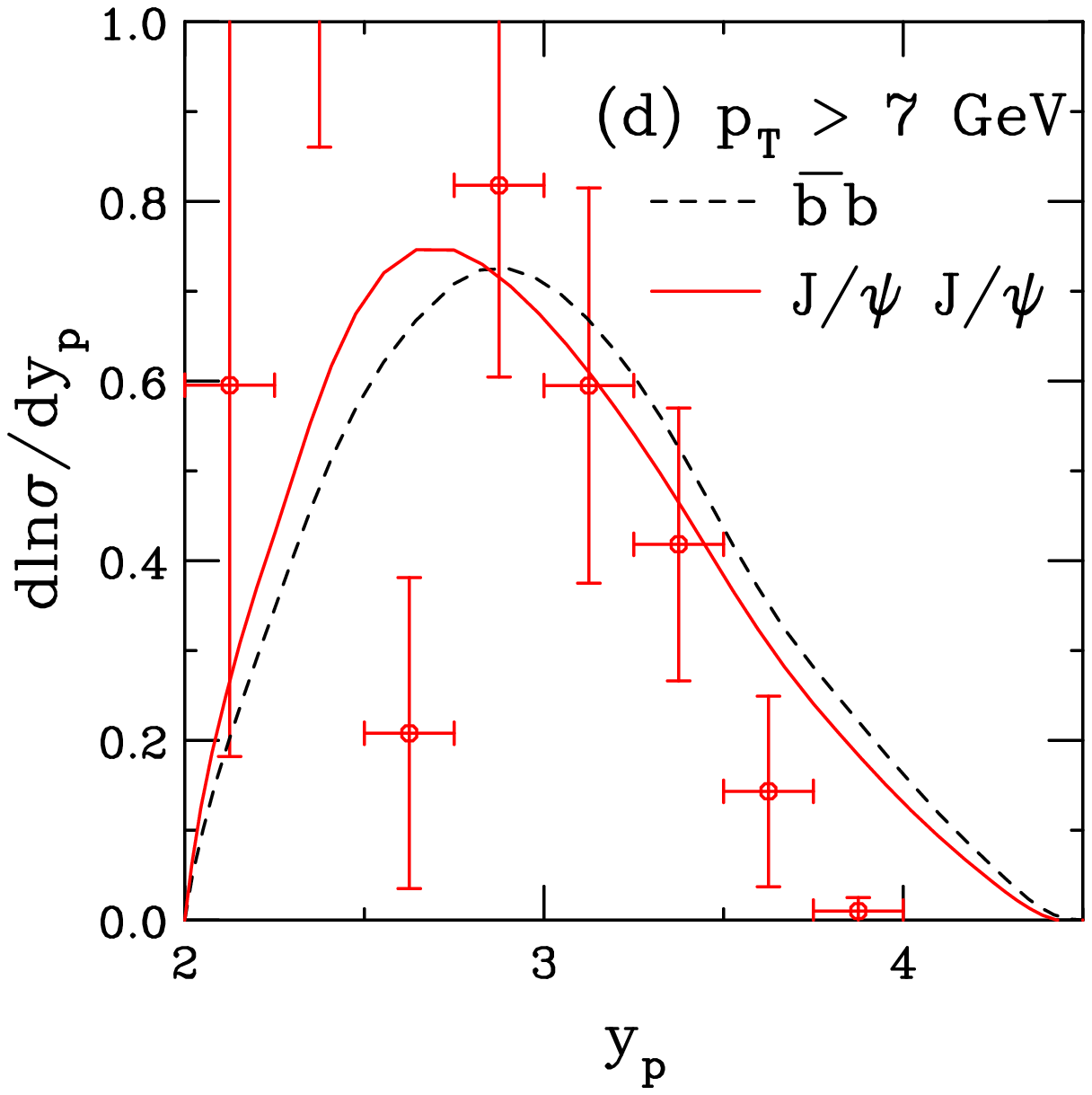} \\
 \end{tabular}
  \caption[]{(Color online) The pair rapidity of the $b$ and
    $\overline b$ (black dashed curves)
    and the $J/\psi$'s resulting from the bottom
    quark decays (red solid curves) are shown compared to the LHCb $J/\psi$
    pair data
    \protect\cite{LHCb} (red circles) for the $p_T$ cuts
    on the $b$ quarks and the $J/\psi$ of 2 (a), 3 (b), 5 (c) and 7~GeV (d).
  }
  \label{fig_yqq}
\end{figure}

The pair rapidity distributions, shown in Fig.~\ref{fig_yqq}, exhibit a
similarly small
decrease in the average $y_p$ with increasing minimum $p_T$, a 2\% decrease
in the average for $b \overline b$ pairs between $p_T > 2$ and $> 7$~GeV, from
3.07 to 3.00 respectively.  There is a 5\% decrease in average $y_p$ for
the $J/\psi$ pairs, from 3.07 with $p_T > 2$~GeV to 2.93 with $p_T > 7$~GeV. 
This small difference on average is sufficient for a small backward shift
between the $b \overline b$ and $J/\psi J/\psi$ curves for $p_T > 7$~GeV, 
especially given the average $p_T$ for the parent $B$ mesons of $J/\psi$s with
the same minimum $p_T$.

\subsection{$A_T$}

\begin{figure}[htpb]\centering
  \begin{tabular}{cc}
  \includegraphics[width=0.5\columnwidth]{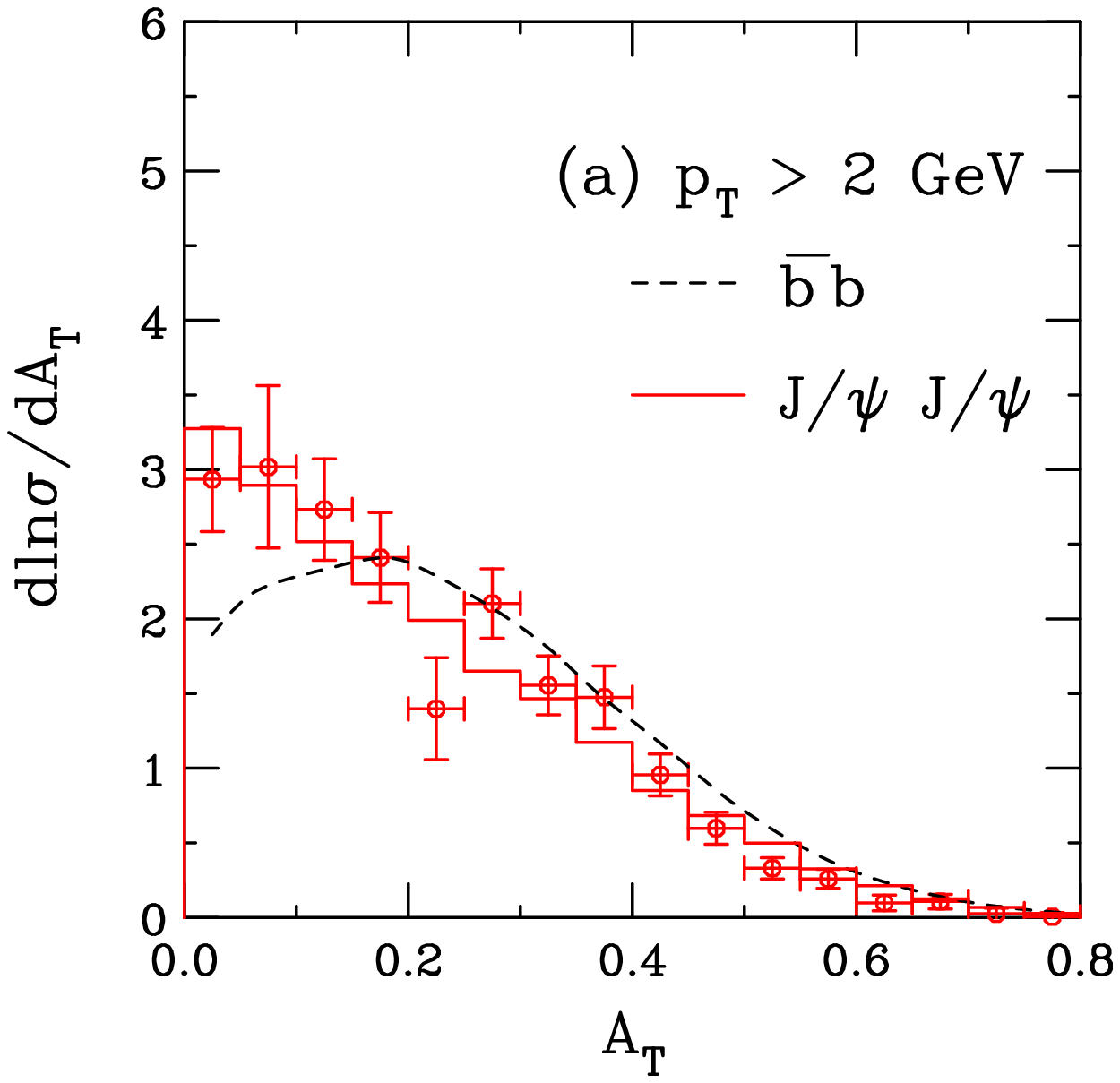} &
  \includegraphics[width=0.5\columnwidth]{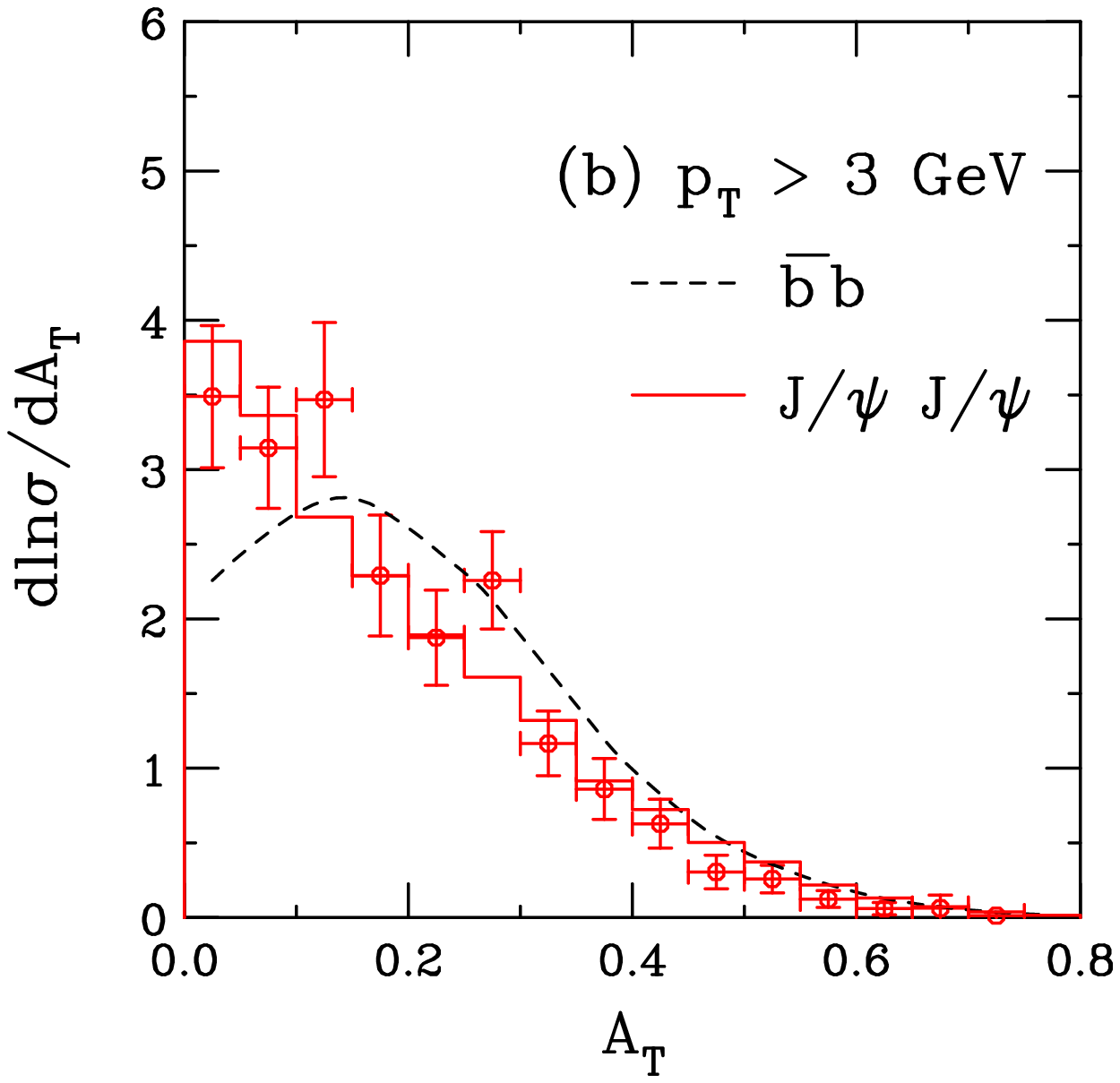} \\
  \includegraphics[width=0.5\columnwidth]{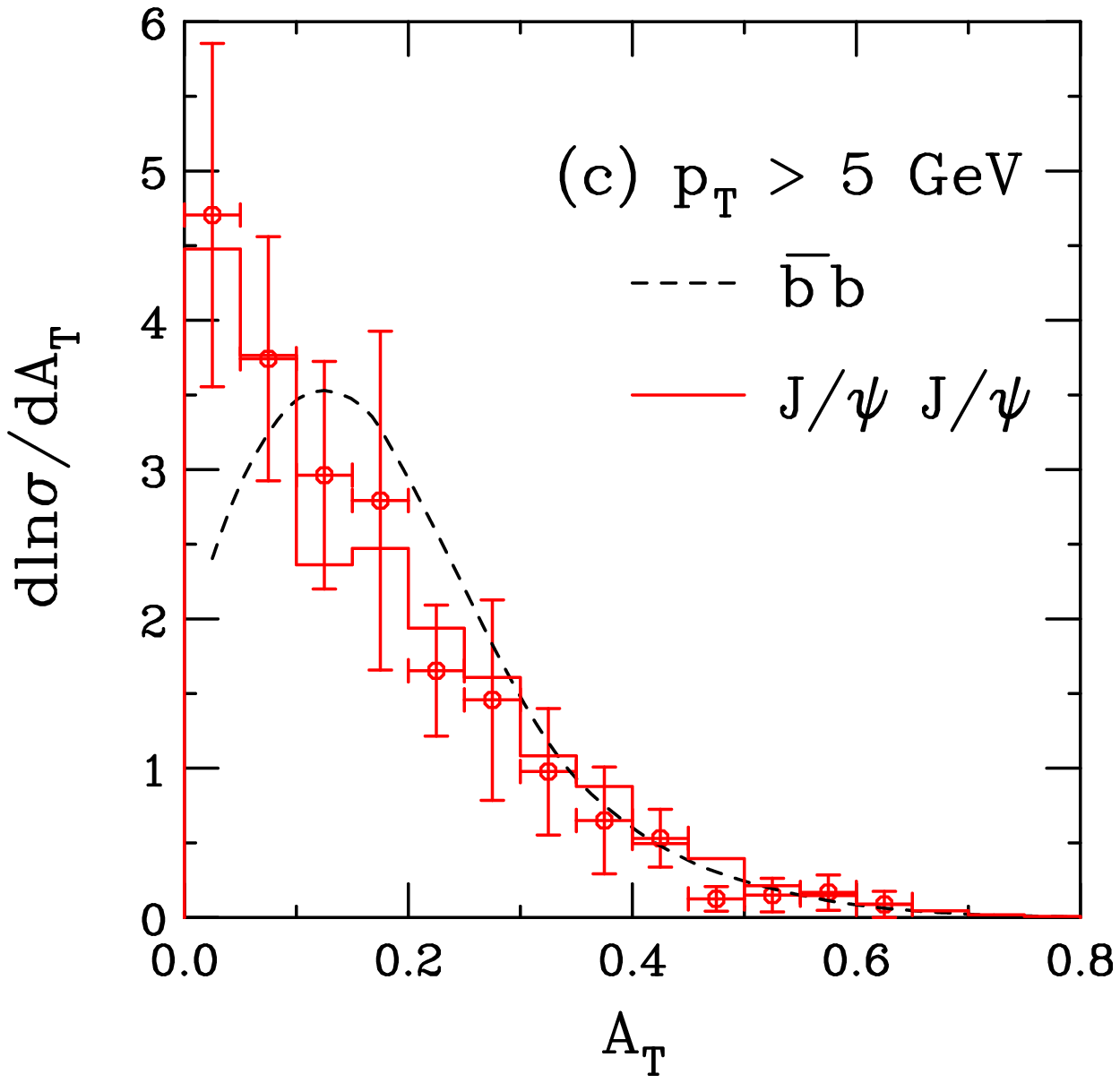} &
  \includegraphics[width=0.5\columnwidth]{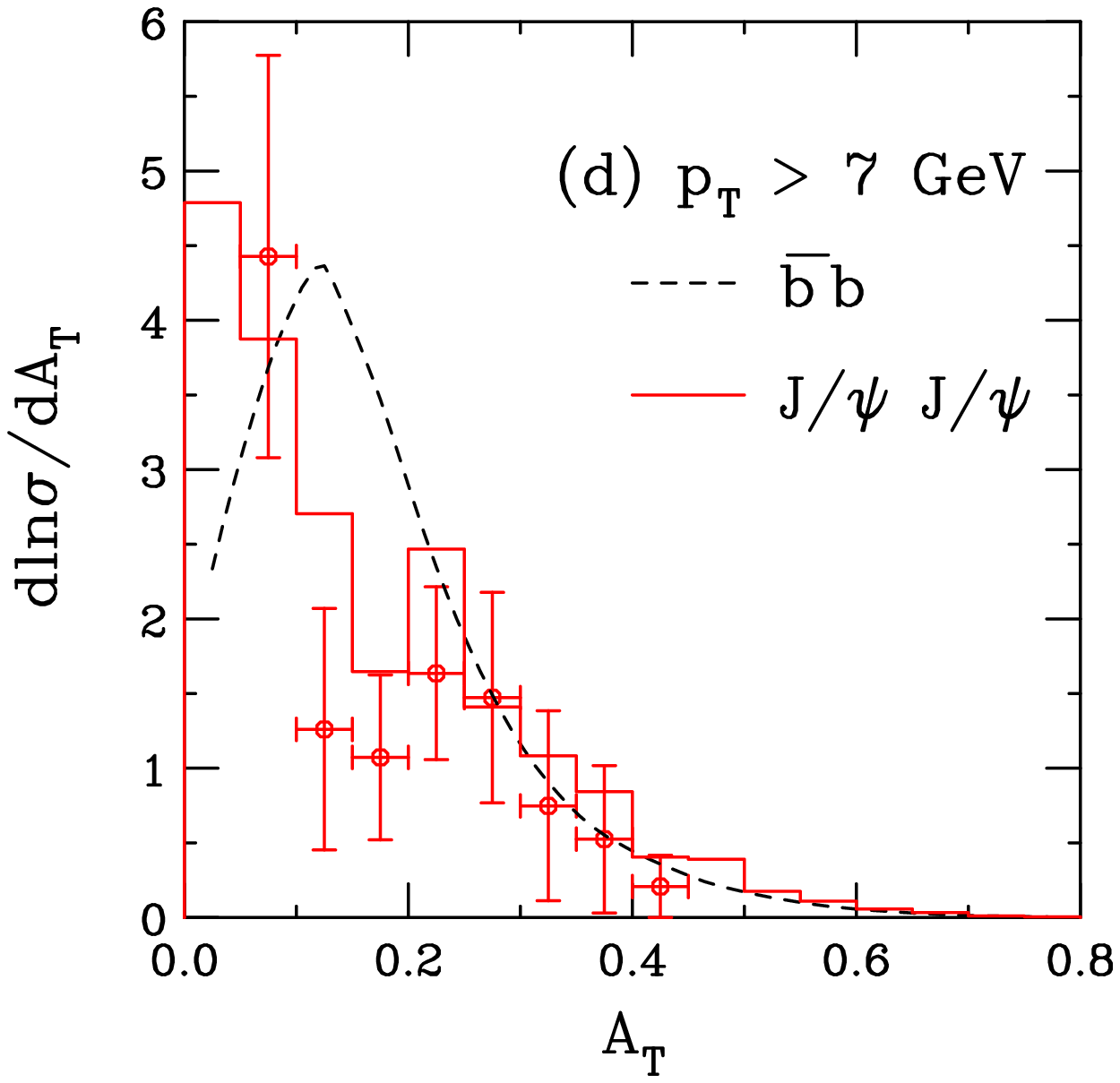} \\
 \end{tabular}
  \caption[]{(Color online) The $p_T$ asymmetry between the $b$ and
    $\overline b$ (black dashed lines) and the
    $J/\psi$'s resulting from the bottom
    quark decays (red histograms) are shown compared to the LHCb $J/\psi$ pair
    data \protect\cite{LHCb} (red circles) for the $p_T$ cuts
    on the $b$ quarks and the $J/\psi$ of 2 (a), 3 (b), 5 (c) and 7~GeV (d).
  }
  \label{fig_asympt}
\end{figure}

The $p_T$ asymmetry, $A_T = |(p_{T1} - p_{T2})/(p_{T1} + p_{T2})|$,
shown in Fig.~\ref{fig_asympt}, would be zero for
$b \overline b$ pairs produced in a back-to-back configuration at leading order.
At NLO, the pairs are no longer
back-to-back and $d\ln \sigma/dA_T$ decreases with increasing $A_T$.  The $A_T$
distribution for the $J/\psi$ pairs is maximal at $A_T = 0$, in accord with the
maximum $|\Delta \phi | \approx \pi$.  This same relation also results in a steeper
$A_T$ distribution for higher minimum $p_T$.  The distribution goes to zero at
$A_T = 1$ if the $p_T$ of one of the $b$ quarks or
$J/\psi$ mesons is very soft or the final states are in alignment.

On the other hand, the $b \overline b$ distributions peak away from $A_T = 0$
due to the inclusion of $k_T$ broadening, as discussed further in
Sec.~\ref{sec:kTcut}.  The peak of the $A_T$ distribution
is at $A_T \approx 0.25$ for $p_T > 2$~GeV.  As the minimum $p_T$
is increased, the distribution for $b \overline b$ pairs becomes narrower
with a higher peak, akin to the $|\Delta \phi^*|$ distributions shown in
Fig.~\ref{fig_azi}.  The average value
of $A_T$ decreases from 0.025 at $p_T > 2$~GeV to 0.17 at $p_T > 7$~GeV.

As previously mentioned, the $J/\psi$ pair $A_T$ distribution is maximum at
$A_T = 0$ instead of a finite $A_T$, as for $b \overline b$.  At lower
minimum $p_T$, the distribution is narrower for the $J/\psi J/\psi$, with an
average of $A_T \approx 0.21$ for $p_T > 2$~GeV.
By the highest minimum $p_T$, $p_T > 7$~GeV, the average is approximately the
same for both, $A_T \approx 0.16$ for $J/\psi J/\psi$.

The trends for the calculated $J/\psi$ pairs are in very good agreement with
the data for all values of minimum $p_T$ studied.  Note also that, above
$A_T \approx 0.4$, the calculated $b \overline b$ and $J/\psi J/\psi$ distributions
are in agreement.

\subsection{$p_{T_p}$ and $M$}

\begin{figure}[htpb]\centering
\begin{tabular}{cc}
  \includegraphics[width=0.5\columnwidth]{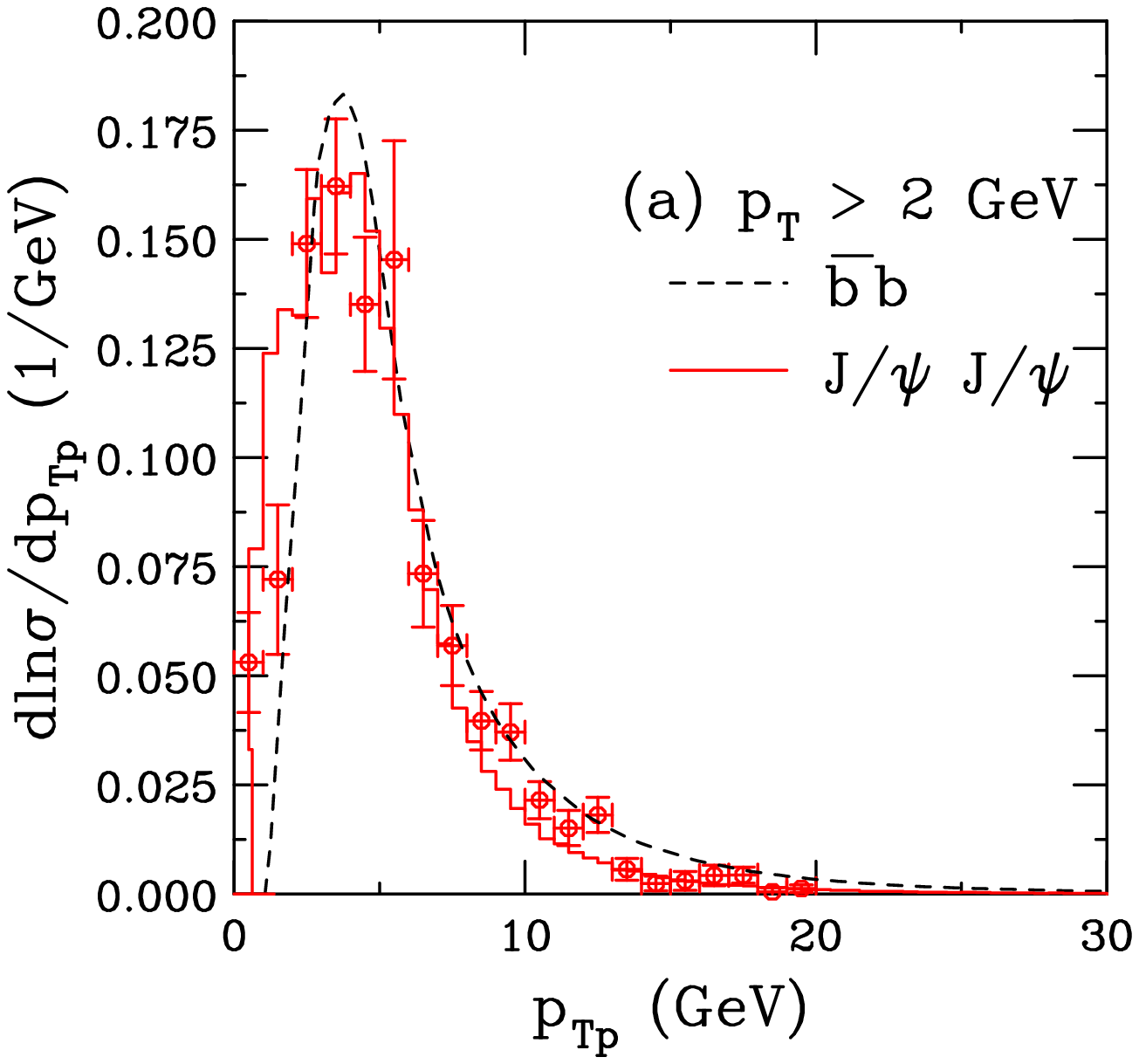} &
  \includegraphics[width=0.5\columnwidth]{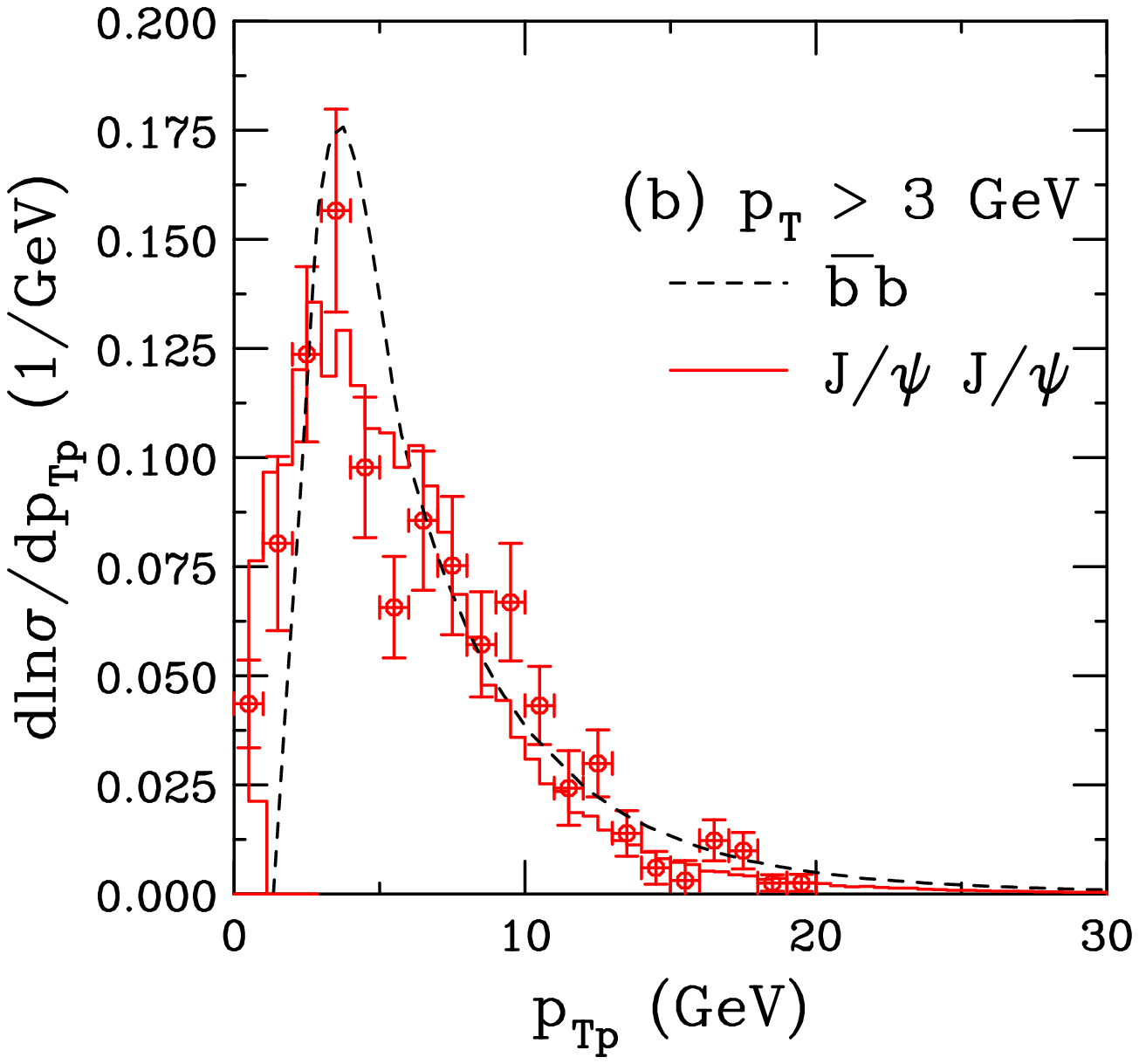} \\
  \includegraphics[width=0.5\columnwidth]{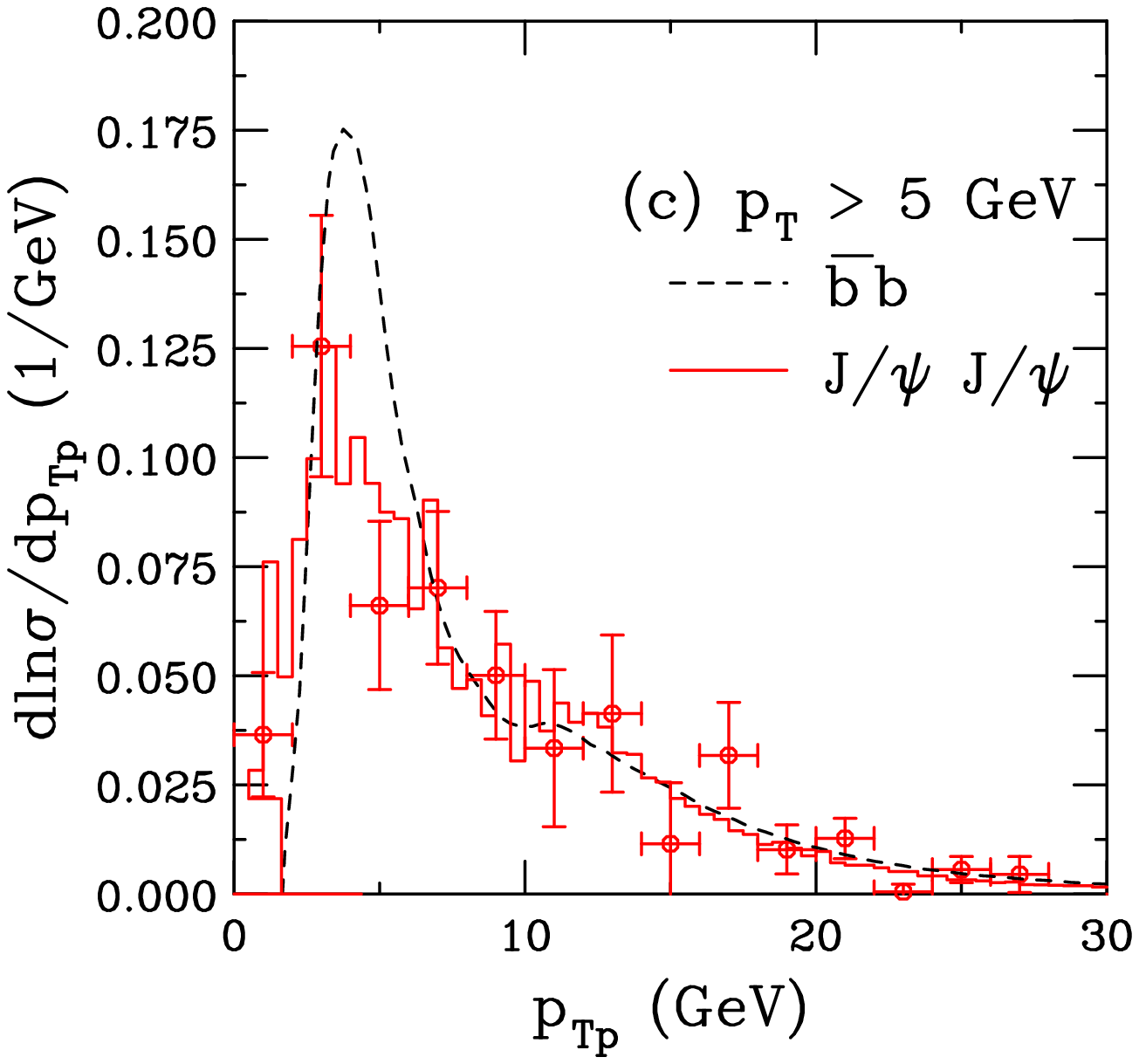} &
  \includegraphics[width=0.5\columnwidth]{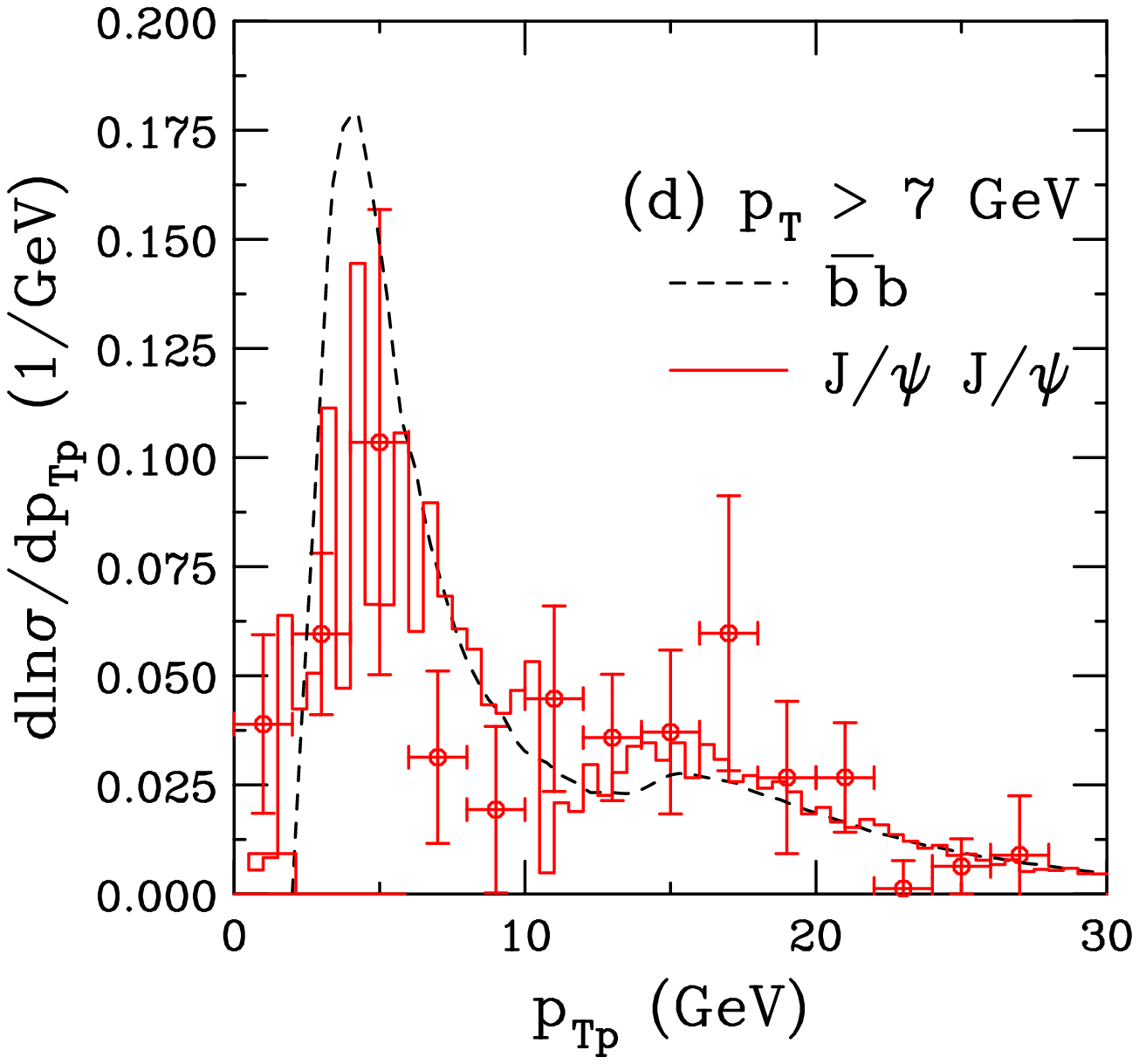} \\
\end{tabular}
  \caption[]{(Color online) The transverse momentum of the $b$ and
    $\overline b$ (black dashed lines) and the $J/\psi$'s resulting from the
    bottom quark decays (red histograms) are shown compared to the LHCb
    $J/\psi$ pair data
    \protect\cite{LHCb} (red circles) for the $p_T$ cuts
    on the $b$ quarks and the $J/\psi$ of 2 (a), 3 (b), 5 (c) and 7~GeV (d).
  }
  \label{fig_ptqq}
\end{figure}

\begin{figure}[htpb]\centering
\begin{tabular}{cc}
  \includegraphics[width=0.5\columnwidth]{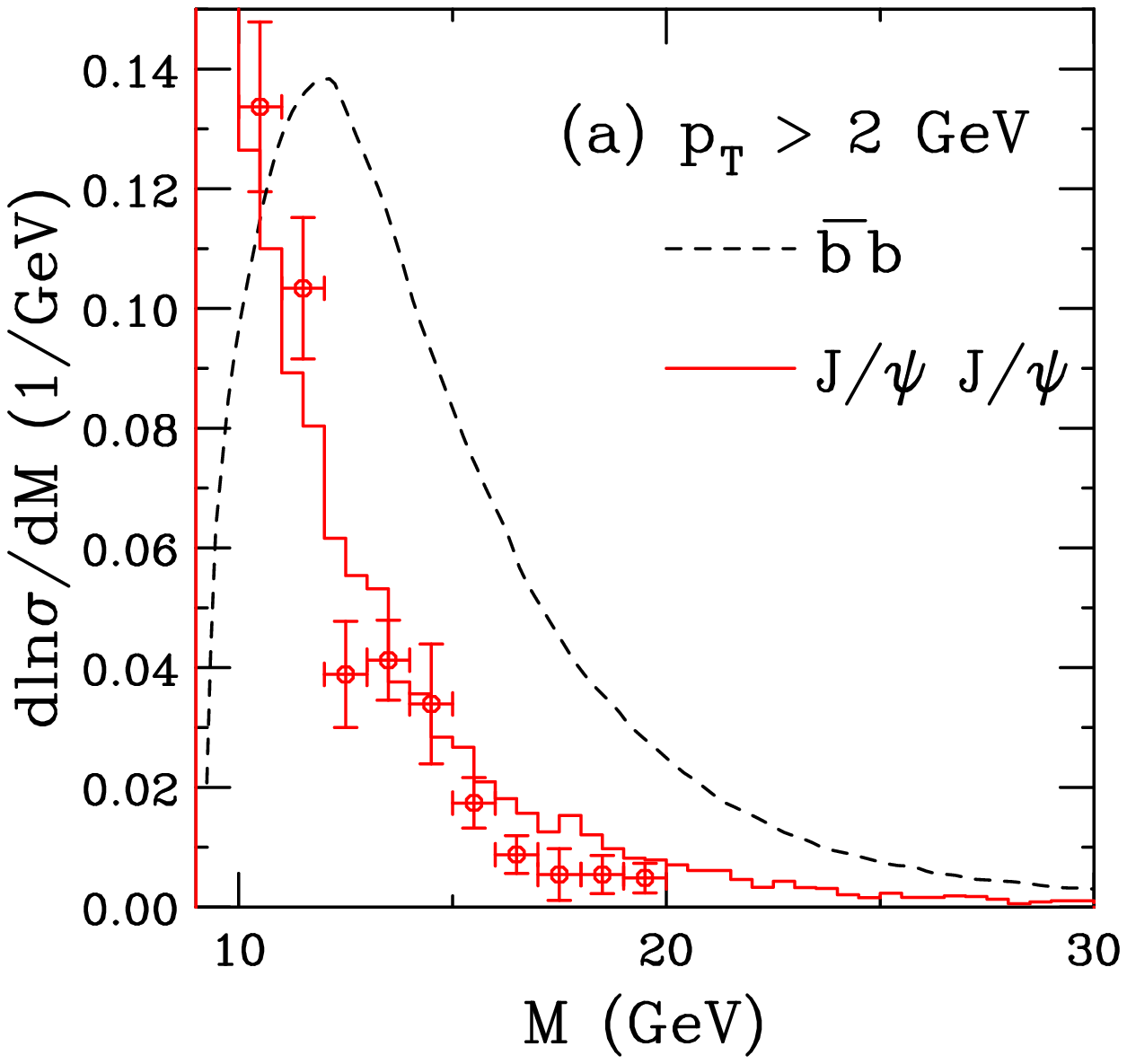} &
  \includegraphics[width=0.5\columnwidth]{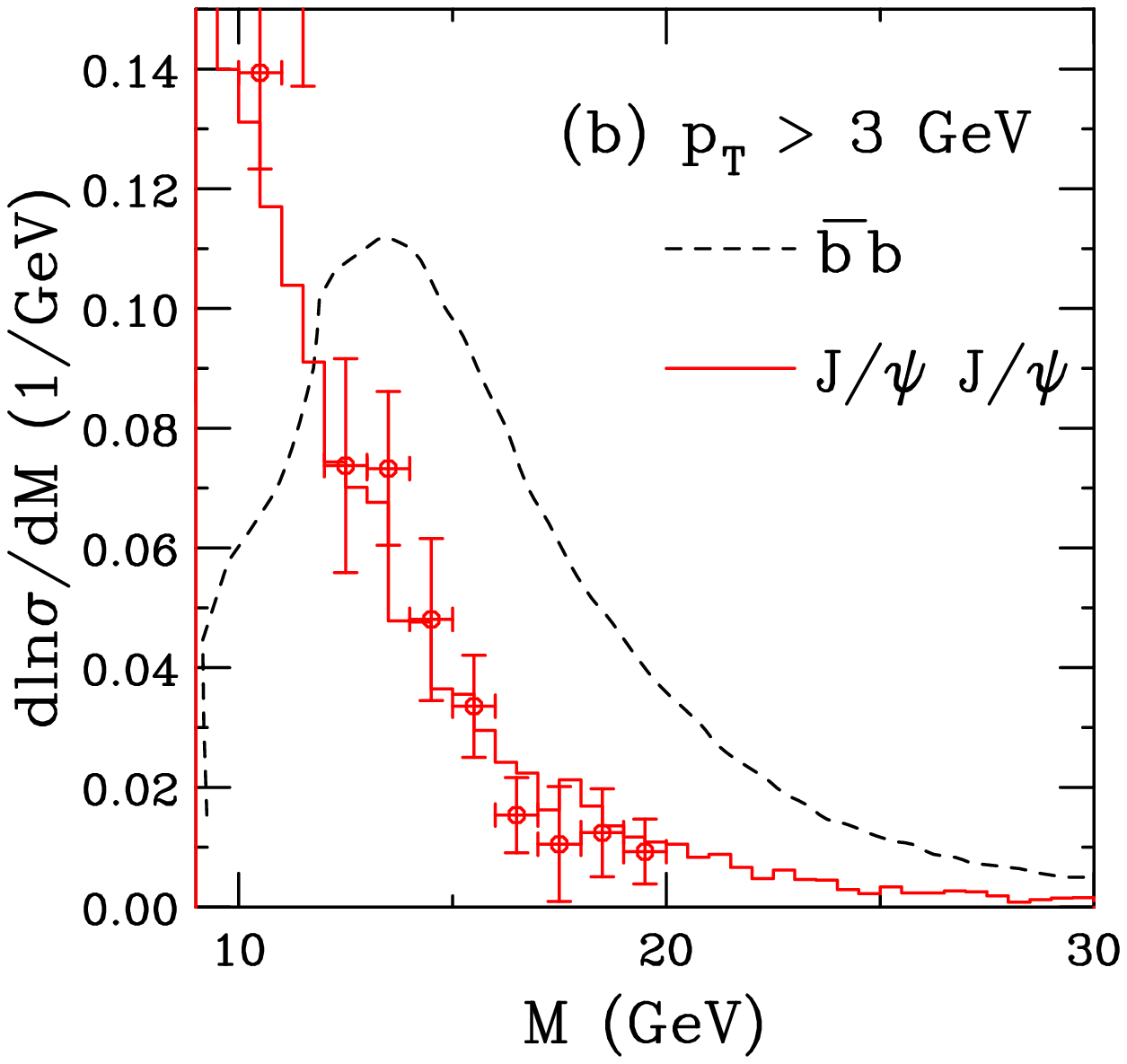} \\
  \includegraphics[width=0.5\columnwidth]{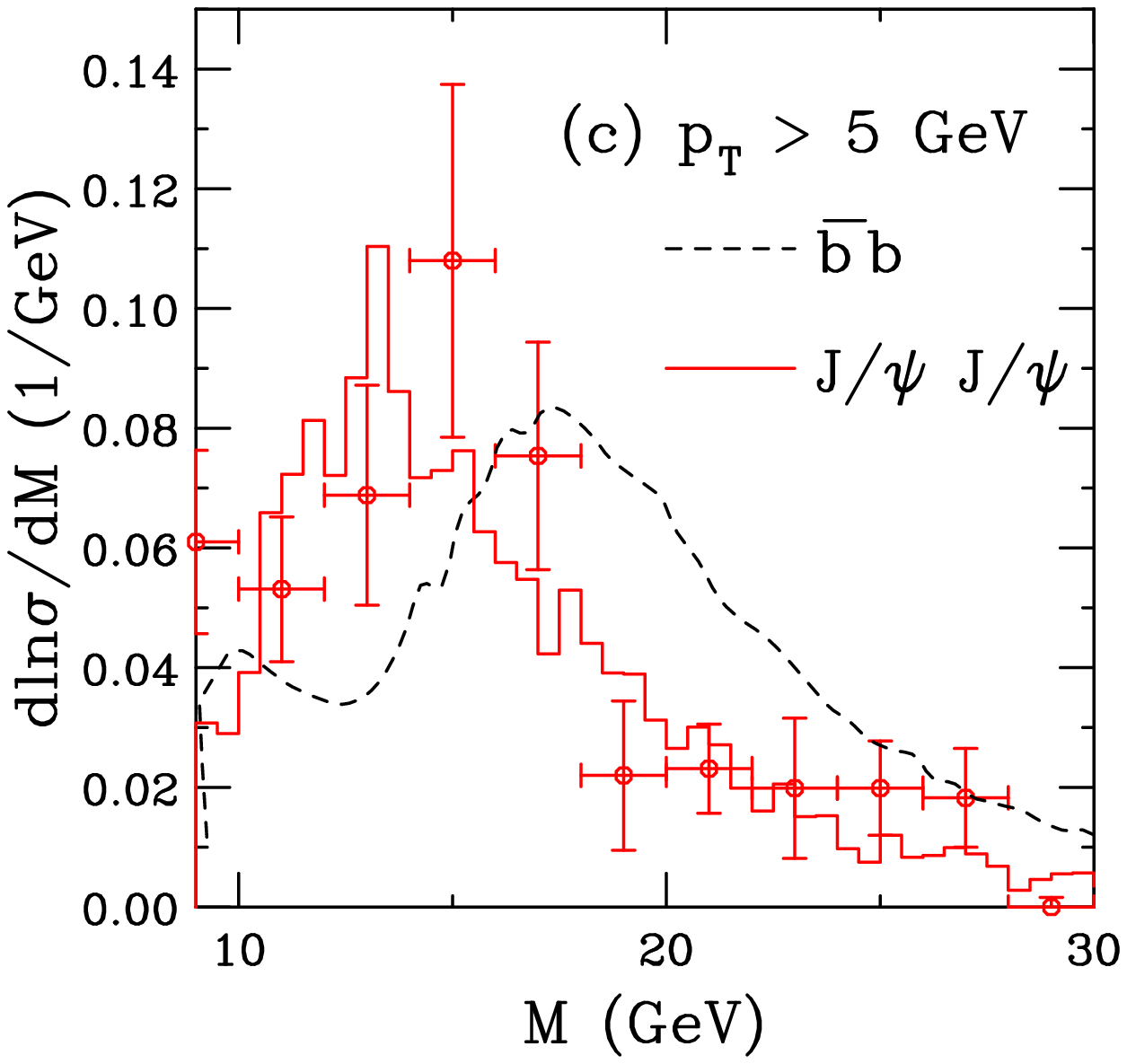} &
  \includegraphics[width=0.5\columnwidth]{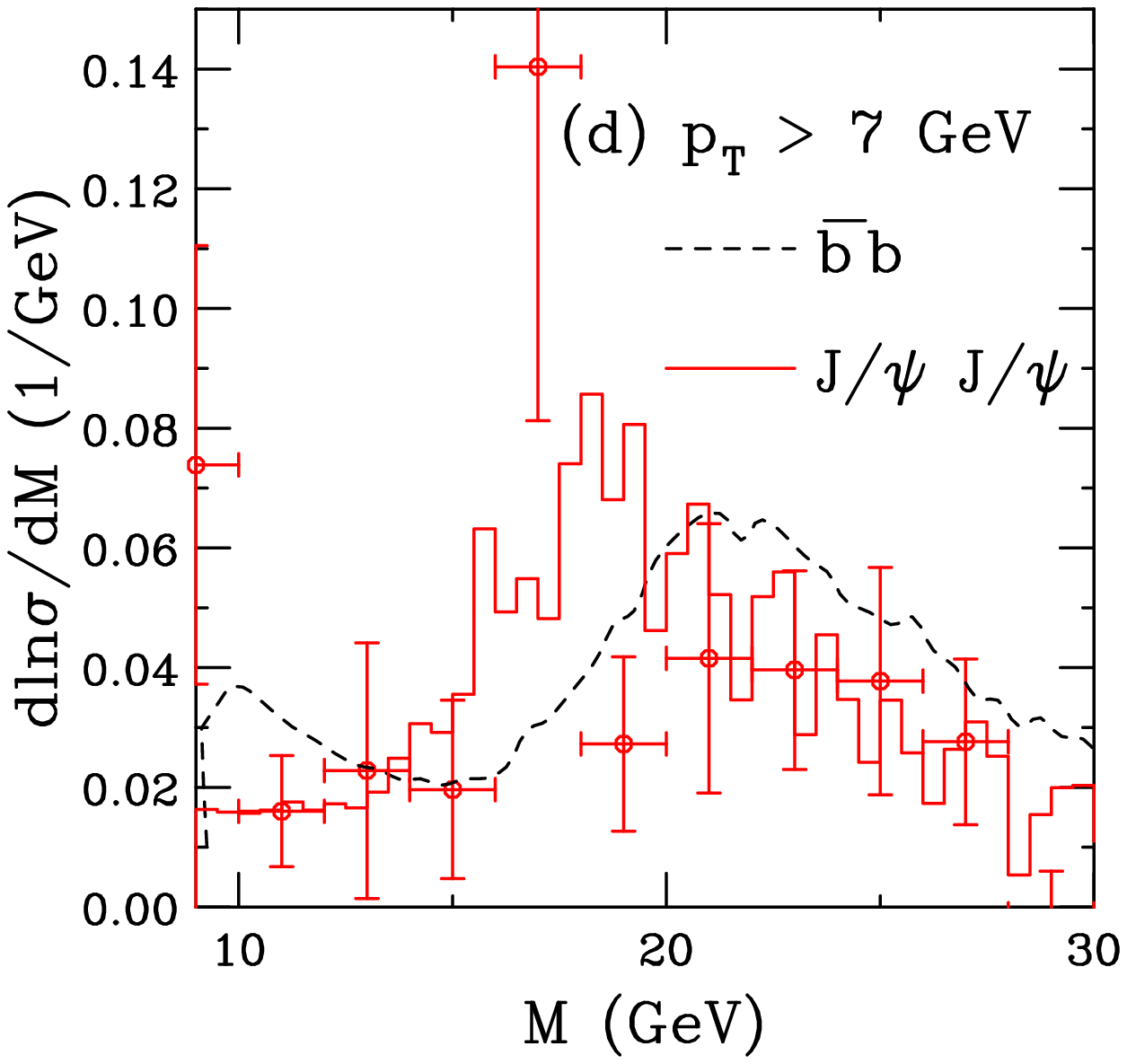} \\
 \end{tabular}
  \caption[]{(Color online) The pair mass of the $b$ and
    $\overline b$ (black dashed lines) and the $J/\psi$'s resulting from the
    bottom quark decays (red histograms) are shown compared to the LHCb
    $J/\psi$ pair data
    \protect\cite{LHCb} (red circles) for the $p_T$ cuts
    on the $b$ quarks and the $J/\psi$ of 2 (a), 3 (b), 5 (c) and 7~GeV (d).
  }
  \label{fig_mqq}
\end{figure}

The last two pair observables measured by LHCb were the pair transverse
momentum, $p_{T_p}$, and pair mass, $M$, distributions, shown in
Figs.~\ref{fig_ptqq} and \ref{fig_mqq} respectively.

The $p_T$ of the pair, shown on a linear scale in Fig.~\ref{fig_ptqq}, peaks
at low $p_{T_p}$ for both the $b \overline b$ and $J/\psi J/\psi$ pairs.  With
the lowest minimum single meson $p_T$, while the peaks of the two calculated
distributions are of similar magnitude, the $J/\psi$ pair peak is shifted
backward by about 1.7~GeV relative to the $b \overline b$, as is evident in
Fig.~\ref{fig_ptqq}(a).  At $p_T > 3$~GeV, while the average $p_{T_p}$ is still
about 1.3~GeV smaller for the $J/\psi$ pairs, most of the difference is at
$p_{T_p} < 5$~GeV.  Above this value, the distribution
is significantly broader than for $p_T > 3$~GeV.

As the minimum $p_T$ is increased, the average values of $p_{T_p}$ for the
initial $b \overline b$ and the final $J/\psi J/\psi$ become more similar.  In
addition, a feature develops at high $p_{T_p}$, a shoulder in the distribution
that appears at effectively twice the minimum $p_T$, independent of whether
the calculation is for the initial $b \overline b$
pairs or the decay $J/\psi$ pairs
although the statistics at high pair $p_T$ is significantly degraded for
$p_T > 7$~GeV.  The rise of this shoulder appears to correspond to the rise of
the peak at $|\Delta \phi | = 0$ in Fig.~\ref{fig_azi} where the
$b \overline b$ and $J/\psi$ pairs are aligned.
In all cases, the calculated $J/\psi J/\psi$ pair
distributions agree quite well with the LHCb data.

A similar trend seen for the pair mass distributions in Fig.~\ref{fig_mqq}.  
The minimum $b \overline b$ pair mass is $2m_b = 9.3$~GeV for 
$m_b = 4.65$~GeV.  Assuming that the $p_T$ of both of the individual mesons
are equal, the square of the pair mass can be written as
$M^2 = 2m_T^2(1 + \cosh(\Delta y))$.  Thus as the minimum single
meson $p_T$ increases, $m_T$ also increases and the average pair mass
moves to higher $M$.  This estimate is accurate for $2 \rightarrow 2$ processes
but is an underestimate for the $2\rightarrow 3$ diagrams that dominate NLO
$b \overline b$ production.  Nonetheless, one can see a clear trend that the
$b \overline b$ peak shifts to higher mass with an increase in minimum $p_T$,
with a residual enhancement at $2m_b$ for the highest minimum $p_T$.

When $J/\psi$ pairs from $b$ decays are considered, the pair mass does not
have a specific threshold any longer.  As shown for $M > 2m_b$, the $J/\psi$
pair mass is steeply decreasing for $p_T > 2$ and 3~GeV while for $p_T > 5$
and 7~GeV, a peak at higher $M$ also develops.  The average mass of the
$J/\psi$ pairs for 
the higher minimum $p_T$ is shifted backward by several GeV: compare the
black curves and the red histograms in Fig.~\ref{fig_mqq}(c) and (d).  The
calculations of the $J/\psi$ pairs follow the LHCb data very closely.

In general, the calculations presented here are in very good agreement for the
pair observables obtained by LHCb for all  values of the minimum $p_T$ chosen.

\section{Sensitivity of $d\sigma/d\phi$ to $\langle k_T^2 \rangle$}
\label{sec:kTcut}

In this section, the sensitivity of observables to the presence of an 
intrinsic $\langle k_T^2 \rangle$
is explored.  While in Ref.~\cite{QQazi}, the sensitivity was studied by
gradually dialing up $\langle k_T^2 \rangle$ to its default value, here only
the results with $\langle k_T^2 \rangle = 0$ and the default 3~GeV$^2$ (at
$\sqrt{s} = 7$~TeV) are 
compared.  The comparison is made for both the $b \overline b$ pairs and
the final $J/\psi J/\psi$.

\begin{figure}[htpb]\centering
\begin{tabular}{cc}
  \includegraphics[width=0.5\columnwidth]{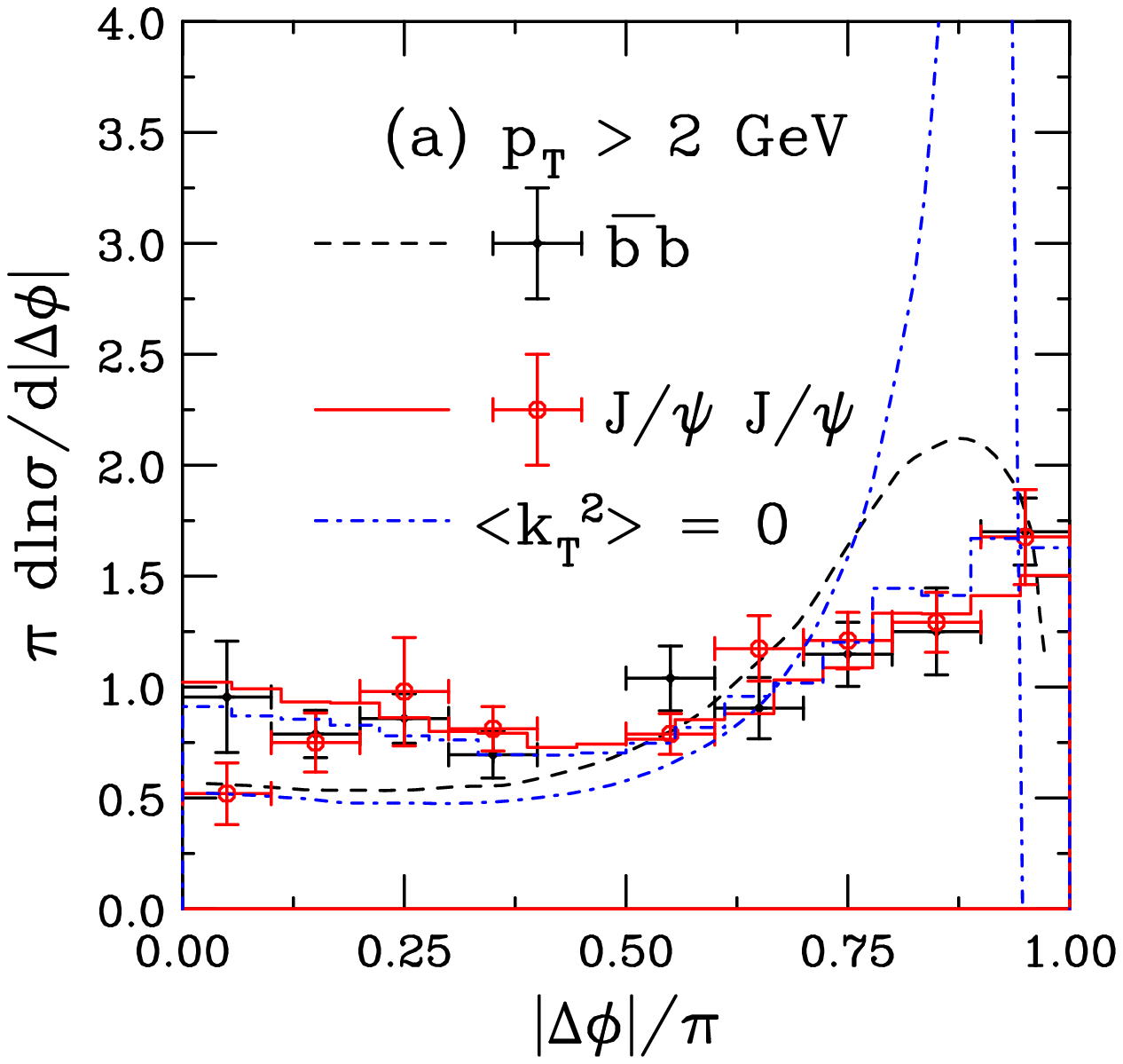} &
  \includegraphics[width=0.5\columnwidth]{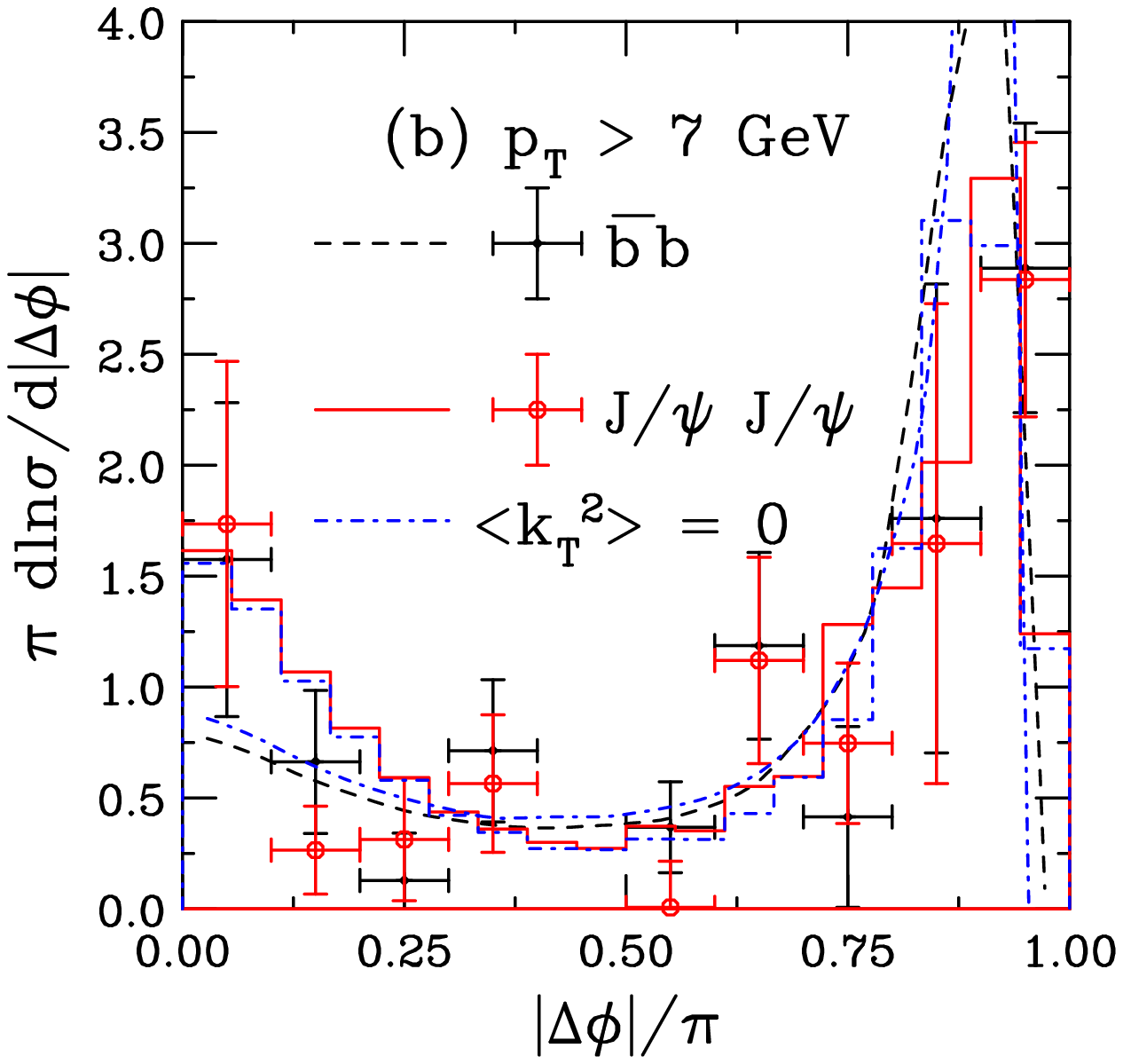} \\
  \includegraphics[width=0.5\columnwidth]{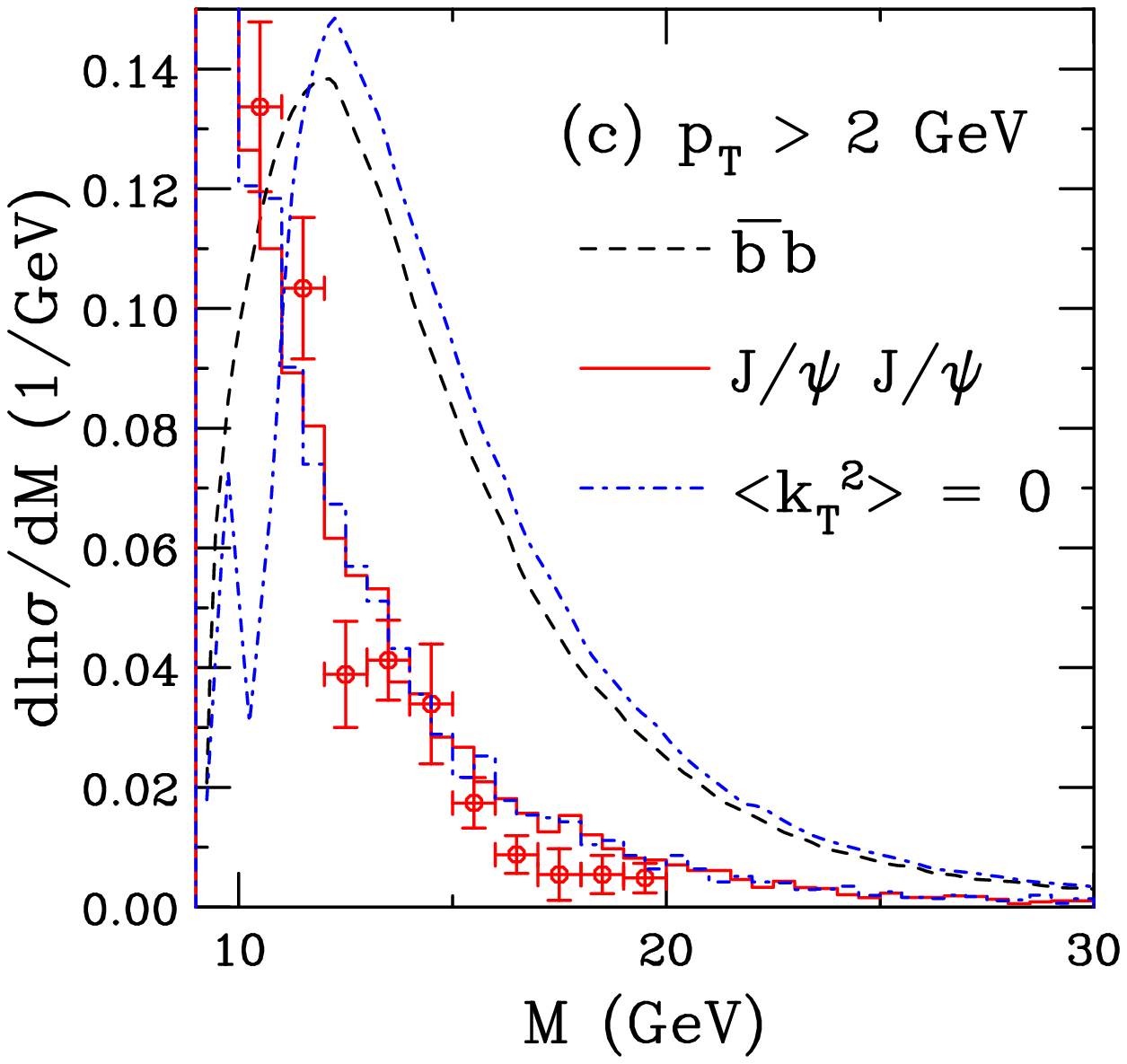} &
  \includegraphics[width=0.5\columnwidth]{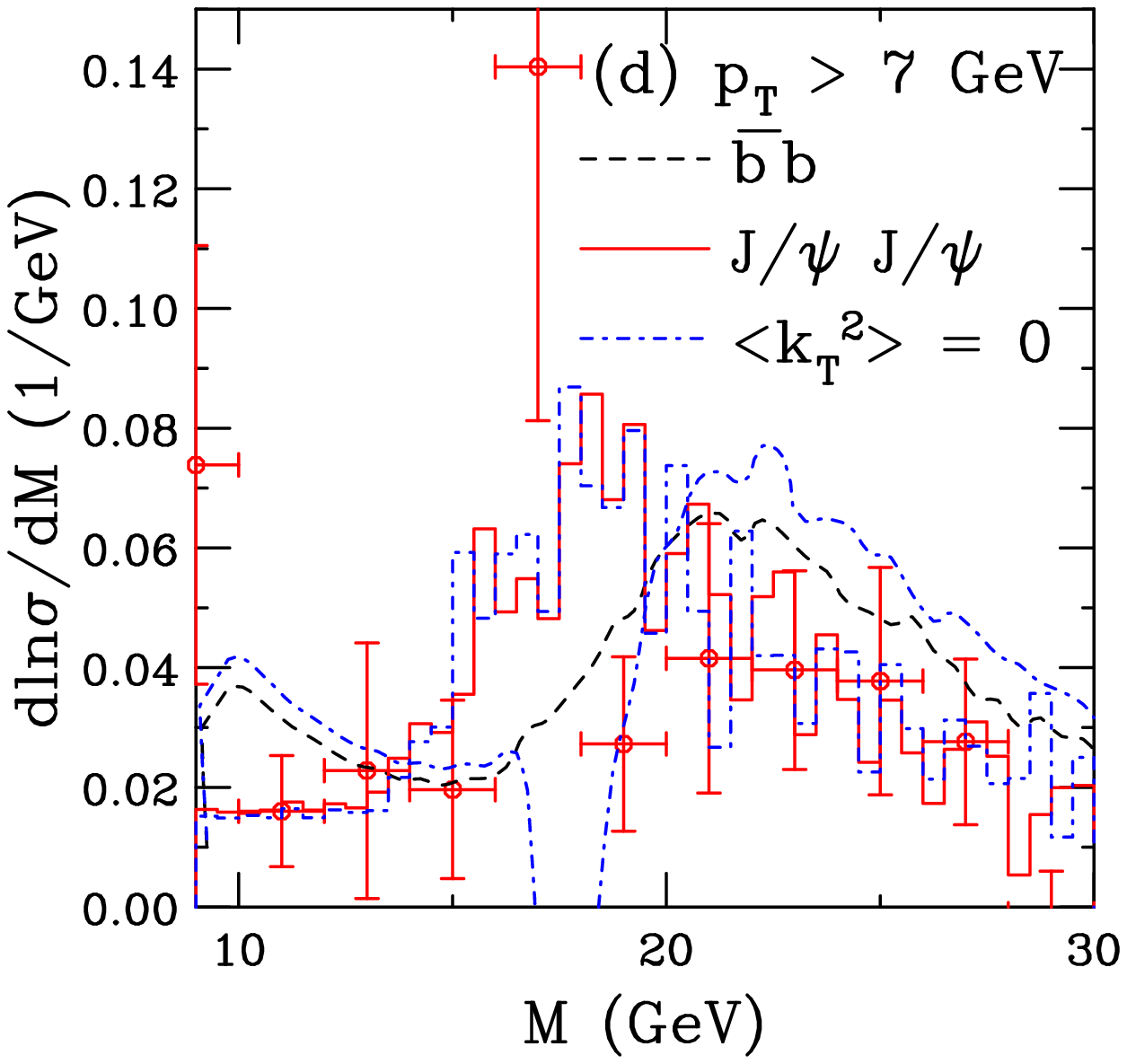} \\
  \includegraphics[width=0.5\columnwidth]{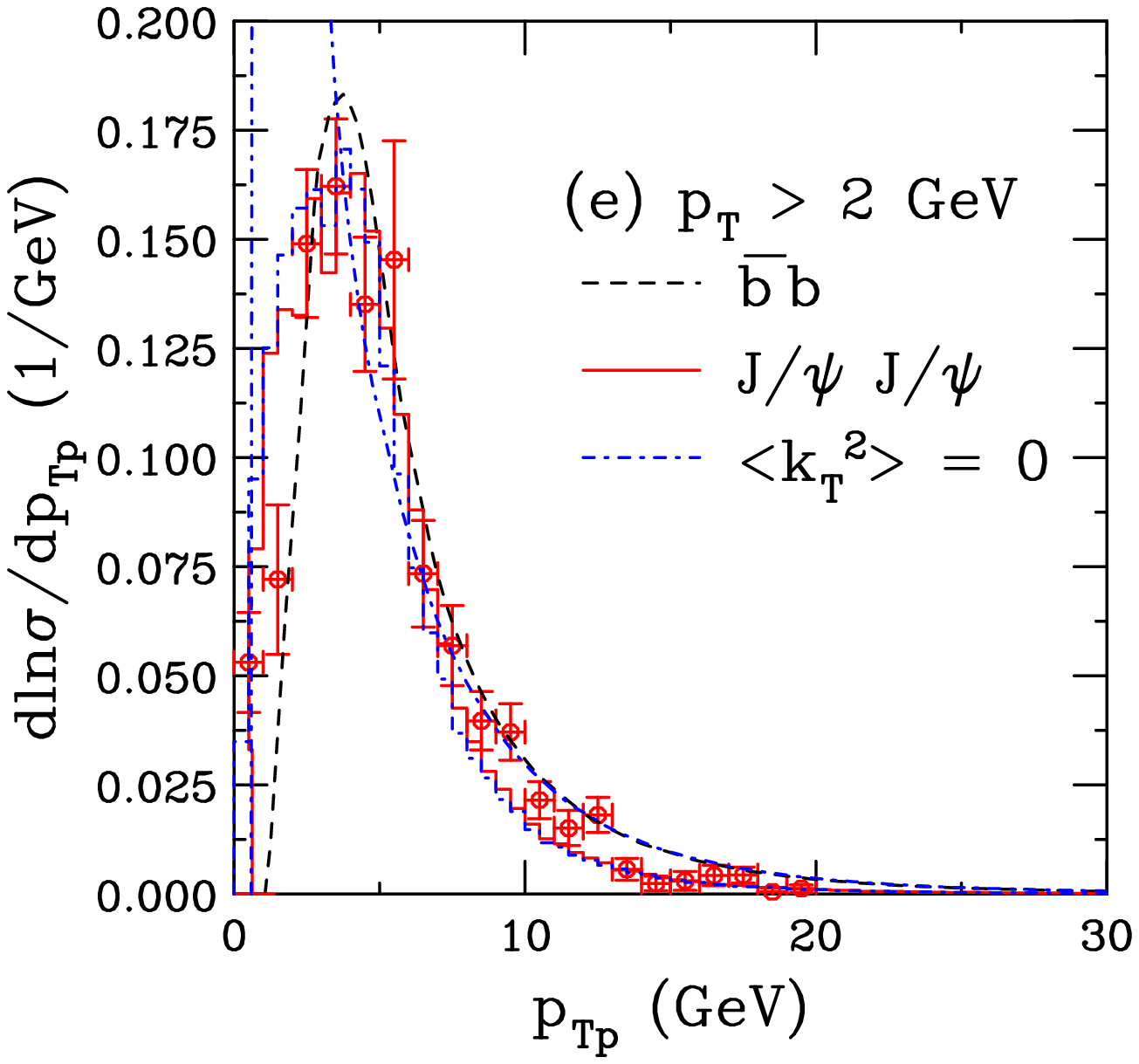} &
  \includegraphics[width=0.5\columnwidth]{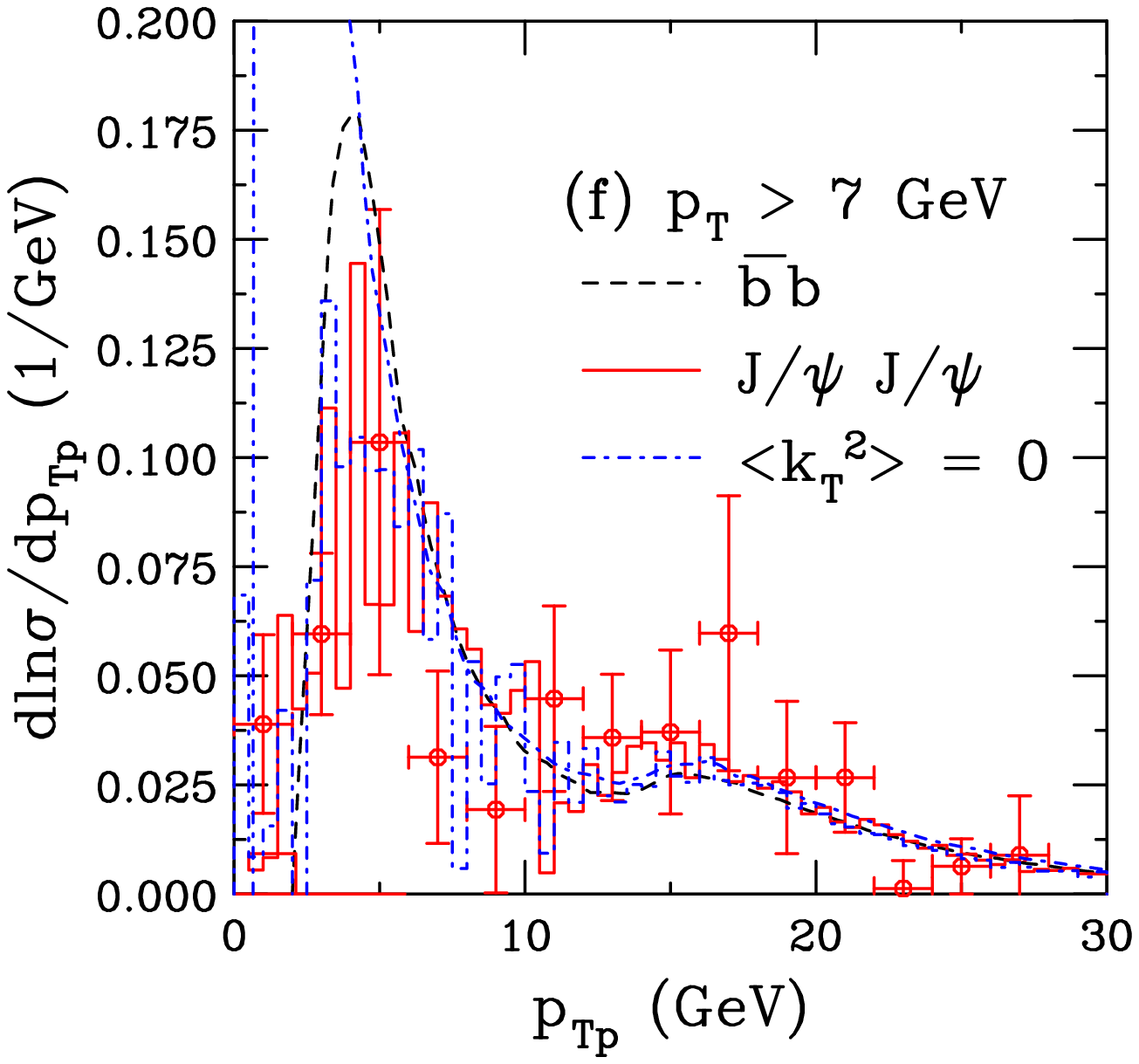} \\
  \includegraphics[width=0.5\columnwidth]{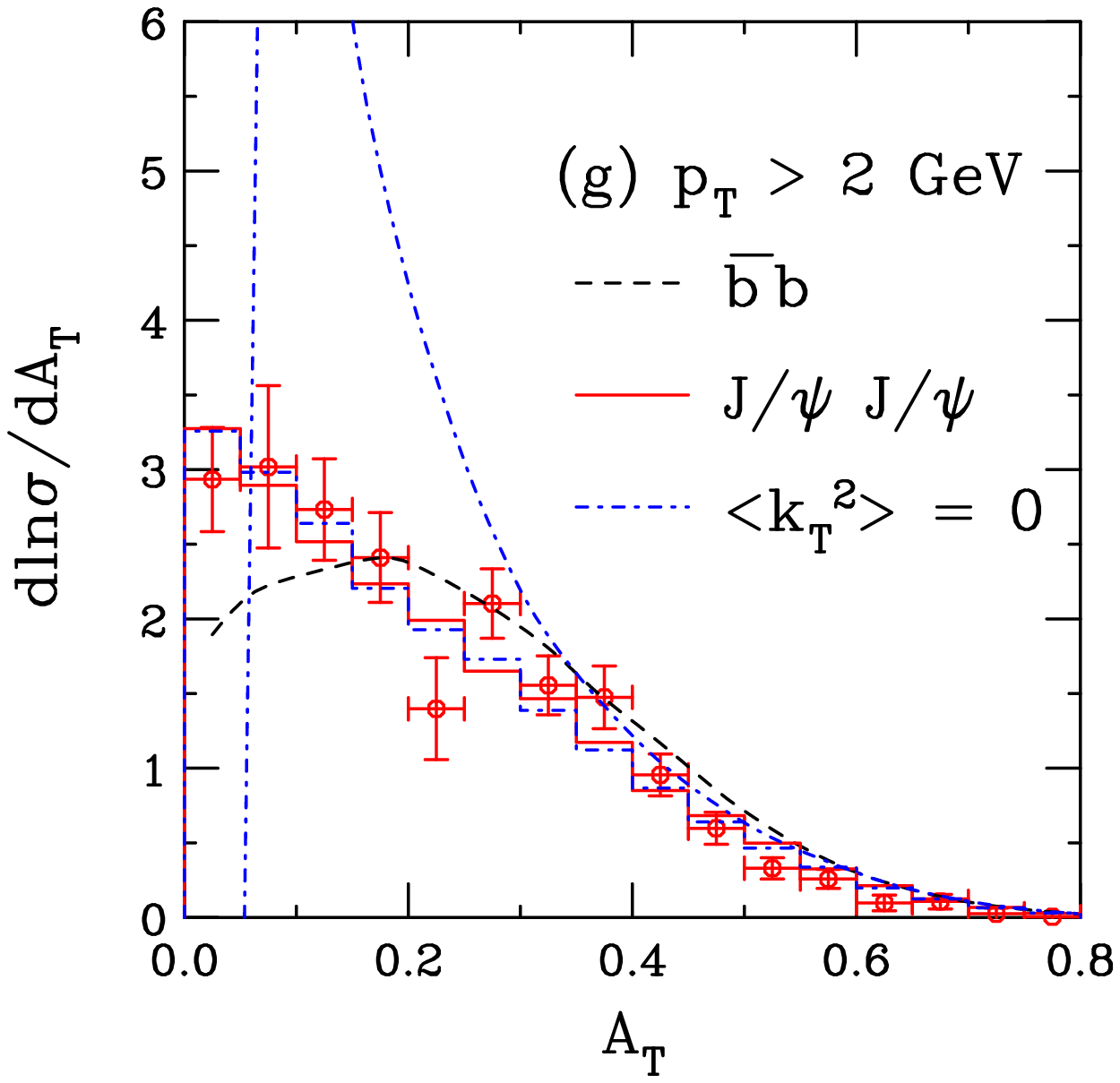} &
  \includegraphics[width=0.5\columnwidth]{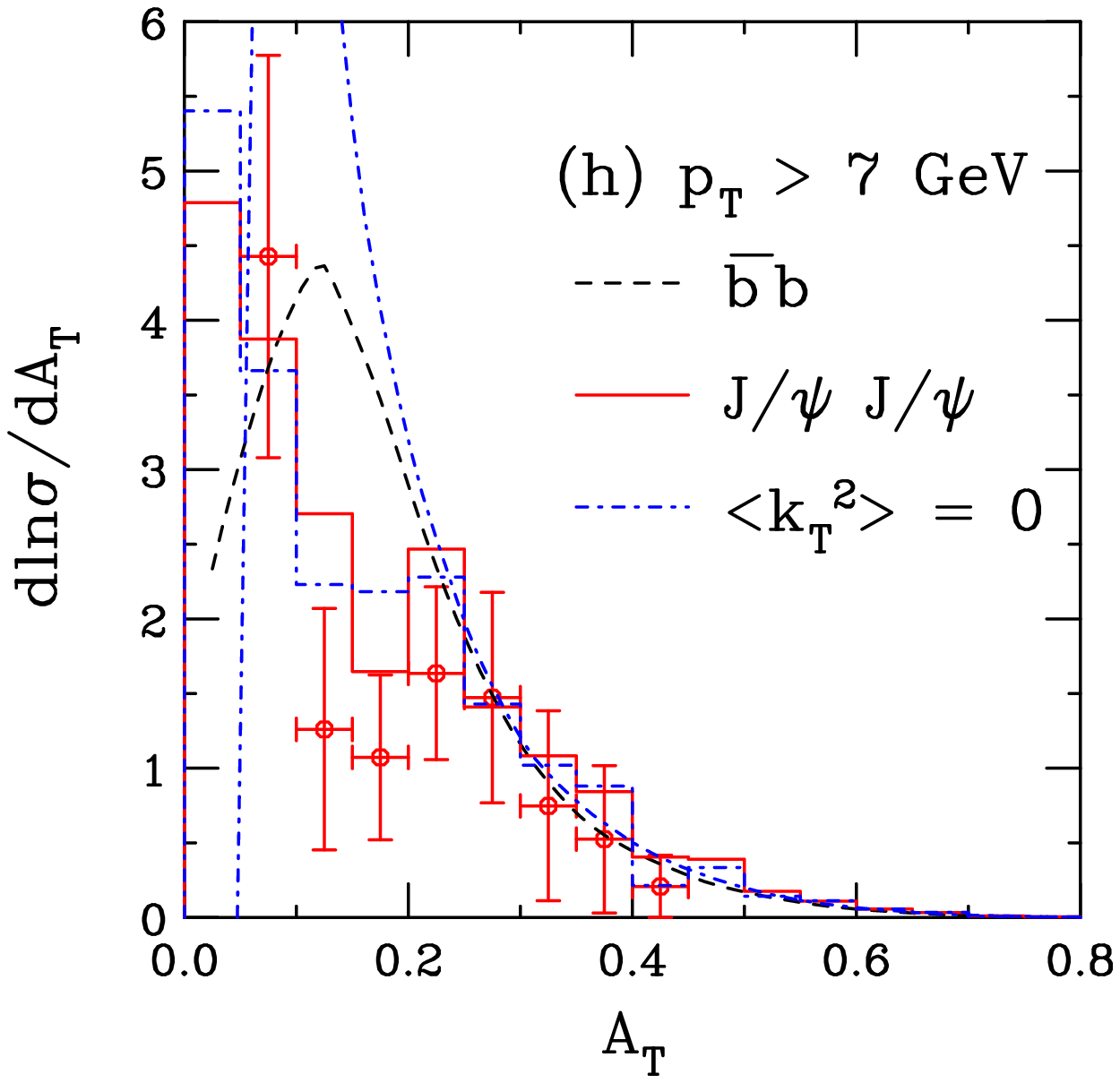} \\
\end{tabular}
\caption[]{(Color online) The difference in the $b \overline b$ and
  $J/psi J/\psi$ pair results for $\langle k_T^2 \rangle = 0$ and the default
  $k_T$ kick.  The $\langle k_T^2 \rangle = 0$ results are shown by the
  blue dot-dashed curves ($b \overline b$) and blue dot-dashed
  histograms ($J/\psi J/\psi$) and
  with the default $k_T$ kick by the black dashed curves ($b \overline b$) and
  red histograms ($J/\psi J/\psi$).  Results are shown for the azimuthal angle
  difference (a) and (b); pair mass (c) and (d); pair transverse momentum
  (e) and (f); and $p_T$ asymmetry (g) and (h).  The results on the left-hand
  side (a), (c), (e) and (g) are shown for $p_T > 2$~GeV while those on the
  right-hand side (b), (d), (f) and (h) are shown for $p_T > 7$~GeV.  The
  LHCb data \protect\cite{LHCb} (black for $b \overline b$, red circles for
  $J/\psi$ pairs) are also shown.
  }
  \label{fig_kt_comp}
\end{figure}

Because $|\Delta y|$ and $y_p$ are independent of $\langle k_T^2 \rangle$,
the comparison is only shown for $|\Delta \phi|$, $M$, $p_{T_p}$ and $A_T$ 
in Fig.~\ref{fig_kt_comp}.  The left-hand side of the figure shows the results
for $p_T > 2$~GeV while calculations for $p_T > 7$~GeV are shown on the 
right-hand side.  The behavior of calculations with $p_T > 3$ and 5~GeV follow
similar trends.  All results with $\langle k_T^2 \rangle = 0$ are given in
black, curves for $b \overline b$ and histograms for $J/\psi J/\psi$.

It is clear that the $b \overline b$ distributions are most affected by the
presence of broadening.  The peaks at $|\Delta \phi^*| \approx \pi$, low
$p_{T_p}$ and low $A_T$ are enhanced.  They are not delta functions
without broadening, as they would be at leading order, but have finite tails
indicative of a NLO process.  There
is no significant change in the pair mass distributions, independent of minimum
$p_T$.

Note that the bins at $|\Delta \phi^*| \approx \pi$, $p_{T_p} \rightarrow 0$
and $A_T \rightarrow 0$ do not go directly to zero but show enhanced peaks
due to the incomplete numerical cancellation of divergences with HVQMNR, as
discussed in Sec.~\ref{sec:model}.  For example, with
$\langle k_T^2 \rangle = 0$, the normalized $A_T$ distribution would still have
a peak at finite $A_T$ but it would be closer to $A_T \approx 0$ and decrease
faster with $A_T$.  The addition of $k_T$ broadening smears out this behavior.
particularly at lower $p_T$.

On the other hand, the $J/\psi$ pair distributions are largely
unaffected by $k_T$
broadening even though the parent $b$ meson pairs are sensitive to
the presence of $k_T$ broadening.  This is because the
decay randomizes the direction of the $J/\psi$ relative to the $b$ meson
parent, independent of the choice of $k_T$.
Thus it is not possible to learn much about broadening in the initial state
by studying the final-state decay products.  It would be better to look at the
$B$ meson pair correlations themselves than studying pair observables
through the $J/\psi$ decay products.

\section{Theoretical Uncertainties}
\label{sec:unc}

\begin{figure}[htpb]\centering
\begin{tabular}{cc}
  \includegraphics[width=0.5\columnwidth]{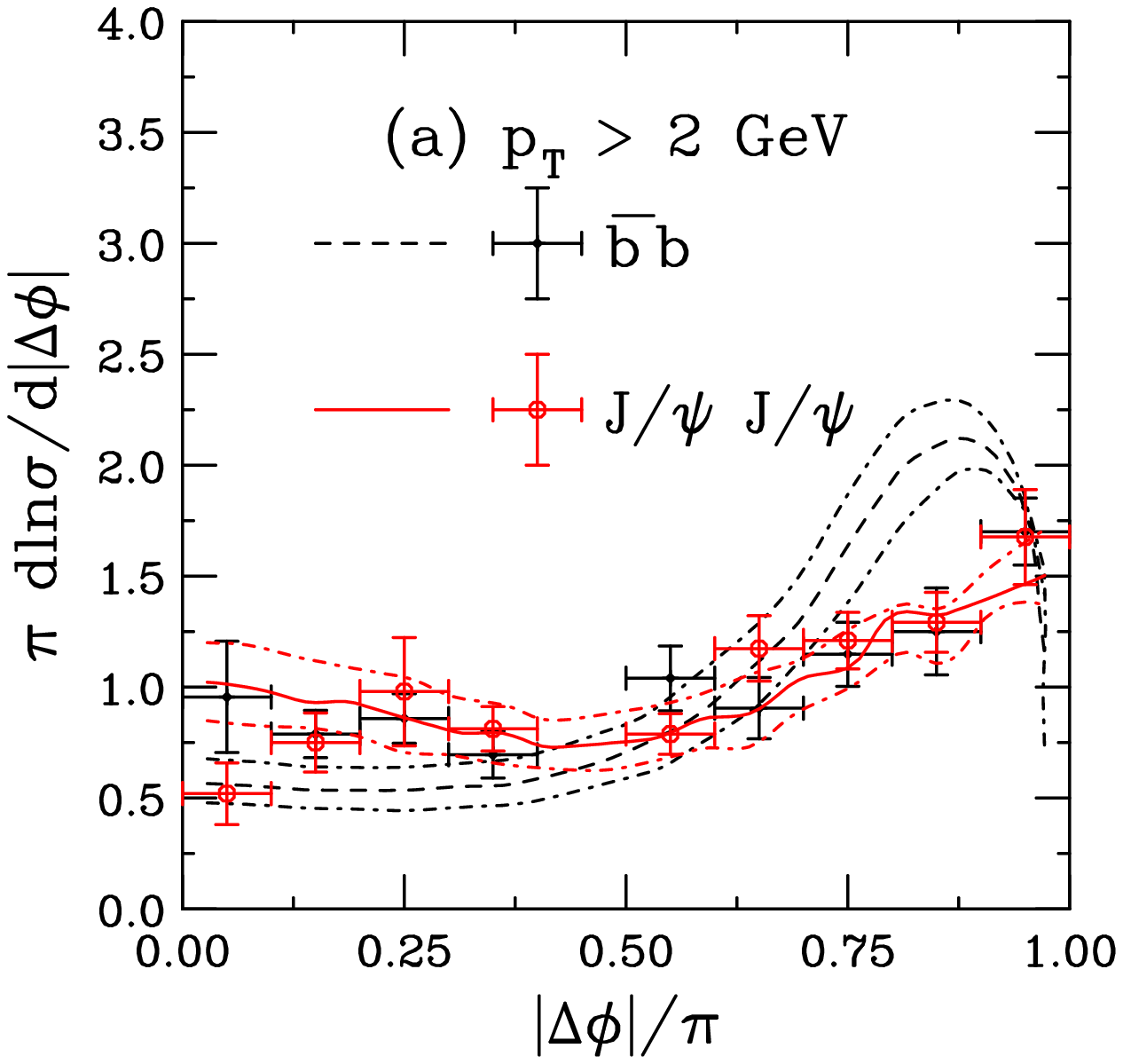} &
  \includegraphics[width=0.5\columnwidth]{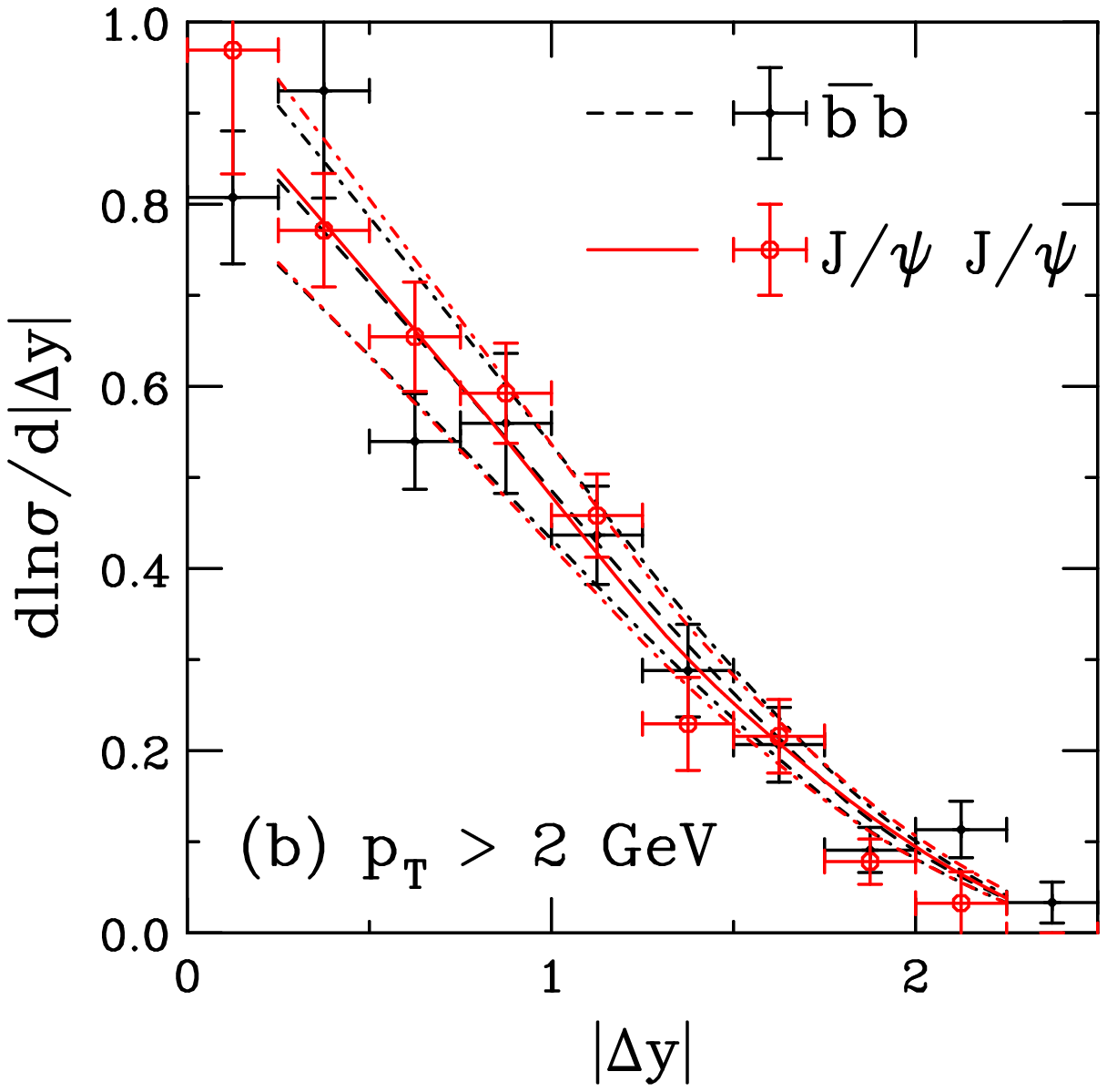} \\
  \includegraphics[width=0.5\columnwidth]{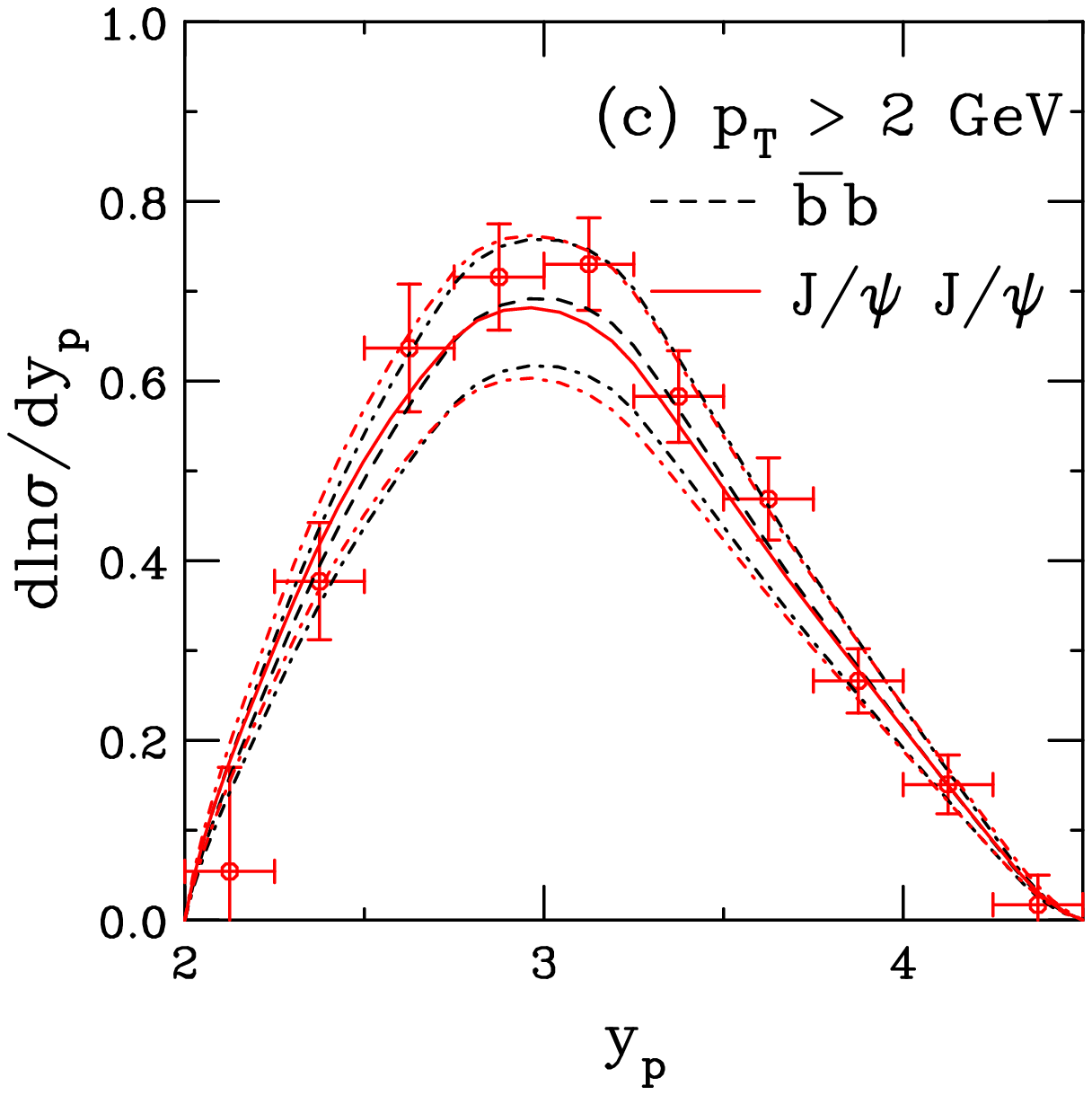} &
  \includegraphics[width=0.5\columnwidth]{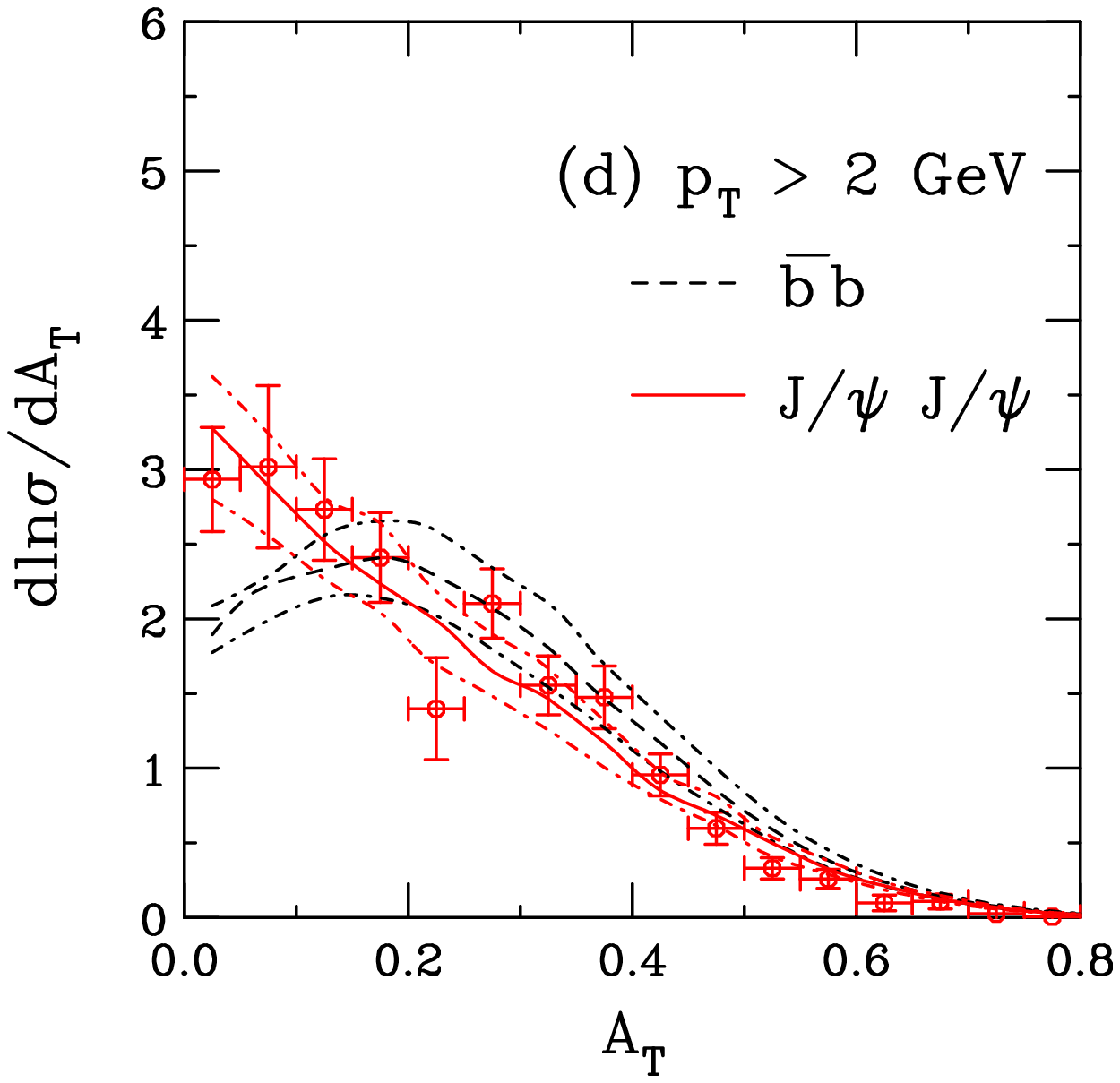} \\
  \includegraphics[width=0.5\columnwidth]{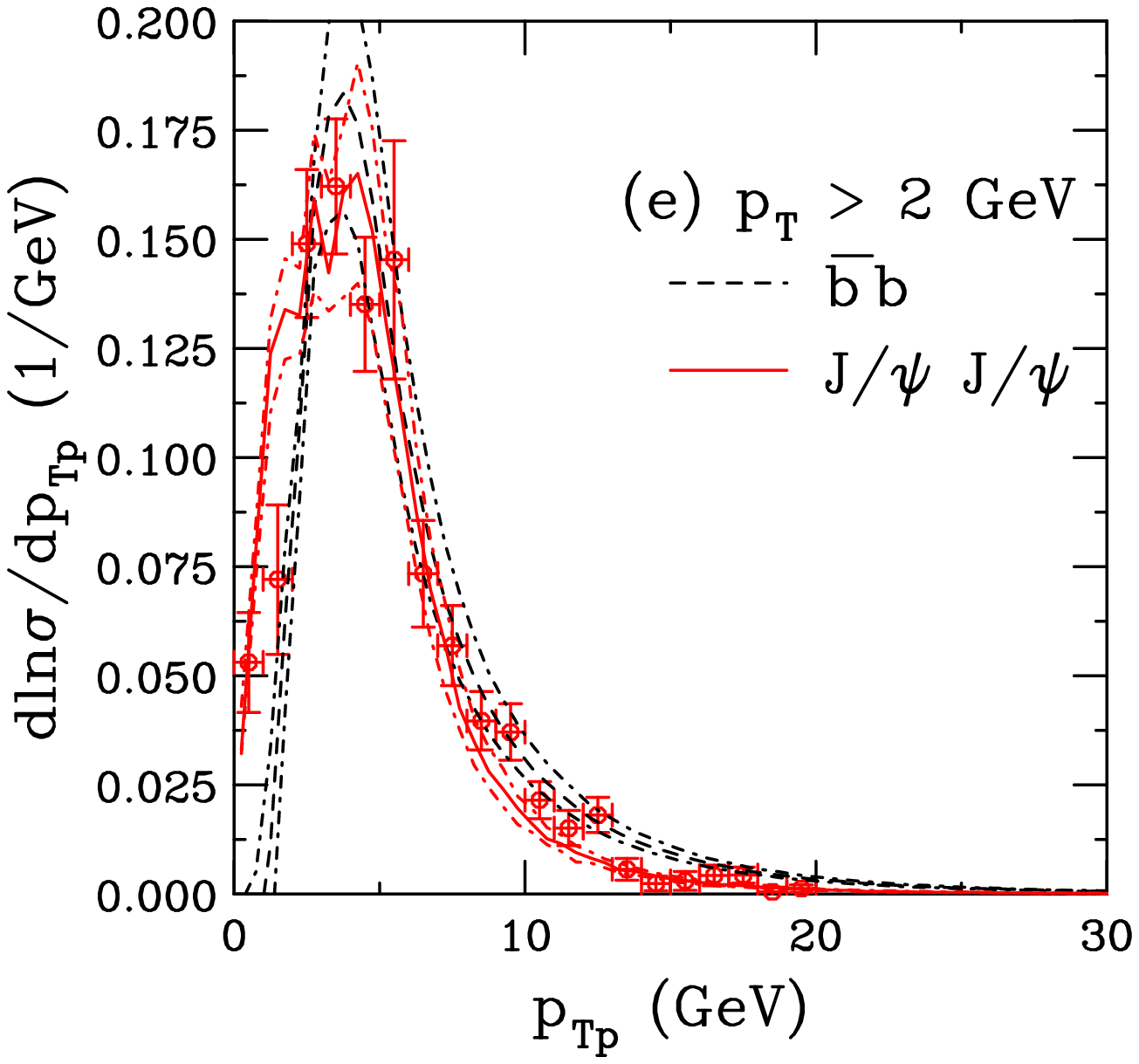} &
  \includegraphics[width=0.5\columnwidth]{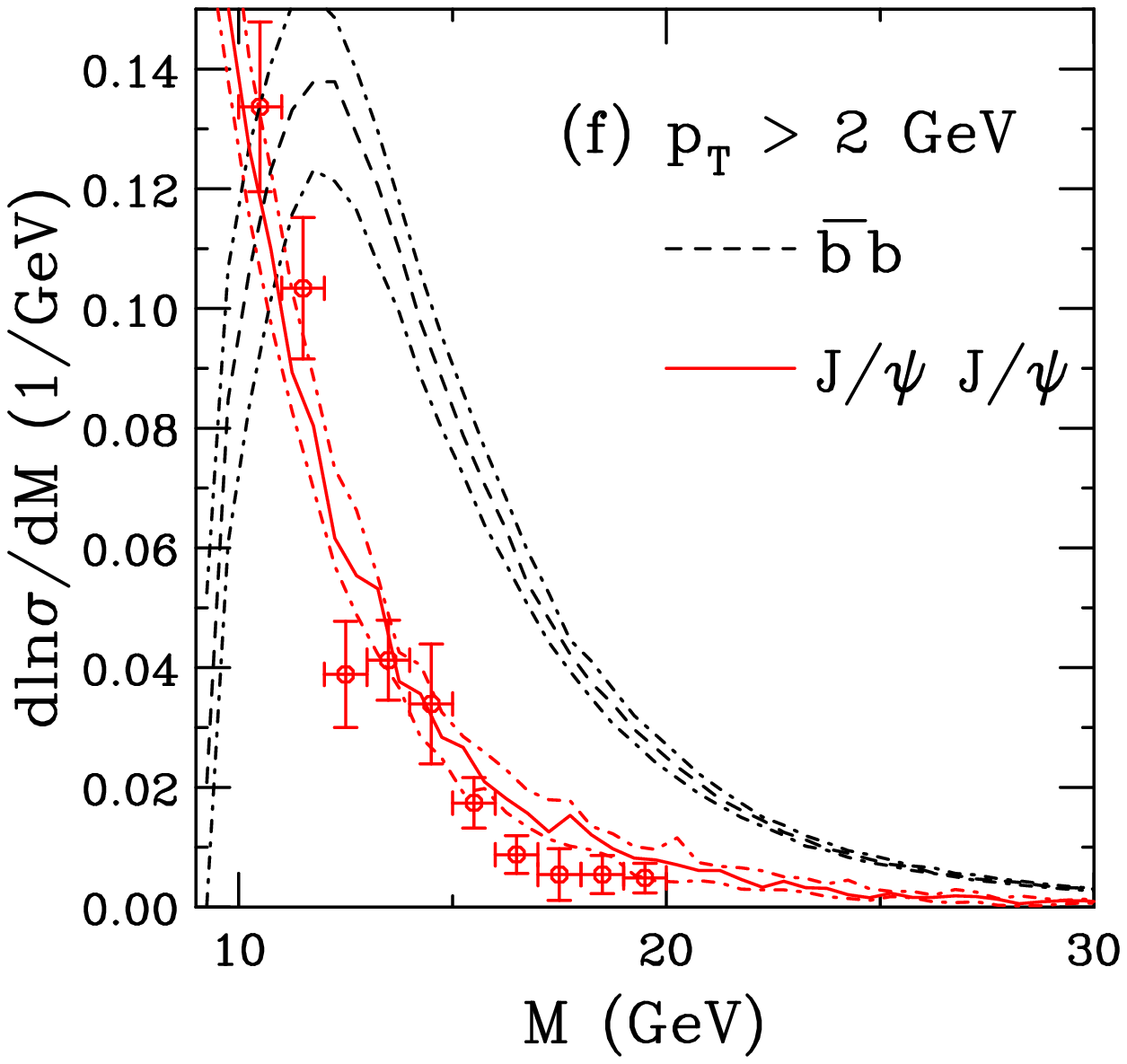} \\
\end{tabular}
\caption[]{(Color online) The mass and scale uncertainty bands are shown for
  the $b \overline b$ pairs (black dashed curves) and $J/\psi J/\psi$ pairs
  (red solid curves)
  and compared to the LHCb data \protect\cite{LHCb} for $p_T > 2$~GeV.
  The limits on the uncertainties are shown by
  dot-dashed curves in both cases.  Results are given for
  the azimuthal difference (a); rapidity difference (b); pair rapidity (c);
  $p_T$ asymmetry (d); pair $p_T$ (e); and pair mass (f).
  }
  \label{fig_bands_2GeV}
\end{figure}

\begin{figure}[htpb]\centering
\begin{tabular}{cc}
  \includegraphics[width=0.5\columnwidth]{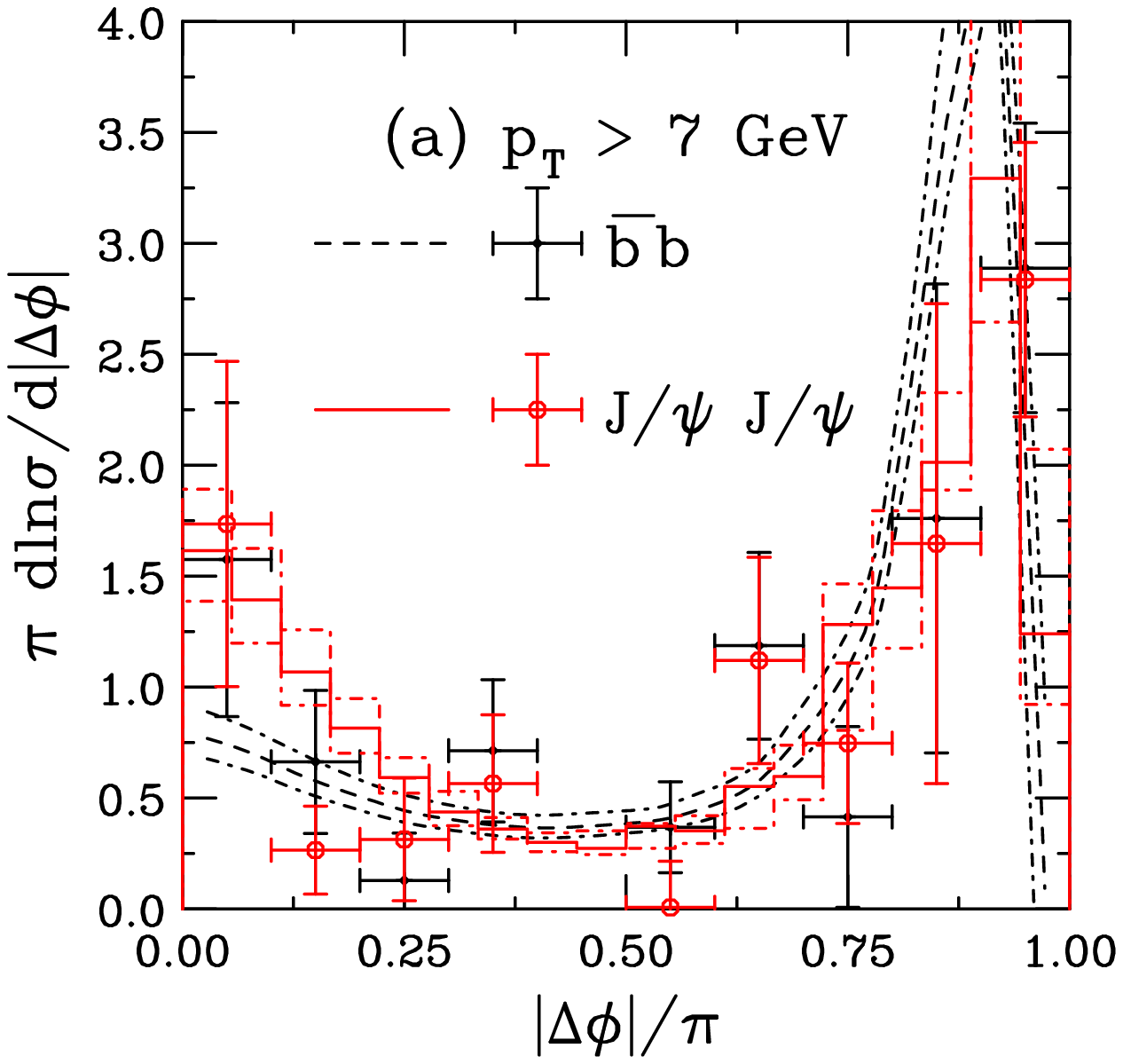} &
  \includegraphics[width=0.5\columnwidth]{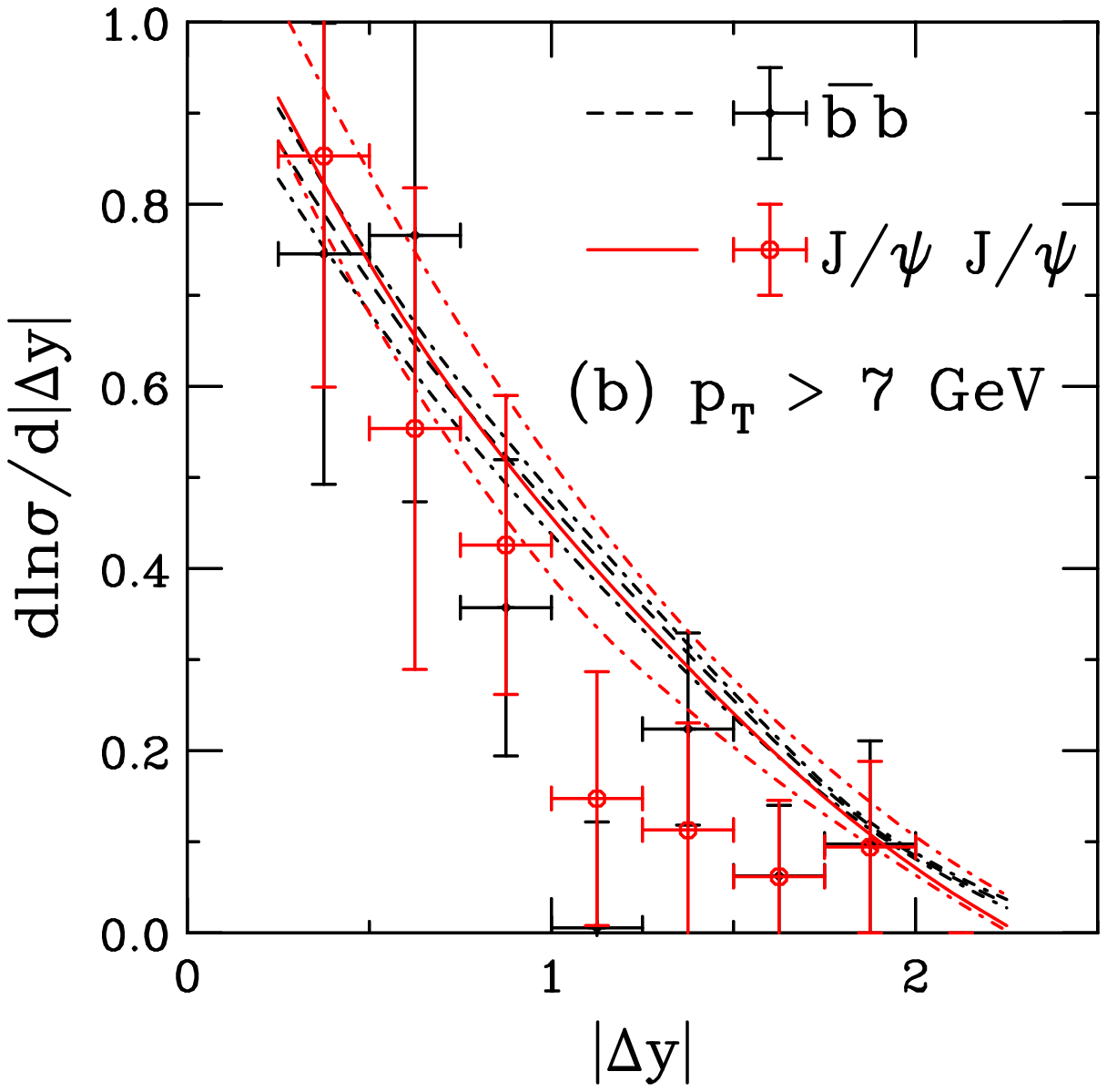} \\
  \includegraphics[width=0.5\columnwidth]{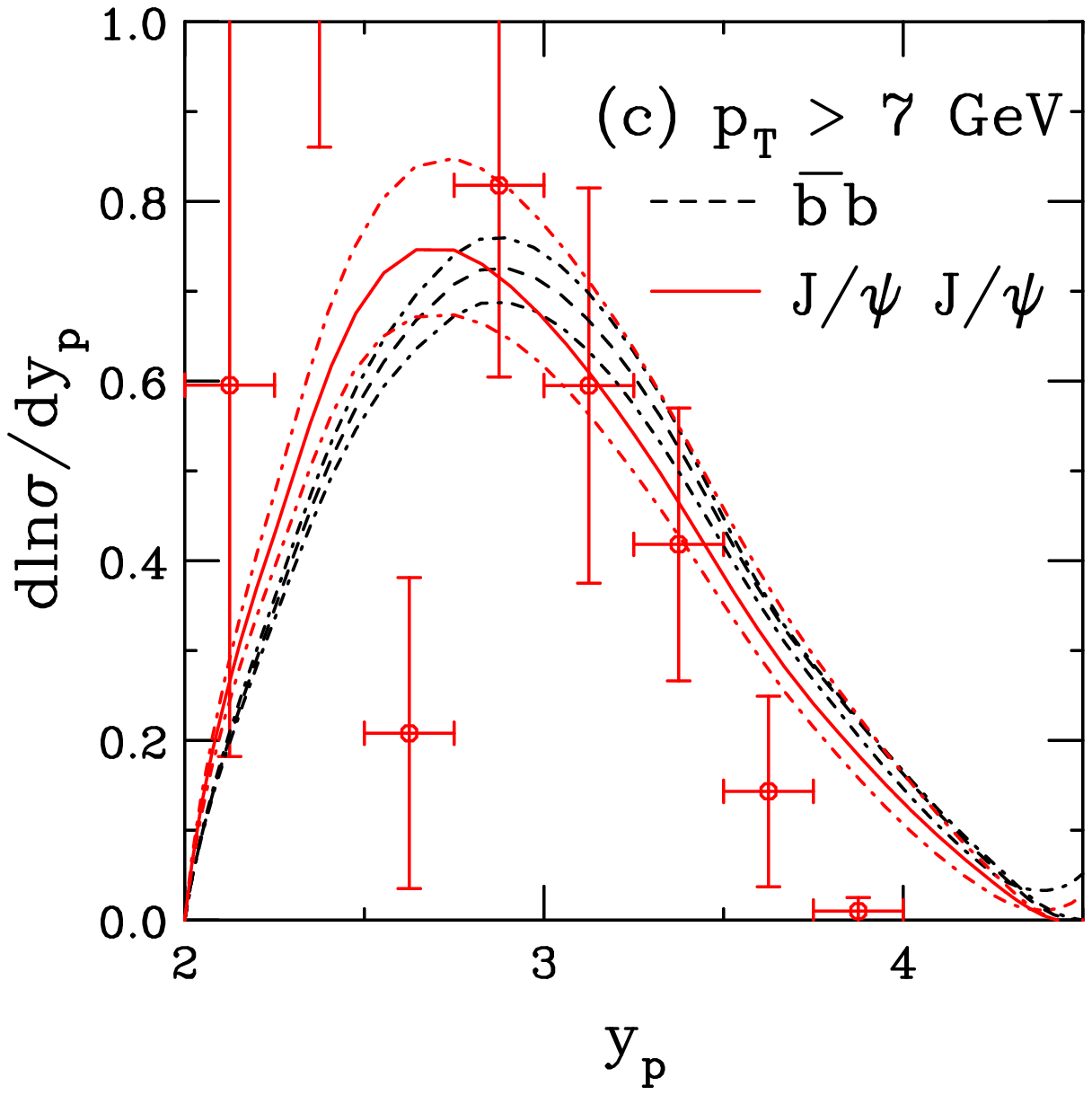} &
  \includegraphics[width=0.5\columnwidth]{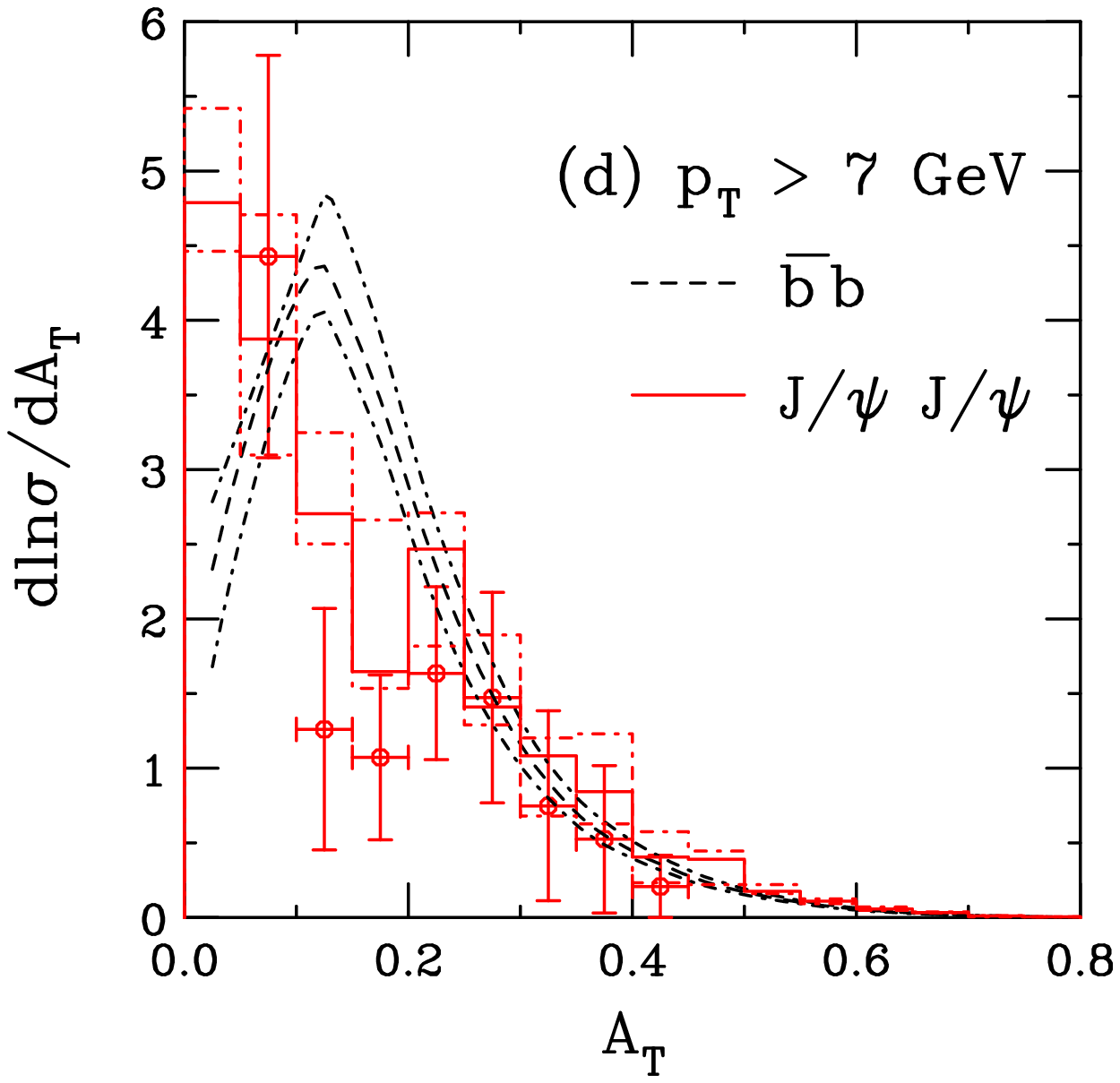} \\
  \includegraphics[width=0.5\columnwidth]{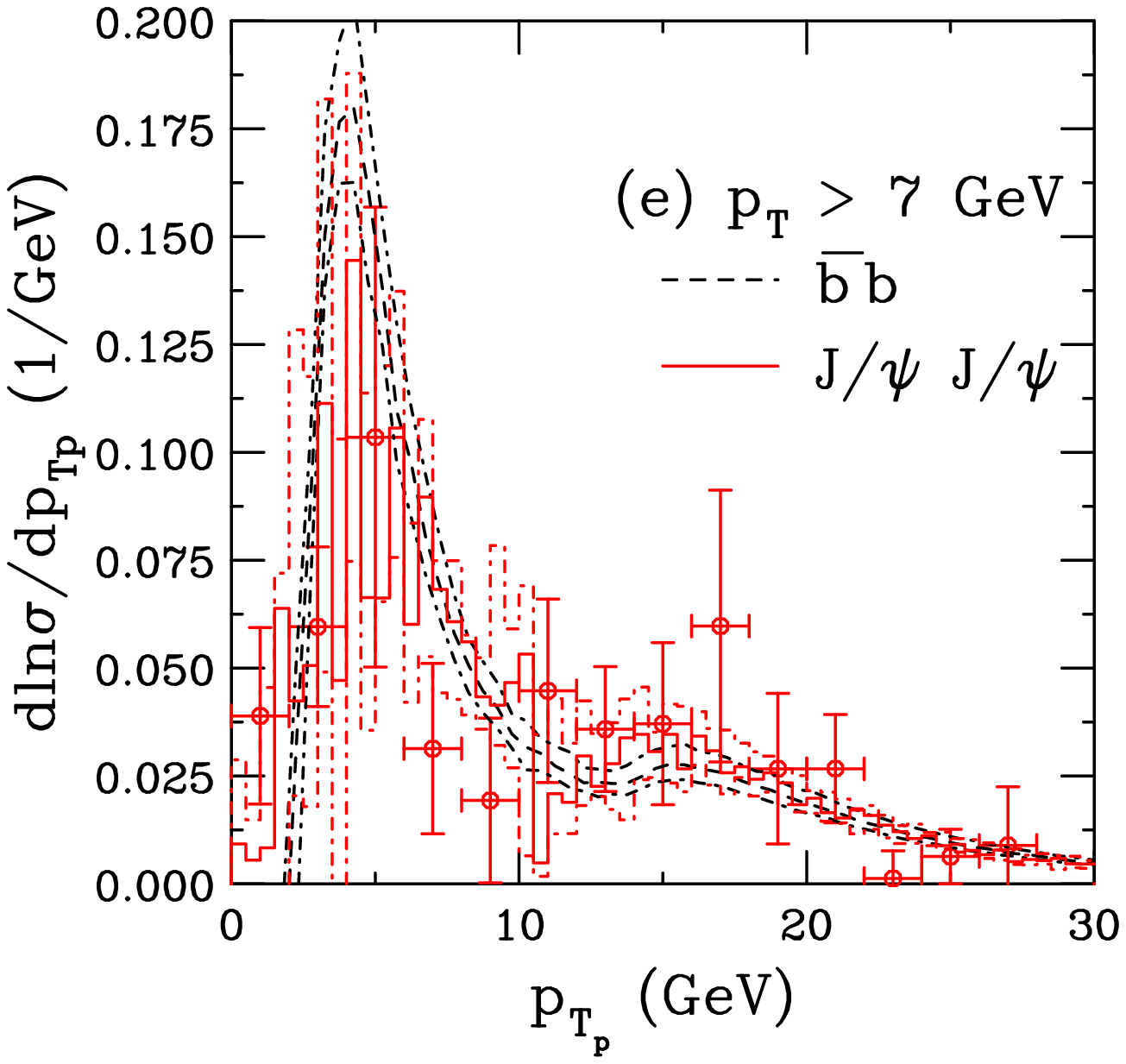} &
  \includegraphics[width=0.5\columnwidth]{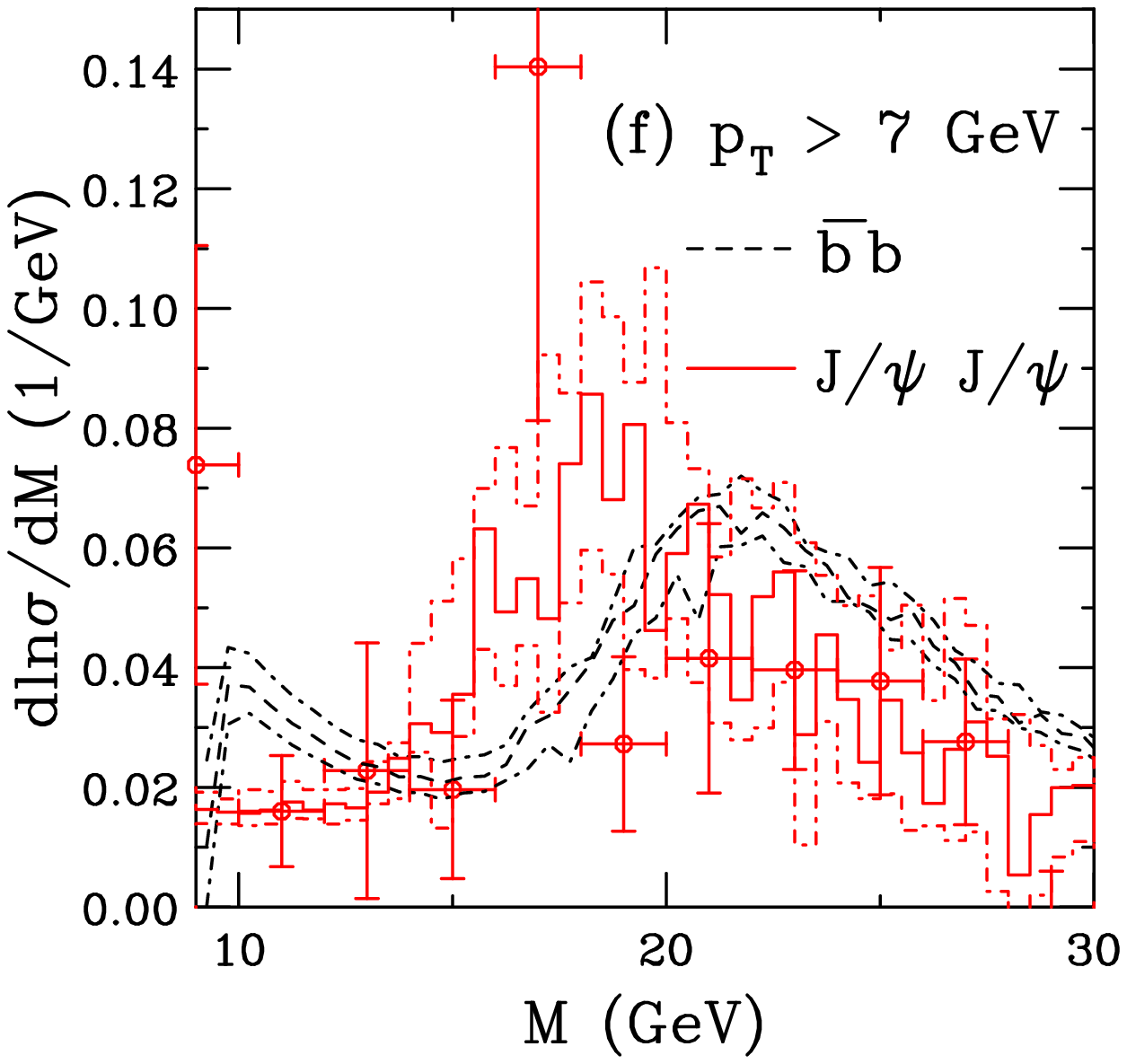} \\
\end{tabular}
  \caption[]{(Color online) The mass and scale uncertainty bands are shown for
the $b \overline b$ pairs (black dashed curves) and $J/\psi J/\psi$ pairs
  (red solid curves)
  and compared to the LHCb data \protect\cite{LHCb} for $p_T > 7$~GeV.
  The limits on the uncertainties are shown by
  dot-dashed curves in both cases.  Results are given for
  the azimuthal difference (a); rapidity difference (b); pair rapidity (c);
  $p_T$ asymmetry (d); pair $p_T$ (e); and pair mass (f).
  }
  \label{fig_bands_7GeV}
\end{figure}

Finally, the mass and scale uncertainties on the $b \overline b$ distributions
and their transition to the $J/\psi$ pair decay products are discussed here.
The  results for all pair observables are shown for the lowest and highest
minimum $p_T$ values, $p_T > 2$~GeV in Fig.~\ref{fig_bands_2GeV} and
$p_T > 7$~GeV in Fig.~\ref{fig_bands_7GeV}.

The mass and scale uncertainties are 
calculated 
based on results using the one standard deviation uncertainties on
the quark mass and scale parameters.  If the central, upper and lower limits
of $\mu_{R,F}/m$ are denoted as $C$, $H$, and $L$ respectively, then the seven
sets used to determine the scale uncertainty are  $\{(\mu_F/m,\mu_F/m)\}$ =
$\{$$(C,C)$, $(H,H)$, $(L,L)$, $(C,L)$, $(L,C)$, $(C,H)$, $(H,C)$$\}$.    
The uncertainty band can be obtained for the best fit sets
\cite{NVF,NVFinprep} by
adding the uncertainties from the mass and scale variations in 
quadrature. The envelope contained by the resulting curves,
\begin{eqnarray}
\frac{d\sigma_{\rm max}}{dX} & = & \frac{d\sigma_{\rm cent}}{dX} \label{sigmax}  \\
& & \mbox{} \!\!\!\!\!\!\!\!\!\!\!\!\!\!\!\!\!\!\!\!\!\!\!\!\!\!\!\!\!\!
+ \sqrt{\left(\frac{d\sigma_{\mu ,{\rm max}}}{dX} -
  \frac{d\sigma_{\rm cent}}{dX}\right)^2
  + \left(\frac{d\sigma_{m, {\rm max}}}{dX} -
  \frac{d\sigma_{\rm cent}}{dX}\right)^2} \, \, , \nonumber
\\
\frac{d\sigma_{\rm min}}{dX} & = & \frac{d\sigma_{\rm cent}}{dX} \label{sigmin}  \\
& & \mbox{} \!\!\!\!\!\!\!\!\!\!\!\!\!\!\!\!\!\!\!\!\!\!\!\!\!\!\!\!\!\!
- \sqrt{\left(\frac{d\sigma_{\mu ,{\rm min}}}{dX} -
  \frac{d\sigma_{\rm cent}}{dX}\right)^2
  + \left(\frac{d\sigma_{m, {\rm min}}}{dX} -
  \frac{d\sigma_{\rm cent}}{dX}\right)^2} \, \, , \nonumber
\end{eqnarray}
defines the uncertainty on the cross section.  Here $X$ is the individual pair
observable for a given minimum $p_T$. In the calculation labeled ``cent'', 
the central values of
$m$, $\mu_F$ and $\mu_R$ are used while in the calculations with subscript
$\mu$, the mass is fixed to the central value while the scales are varied and
in the calculations with subscript $m$, the mass is varied while the scales
are held fixed.  The central values of the bottom quark mass, $\mu_F/m$ and
$\mu_R/m$, as well as their one standard deviation uncertainties, can be found
in Sec.~\ref{sec:model}.

Note that in the calculation of the uncertainites in the normalized ratios,
all distributions are divided by the central value of the total cross section
before calculating the uncertainty as in Eqs.~(\ref{sigmax}) and (\ref{sigmin}).
This is consistent with calculating the uncertainty on the distributions via
these equations and then dividing by the central value of the integrated cross
section and is more consistent with the uncertainties obtained on the
distributions themselves \cite{lhc_ppb}.
If one instead divided by the total cross section for each mass and
scale combination, the uncertainties would be underestimated \cite{lhc_ppb}.

The mass and scale variations do not significantly change the shapes of the
distributions relative to the shape of the central distribution.  In the case
of bottom quark production, the mass uncertainties on the integrated cross
section are smaller than those due to the scales by a few percent.  The
uncertainties on the integrated $b \overline b$ cross section are smaller for
the higher $p_T$ cuts, decreasing by about a factor of 10 between $p_T > 2$~GeV
and 7~GeV.

The $J/\psi$ pair cross sections generally reflect these trends.  The $J/\psi$
pair integrated cross sections are smaller and decrase faster with $p_T$ cut,
resulting in a factor of $\approx 100$ decrease between minimum $p_T$ of 2 and
7~GeV.  This relative difference in cross section is due to the fact, as
mentioned previously, that
a $J/\psi$ satisfying the same minimum $p_T$ originates from a higher $p_T$
bottom quark.  In addition, due to the decay kinematics, some of the $J/\psi$'s
will no longer fall within the acceptance and $J/\psi$
pairs will not be measured.

\section{Rapidity dependence}
\label{sec:rap}

\begin{figure}[htpb]\centering
\begin{tabular}{cc}
  \includegraphics[width=0.5\columnwidth]{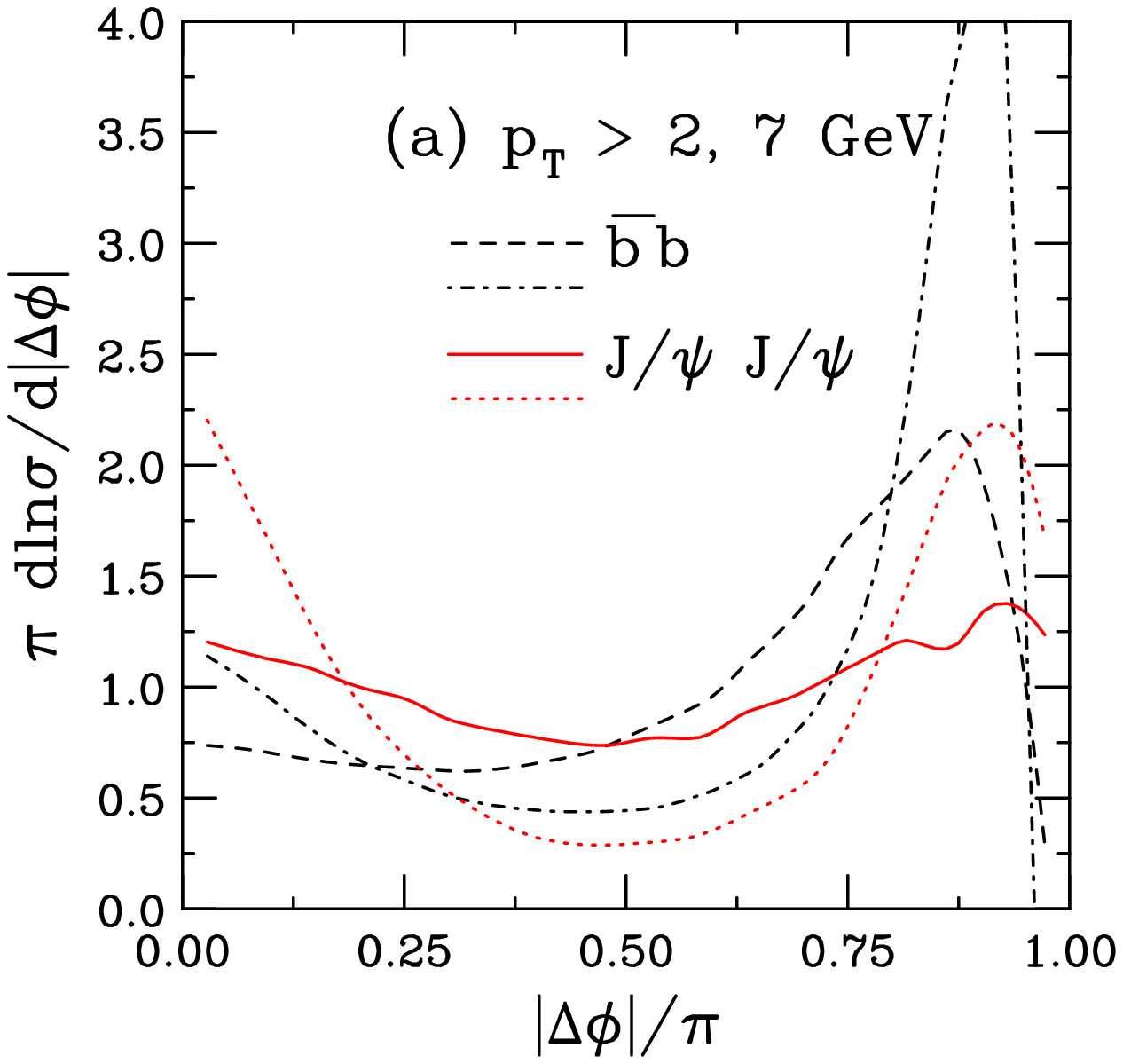} &
  \includegraphics[width=0.5\columnwidth]{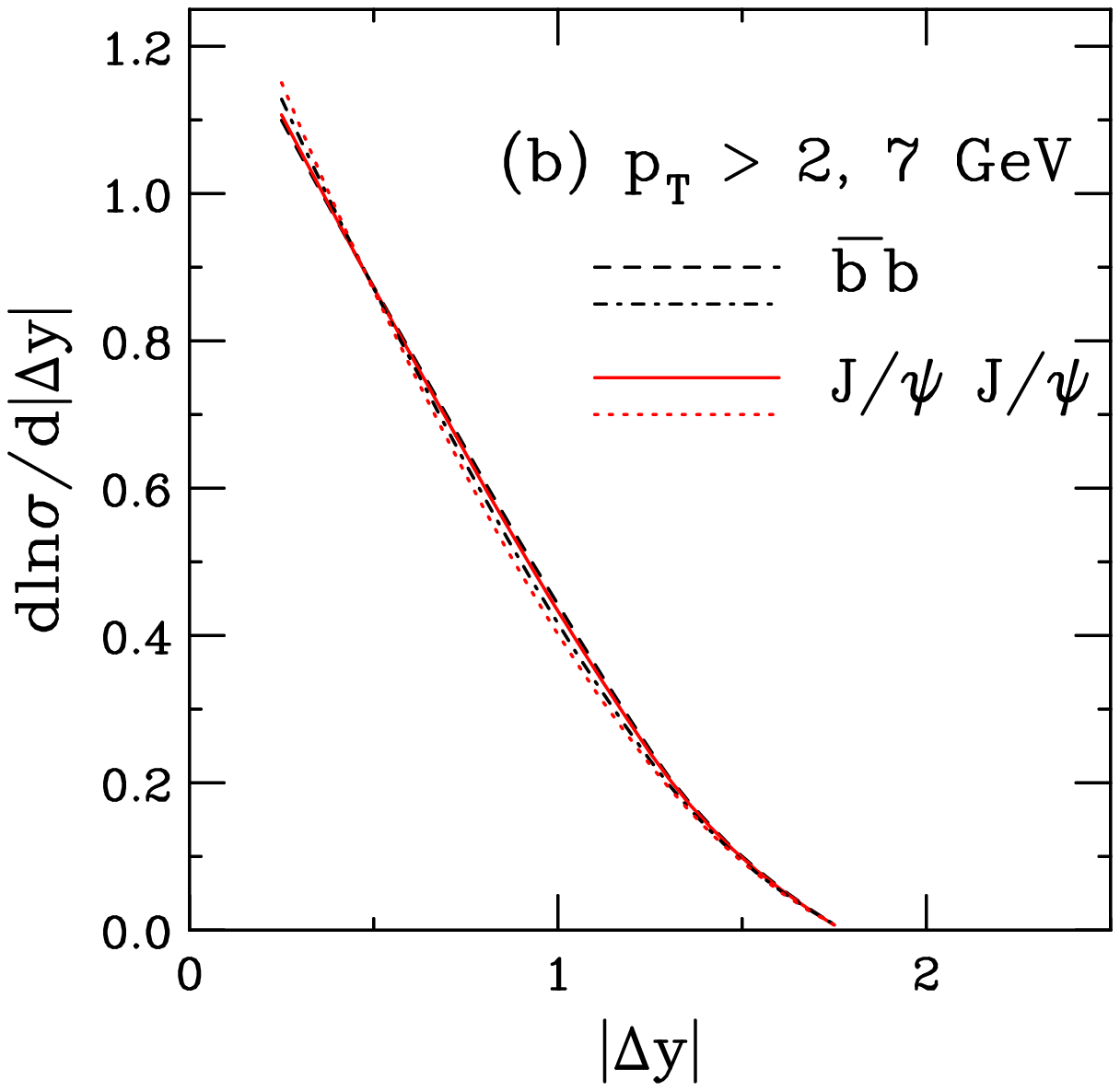} \\
  \includegraphics[width=0.5\columnwidth]{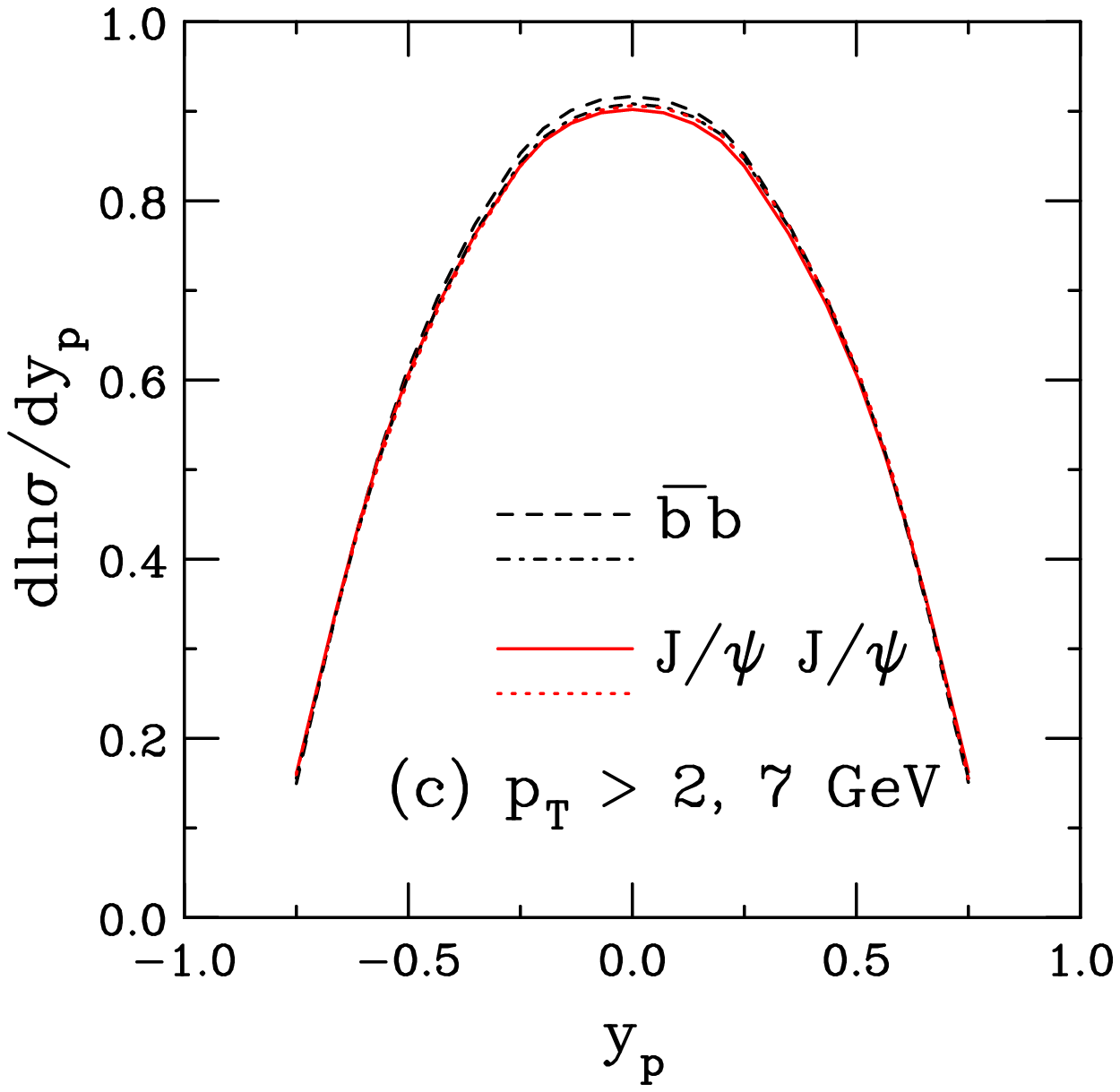} &
  \includegraphics[width=0.5\columnwidth]{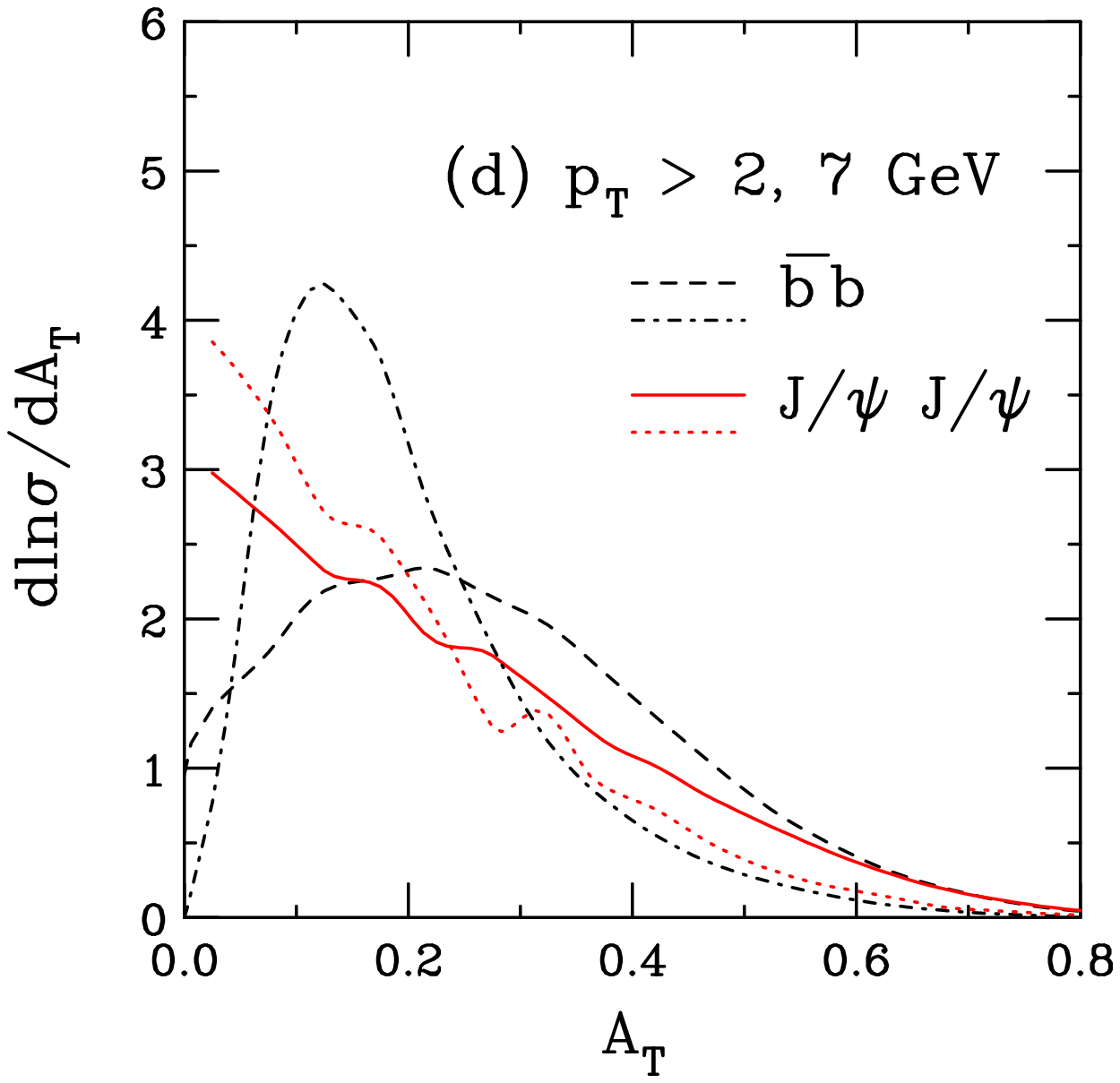} \\
  \includegraphics[width=0.5\columnwidth]{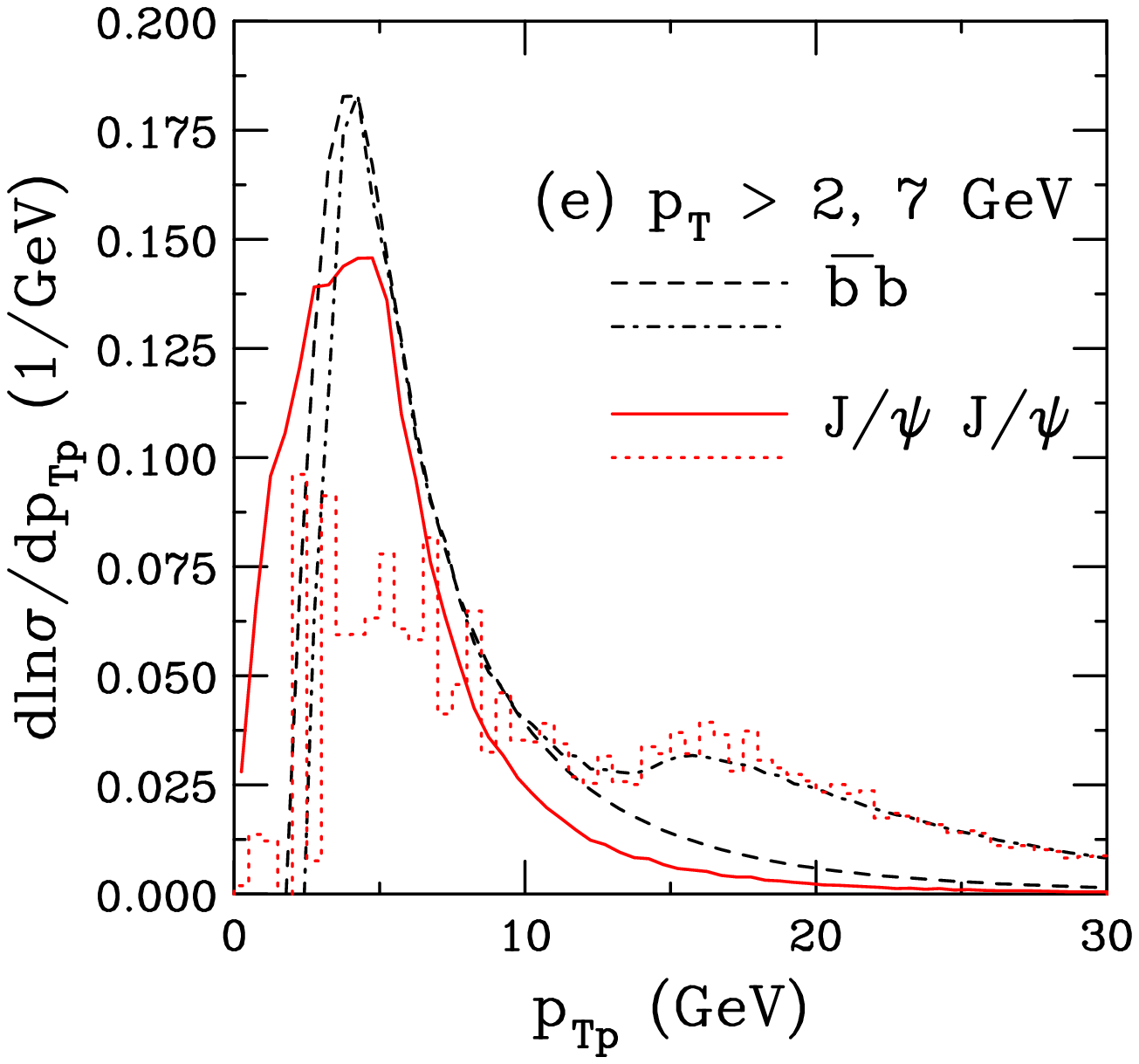} &
  \includegraphics[width=0.5\columnwidth]{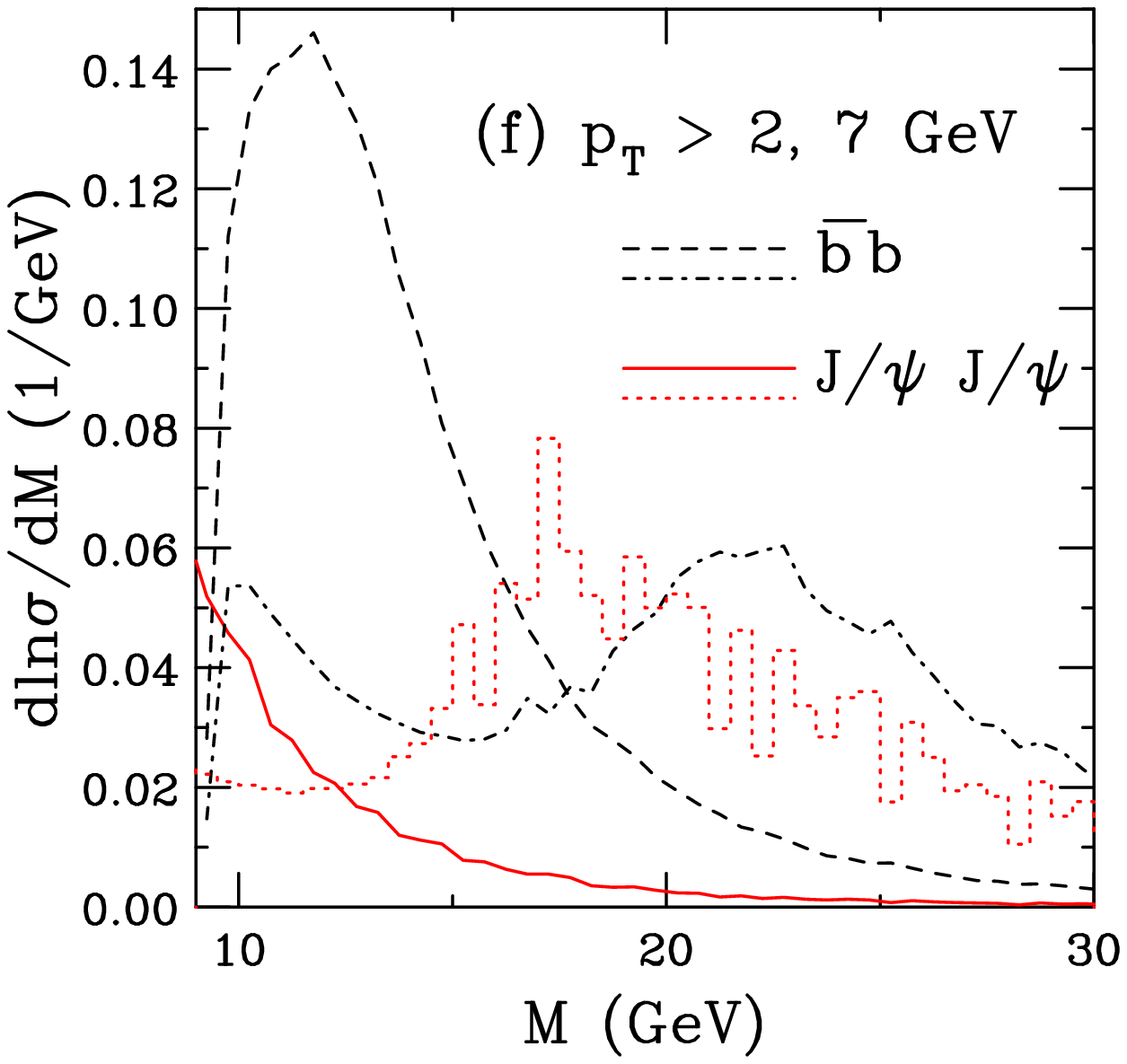} \\
\end{tabular}
  \caption[]{(Color online) The results are shown for
    $b \overline b$ pairs (black curves) and $J/\psi J/\psi$ pairs
    (red curves)
    for $p_T > 2$ (dashed for $b \overline b$ and solid $J/\psi$ ) and 7~GeV
    (dot dashed for $b \overline b$ and dotted $J/\psi$).
  Results are given for
  the azimuthal difference (a); rapidity difference (b); pair rapidity (c);
  $p_T$ asymmetry (d); pair $p_T$ (e); and pair mass (f).
  }
  \label{fig_cent_y}
\end{figure}

The rapidity dependence of the correlation is studied by calculating the same
pair quantities considered by LHCb in the midrapidity region, $|y| \leq 0.8$.
The results for both the parent $b \overline b$ and the $J/\psi$ pair decay
productions are shown in Fig.~\ref{fig_cent_y} for $p_T > 2$ and 7~GeV.

Generally, the shapes of the distributions at mid and forward rapidity are
rather similar.  While there are some differences between the shapes in the
two rapidity regions, they are not large.  Most of the differences are due to
the narrower rapidity acceptance employed at midrapidity, 1.6 units rather than
2.5 units at forward rapidity.

The main difference in the azimuthal distributions is the behavior at small
$|\Delta \phi|$.  The peak at $|\Delta \phi| \rightarrow 0$ is higher at
midrapidity, particularly for larger $p_T$ $b$ mesons and $J/\psi$s.  Thus the
narrower rapidity distribution covered at midrapidity seems to favor production
topologies where the $b \overline b$ is produced at smaller angles, {\it i.e.}
$gg \rightarrow b \overline b g$ where the final-state gluon is hard and
balanced against the $b \overline b$ pair.  In addition, at
$|\Delta \phi| \rightarrow \pi$, the back-to-back peak is narrower.  The
ratios of the distributions at $|\Delta \phi| = \pi$ to that at
$|\Delta \phi| = 0$ are reduced at midrapidity, particularly for the $J/\psi$
pairs where the results at $|\Delta \phi| = \pi$ and 0 are $\approx 1:1$ for both
$p_T$ values while it is $\approx 2:1$ for the azimuthal separation between
$J/\psi$ pairs at forward rapidity.  The differences between the peaks for
$b \overline b$ pairs are less pronounced between central and forward rapidities
but are still visible.

The rapidity gap distributions, $|\Delta y|$, are steeper in the chosen central
rapidity region, simply because the rapidity range is $\approx 1$ unit of rapidity
narrower than the forward region studied so far.  The pair rapidity
distribution, $y_p$, is now symmetric around $y_p = 0$ at midrapidity where
there is ample phase space for production.  On the other hand, at forward
rapidity, the average pair rapidity is not symmetric around the center of the
rapidity bin ($\langle y_p \rangle = 3.25$) but closer to the lower end of
the range with the calculated $\langle y_p \rangle = 3.08$. Regardless of the
rapidity region, the $|\Delta y|$ and $y_p$ distributions are independent of
$p_T$ minimum and whether $b \overline b$ or $J/\psi J/\psi$ pairs are
considered.

The average value of the $p_T$ asymmetry, $A_T$, is somewhat larger for central
rapidity.  On the other hand, the average pair $p_T$, $p_{T_p}$, is higher for
midrapidity.  This is not unexpected because the average $b$ hadron $p_T$ is
reduced at forward rapidity relative to central.  The average $p_{T_p}$ is
$\approx 1$~GeV higher at central rapidity for $p_T > 2$~GeV and 3~GeV higher for
$p_T > 7$~GeV.  Finally, the average pair mass does not vary significantly with
rapidity.  It increases slightly for forward rapidity, especially for
$p_T > 7$~GeV, likely because of the smaller $\Delta y$ for the midrapidity
interval chosen.

\section{Cold nuclear matter effects}
\label{sec:cnm}

In Ref.~\cite{QQazi}, the effects of shadowing and enhanced $k_T$ broadening
on $c \overline c$ production in cold nuclear matter, with the focus on
5.02~TeV $p+$Pb collisions, was studied.  As discussed there, it has been
suggested \cite{gossiaux,younus1,younus2,younus3}
that energy loss by heavy quarks in heavy-ion collisions could
change the azimuthal correlations.

However, it must first be ascertained how the
heavy flavor pair distributions are influenced by the presence of
cold nuclear matter.
For example, additional $k_T$ broadening may be present with a nuclear target
due to multiple scattering with nucleons along the path of the initial proton
(or nucleus).  The strength of the effect depends on the impact parameter of
the collision.  Energy loss in matter, on the other hand, may result in a
shift of the transverse momentum of the heavy quark, akin to a change in the
fragmentation function.  These effects would be in addition to the modification
of the parton densities in the nucleus, referred to as shadowing or nPDF
effects, calculated assuming collinear factorization.

Here the focus is on the parent $b \overline b$ correlations instead of their
decays to $J/\psi$ which do not convey as clear of an effect because the decay
is isotropic in the rest frame of the $b$ meson.
For illustrative purposes, two particular pair
variables are studied out of
the six discussed previously: the pair rapidity and the azimuthal separation.
Several different scenarios are studied: shadowing alone for both $p+$Pb and
Pb+Pb collisions with $\langle k_T^2 \rangle$ and $\epsilon_P$ given in
Eqs.~(\ref{eq:epsP}) and (\ref{eq:kt2});
enhanced broadening in $p+$Pb collisions; and enhanced
broadening with energy loss, represented by an increase in the Peterson function
parameter, in Pb+Pb collisions.  Shadowing is represented by the central
EPS09 NLO shadowing ratio \cite{EPS09} for each of the LHCb $p_T$ cuts.  As
noted in Ref.~\cite{QQazi} and in Figs.~\ref{fig_bands_2GeV} and
\ref{fig_bands_7GeV}, if $\langle k_T^2 \rangle$ is kept fixed, the mass and
scale uncertainties do not substantially change the shapes of the distributions.

The central gluon modification of the latest nPDF set, EPPS16
\cite{EPPS16}, is not significantly different from EPS09.  However, EPPS16 has
five additional parameters relative to EPS09, resulting in larger uncertainty
bands with an uncertainty of 25-30\% due to shadowing \cite{816TeV_pred}.
Although the uncertainty due to shadowing is significant, it is smaller than the
uncertainties due to the heavy quark mass and scale variations, particularly
for charm quarks \cite{lhc_ppb}.  The larger bottom quark mass and comparably
larger scales reduce both the overall uncertainty in the baseline $p+p$
cross section and the shadowing effect in $p+$Pb and Pb+Pb collisions
because of the larger parton momentum
fraction accessed and the evolution of the shadowing due to the larger
factorization scale.  To avoid overlapping
ratios in the following figures and better illustrate the effects, only results
with the central nPDF set are shown.

In Ref.~\cite{QQazi}, the sensitivity to the magnitude of $k_T$  broadening was
studied, varying the factor $\Delta$ in Eq.~(\ref{eq:kt2}) between 0 and 1.
So far, in this work, $\Delta = 1$ has been used as a default.
Here, to model broadening in medium, $\Delta = 2$ is used for $p+$Pb collisions
and $\Delta = 4$ is used in the Pb+Pb calculations relative to $p+p$ collisions
with $\Delta = 1$.  In the case with
`shadowing only', $\Delta = 1$ is still employed.  In addition, energy loss in
Pb+Pb collisions is modeled by changing the Peterson function parameter,
$\epsilon_P$ from the value used in these calculations heretofore,
$\epsilon_P = 0.0004$ \cite{QQazi}, to the default value used previously,
$\epsilon_P = 0.006$ \cite{Peterson}.  This change reduces the average $z$
in the Peterson function from 0.93 to 0.83, a difference of about 10\%.
See Ref.~\cite{QQazi} for the sensitivity of the single
$B$ meson $p_T$ distribution to $\epsilon_P$.

The calculations shown here are done at 8.16~TeV for $p+$Pb collisions and 5~TeV
for Pb+Pb collisions.  The $p+p$ results used to calculate the nuclear
modification factors, $R_{p {\rm Pb}}$ and $R_{\rm PbPb}$ respectively, are
calculated at the same energies.  The results are calculated both at central
$(|y| \leq 0.8)$ and
forward ($2 < y < 4.5$) rapidity.

Note that there is a
rapidity shift in $p+$Pb collisions due to the requirement of equal velocity
beams at the LHC.  The calculations shown assume the proton is moving in the
positive $y$ direction so that the parton
momentum fraction, $x$, probed by the nucleus, is low.  The change in the
shadowing ratios is then
small.  If the beam directions were switched, the momentum
fraction in the nucleus is in the antishadowing region.  In Pb+Pb collisions,
the parton from the forward-going nucleus is large
(in the antishadowing region) while that in the backward-going direction is
small (shadowing) and the collisions are again forward-backward symmetric, as
in $p+p$.

First, the $p+p$ distributions calculated at 5 and 8.16~TeV were checked to see
if the shapes of the pair distributions were modified at different energies.
The shapes
remain the same for all $p_T$ cuts, even at the lowest energy and highest
minimum $p_T$.  Note that this agreement will eventually break down at lower
energies, especially for higher $p_T$, as one reaches the edge of available
phase space, particularly at forward rapidity.  Because the shapes of the
distributions remain unchanged between 5 and 8.16~TeV, these results are not
illustrated.

There will be some uncorrelated background to the correlated
calculations shown here, particularly in ion-ion collisions.  The background
would be larger for $c \overline c$ pairs due to the larger production cross
section. Scaling $p+p$ production by the number of binary nucleon-nucleon
collisions, several
hundred $c \overline c$ pairs can be produced in a single Pb+Pb collision
at the LHC \cite{RVcent}.  This uncorrelated background would be reduced for
$b \overline b$ production because of its relatively smaller production cross
section: only a few $b \overline b$ pairs would be produced in a given Pb+Pb
event.  Even so, the correlated pair signals suggested here could be
substantially washed out if they are not seen to be arising from a common
decay vertex.  Uncorrelated production may also arise in high multiplicity
$p+p$ and $p+$Pb events which could also
affect the proposed correlation in these collisions.
In lepton pair channels, uncorrelated background could be
removed by like-sign pair subtraction \cite{RVdilep2} although, for
$b \overline b$ production, correlated $b \overline b$ pair decays can also
lead to like-sign lepton pairs \cite{Andre}.

Aside from independent uncorrelated production, two relatively independent
$Q \overline Q$ pairs can be produced in double parton scattering in all these
collision systems.  The double parton scattering contribution has been
calculated in Ref.~\cite{KPfins} for $D \overline D$ and $D D$ production.  The
probability of such contributions should be reduced for bottom pair production
due to the larger bottom quark mass and higher associated scales.

This section is divided into three subsections.  To set the stage, first the
single $b$ meson modifications are shown as a function of $p_T$ for all four
cases ($p+$Pb with shadowing alone; $p+$Pb with shadowing and $\Delta = 2$;
Pb+Pb with shadowing only; and Pb+Pb with shadowing, $\Delta = 4$ and
$\epsilon_P = 0.006$) at both forward and central rapidity.  The pair results
are then shown for
the pair rapidity and the azimuthal separation between the heavy mesons.
Here the nuclear modifications are shown for forward and central
$p+$Pb and Pb+Pb collisions, both with shadowing alone and then including
enhanced $k_T$
broadening, as well as modification of the fragmentation function in Pb+Pb
collisions.  However, now the results are shown for the same minimum $p_T$ cuts
on the $b$ mesons used by LHCb for their
$b \overline b \rightarrow J/\psi J/\psi$ analysis.  All results will be
presented as the modification factors, $R_{p {\rm Pb}}$ and
$R_{\rm PbPb}$, calculated as the per nucleon cross section in $p+$Pb and Pb+Pb
collisions respectively relative to the $p+p$ result at the same energy.

It has already been
noted that there is no modification of
the $p+p$ distributions, $d\ln \sigma/dX$,
as a function of center of mass energy.  However, some
modification of the distributions 
can be observed in $p+$Pb and Pb+Pb collisions relative to $p+p$,
as will also be shown.
Differences can arise with nuclear beams because
the momentum fraction probed changes with changing $\sqrt{s}$.  The $\approx 40$\%
increase in $\sqrt{s}$ between 5 and 8.16~TeV reduces the $x$ values
correspondingly.  Thus the shadowing effect could potentially lead to a
modification, especially in Pb+Pb collisions where one of the lead nuclei is
probed
at relatively high momentum fraction, $x \approx 0.02$, in the forward rapidity
region.  Choosing a higher $p_T$ cut also probes higher $x$ and larger
scales.  Nonetheless, shadowing alone does not modify the shapes of the
distributions at different energies.  Observable differences only appear with
enhanced broadening or modification of the fragmentation function, as is
discussed in the remainder of this section.

\subsection{Modification of single $b$ meson $p_T$ spectra}
\label{sec:single_b}

\begin{figure}[htpb]\centering
\begin{tabular}{cc}
  \includegraphics[width=0.5\columnwidth]{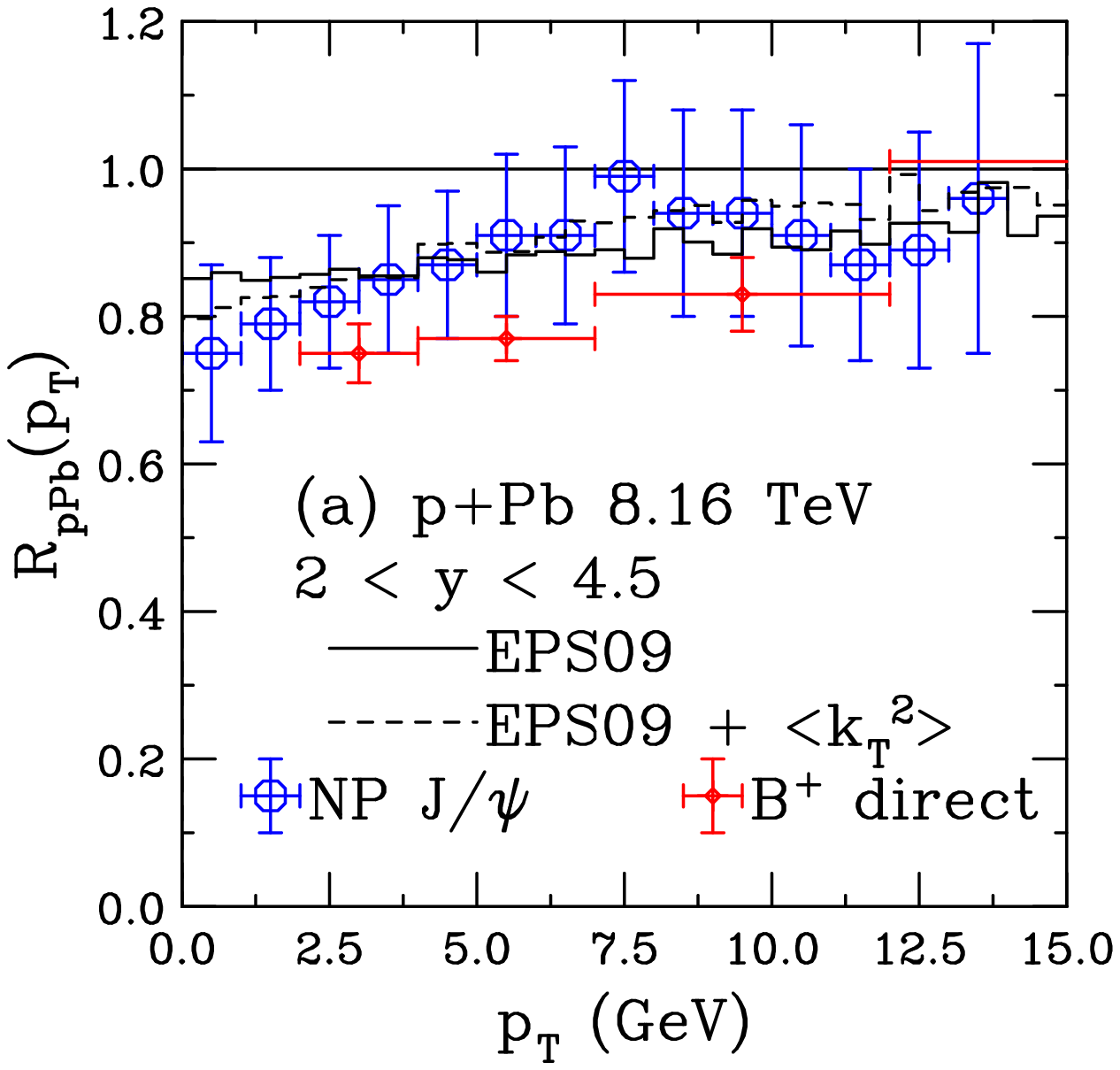} &
  \includegraphics[width=0.5\columnwidth]{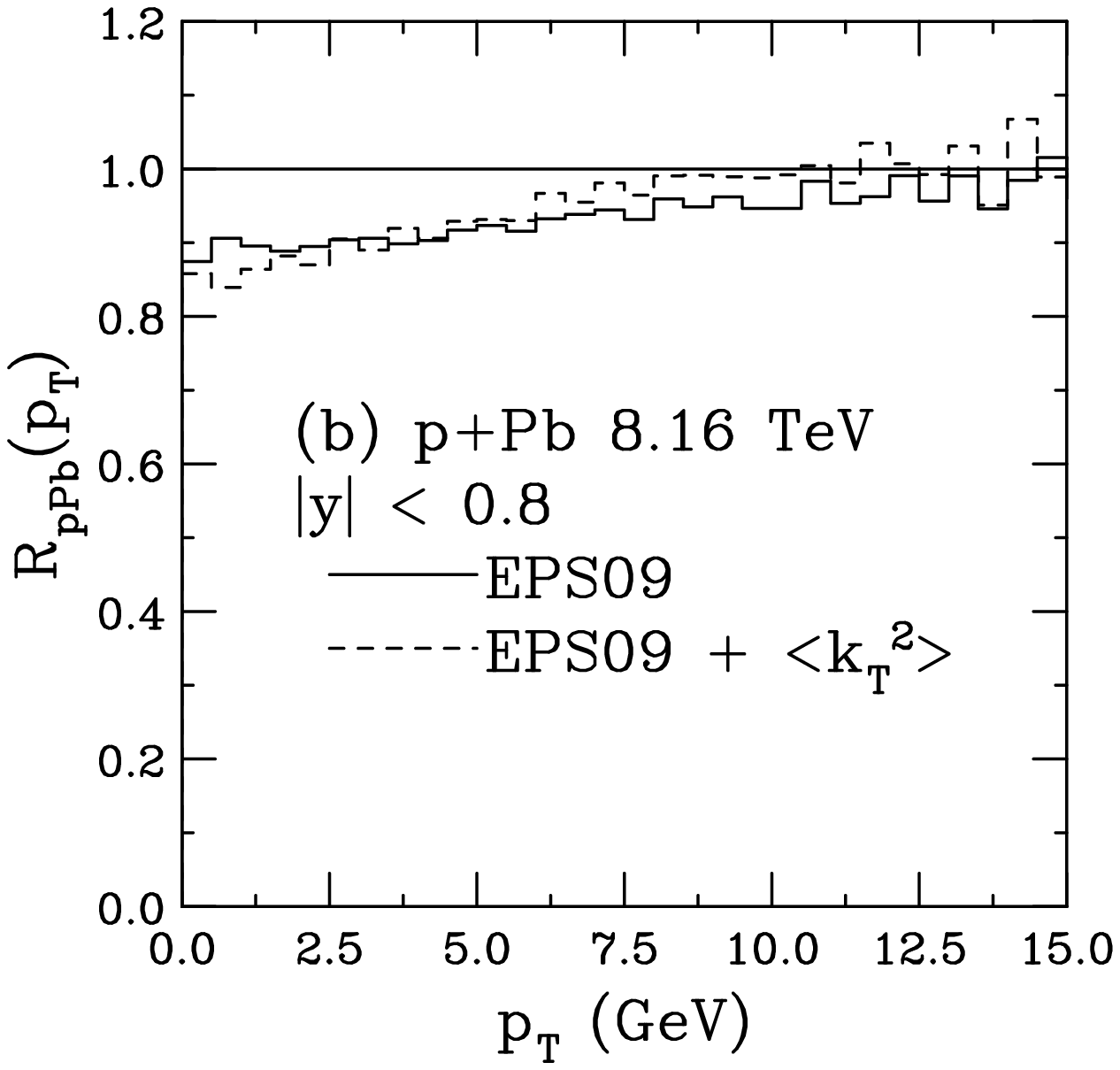} \\
  \includegraphics[width=0.5\columnwidth]{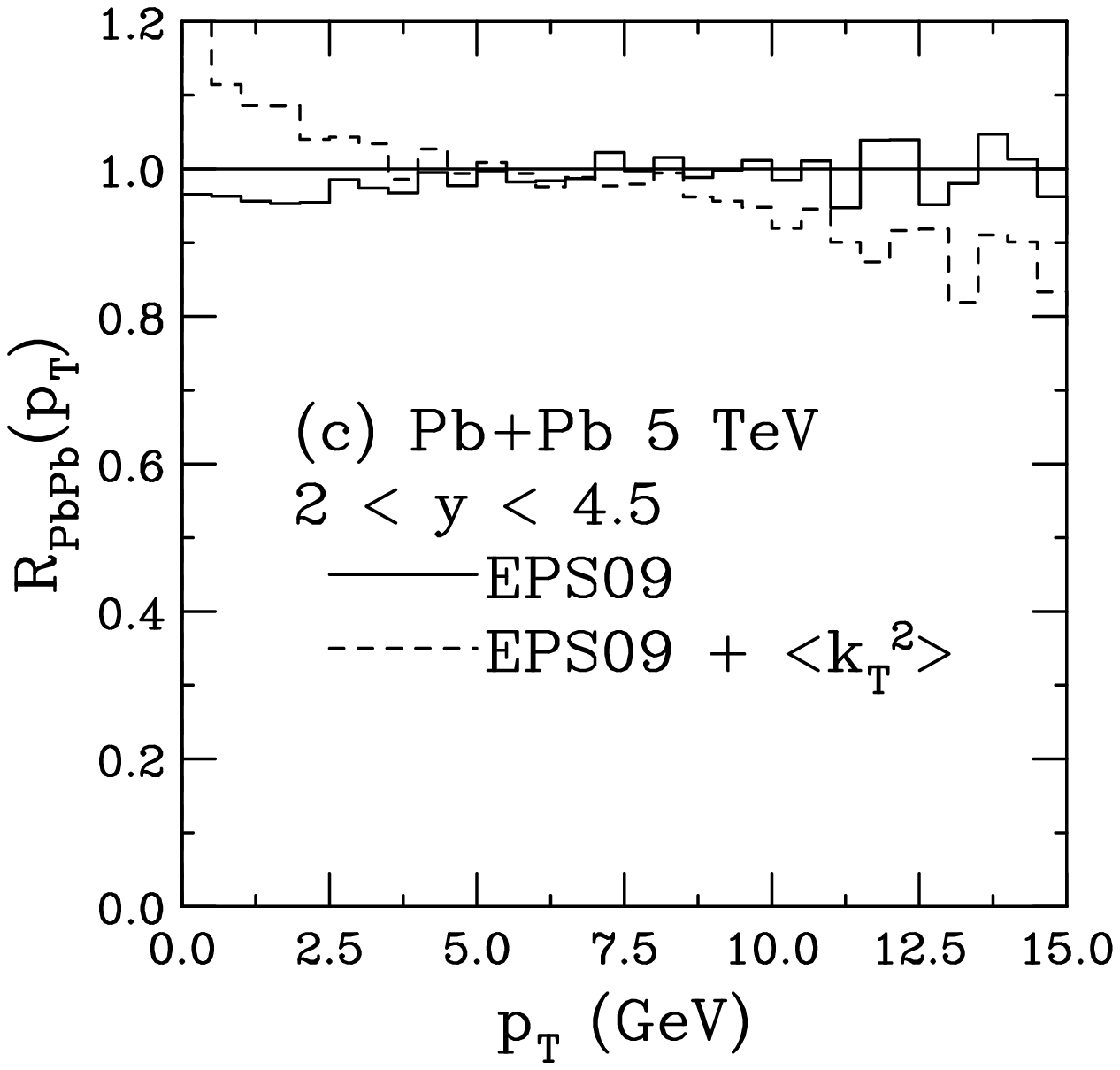} &
  \includegraphics[width=0.5\columnwidth]{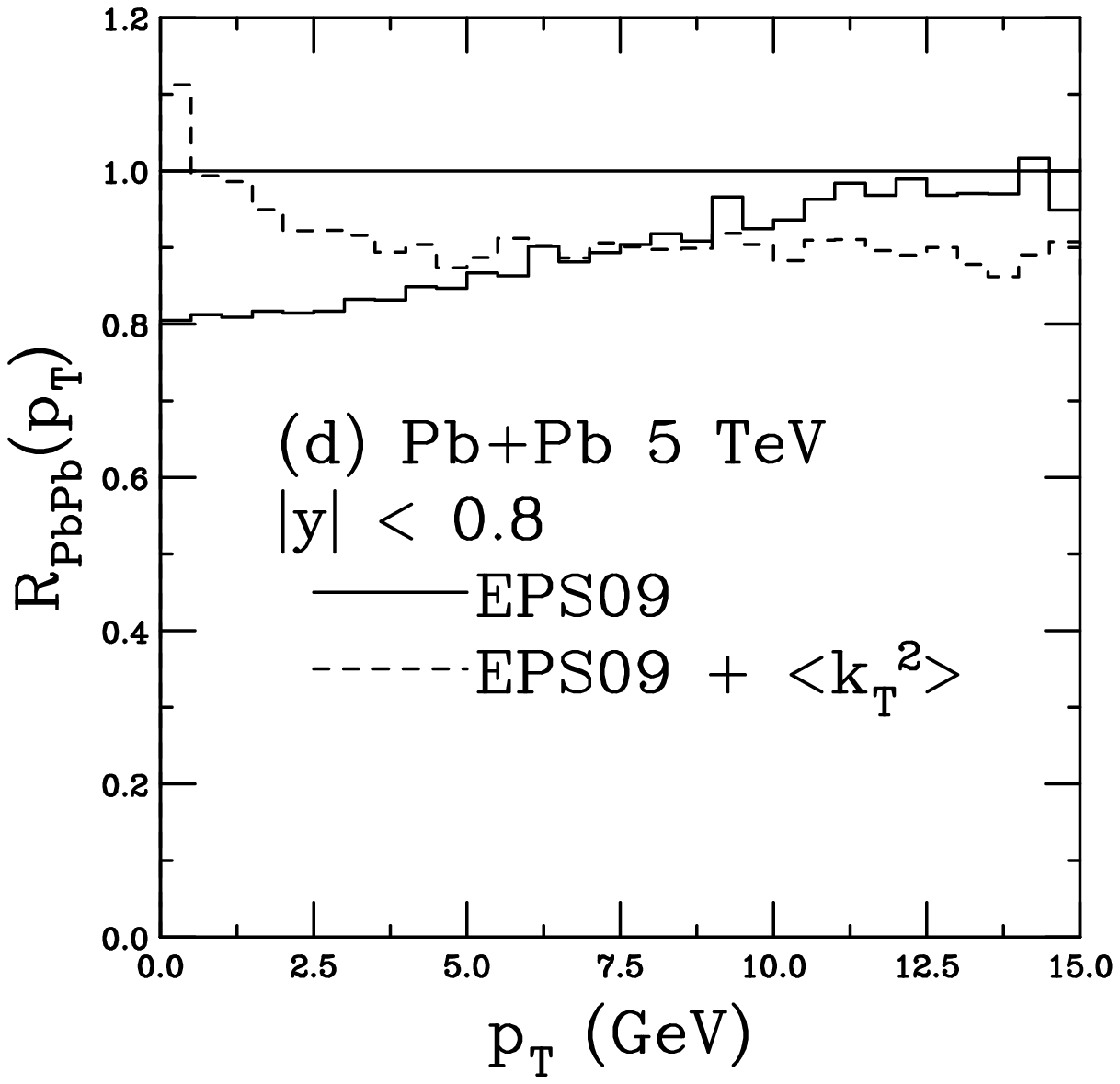} \\
 \end{tabular}
\caption[]{(Color online) Cold nuclear matter effects on $b$ quark $p_T$
  distributions for (a) and (b) $p+$Pb collisions
  at 8.16~TeV with central EPS09 and the same $k_T$ kick as in $p+p$ (solid)
  and additional $k_T$ broadening in Pb (dashed); (c) and (d) Pb+Pb collisions
  at 5~TeV with central EPS09 with the same $k_T$ kick in $p+p$ and Pb+Pb
  (solid) and additional $k_T$ broadening in the Pb nuclei with a modified
  fragmentation function in Pb (dashed).  Results are shown for forward
  rapidity in (a) and (c), central rapidity in (b) and (d).  In (a) the
  calculations are compared to the LHCb data on nonprompt $J/\psi$s
  \protect\cite{LHCb_NPpsi} and direct $B^+$ \protect\cite{LHCb_Bp}.
  }
  \label{fig_cnm_bptdist}
\end{figure}

Figure~\ref{fig_cnm_bptdist} shows the ratios $R_{p {\rm Pb}}(p_T)$ (a), (b) and
$R_{\rm PbPb}(p_T)$ (c), (d) at forward (a), (c) and central (b), (d) rapidities
for single $b$ mesons.  These calculations can inform the results shown later
for $b \overline b$ pairs.

The results for $p+$Pb collisions at forward rapidity
are compared to LHCb data from nonprompt $J/\psi$ \cite{LHCb_NPpsi} and
direct $B^+$ mesons \cite{LHCb_Bp}.  The calculations, using the central EPS09
NLO set only, with or without any additional $k_T$ kick, agree very well with
the LHCb data, especially that for nonprompt
$J/\psi$s, shown in blue.  While the direct $B^+$ data, shown in red, are
within one standard deviation of the nonprompt $J/\psi$ uncertainty, they are
below the central EPS09 NLO calculation for $p_T < 7$~GeV.  If the full
EPS09 NLO uncertainty band was shown, however, the $B^+$
data should be within the limits of the calculated band.  The
ratios including a higher $k_T$ kick for $p+$Pb collisions are very similar to
those with shadowing only, similar to the midrapidity calculations shown for
single $D$ mesons at midrapidity in Ref.~\cite{QQazi}.
The maximum shadowing effect on $b$ mesons at low $p_T$
is $\approx 15$\% at forward rapidity and $\approx 10$\% at midrapidity.

The single bottom $R_{p {\rm Pb}}(p_T)$ at forward, backward and mid-rapidity
was also calculated
for shadowing only at 8~TeV \cite{LandsbergShao} employing a $p_T$ and rapidity
dependent parameterization of the $p+p$ cross section.  Bands were shown for
several nuclear parton densities and compared to calculations with $k_T$
broadening and energy loss \cite{Vitev} in Ref.~\cite{816TeV_pred}.  The
shadowing parameterizations employed in Ref.~\cite{816TeV_pred} ranged from 
$\approx 5-40$\% effects at $p_T \approx 5$~GeV at forward rapidity with a slightly
weaker effect at central rapidity.  On the other hand, for the same $p_T$, a
$0-8$\% enhancement was seen for the calculations with broadening and energy
loss \cite{816TeV_pred}.

While not shown, the calculated ratio at backward rapidity is subject
to $5-20$\% antishadowing at low $p_T$.  This is consistent with the
$p_T$-integrated modification
factor $R_{p {\rm Pb}}$ shown as a function of rapidity in
Ref.~\cite{816TeV_pred}: antishadowing at backward rapidity and increasing
shadowing at central and forward rapidity.  The LHCb $B$ meson data at forward
and backward rapidity are consistent with this trend \cite{LHCb_NPpsi,LHCb_Bp}.

The NLO results calculated for $p+$Pb collisions shown here with the
central EPS09 NLO set at forward and central rapidity are in good agreement
with the bands shown for the parameterizations with EPS09 NLO shadowing applied
to the parameterization of the cross section in
Refs.~\cite{LandsbergShao,816TeV_pred}.  

The results shown in Fig.~\ref{fig_cnm_bptdist}(c) and (d) for Pb+Pb collisions
are, perhaps, somewhat more surprising.  First, for shadowing only,
it is notable that, at forward
rapidity, $R_{\rm PbPb}(p_T)$ is approximately unity with a negligible $p_T$
dependence while the modification factor is a rather strong function of $p_T$
at midrapidity.  These results are easily explained, however.  The Pb+Pb
modification factor at forward rapidity is the product of the
$p+$Pb results at forward and backward rapidity,
$R_{AA}(y; X) = R_{p A}(y; X) \times R_{p A}(-y; X)$ \cite{Andronic} where $X$ is
another kinematic quantity such as $p_T$ or $|\Delta \phi|$.

The combination of $\approx 15$\%
suppression at forward rapidity with a $\approx 15$\% enhancement at backward
rapidity is effectively unity.  (Because Pb+Pb collisions are symmetric about
midrapidity, one would see a similar modification factor for Pb+Pb collisions at
backward rapidity.)  On the other hand, the midrapidity shadowing results are
effectively the $R_{p {\rm Pb}}(p_T)$ result in Fig.~\ref{fig_cnm_bptdist}(b)
squared.  Note that the analogy between $R_{\rm PbPb}$ and the product of
$R_{p {\rm Pb}}$ at forward and backward rapidity here is not exact because of the
different energies of the two collisions: the lower energy Pb+Pb collisions
would be at higher $x$ than the corresponding rapidity in $p+$Pb collisions.
However, the difference between the two should be small.

The Pb+Pb calculations shown in the dashed histograms, including increased
broadening and a modified fragmentation function parameter, exhibit quite
different behavior.  As shown in Figs.~\ref{fig_cnm_bptdist}(a) and
(b), increasing the relative $k_T$ broadening in the lead nucleus does not
strongly change the modification factor.  Therefore the change in slope seen in
these results is due to the change in the fragmentation parameter $\epsilon_P$.
Increasing $\epsilon_P$ changes the slope of the $p_T$ distribution, enhancing
the low $p_T$ part of the spectrum and depleting the high $p_T$ contribution,
see the $b$ meson $p_T$ distributions in
Ref.~\cite{QQazi}.  Thus, this behavior, while perhaps initially surprising,
is easily understood.

In the following subsections, the pair results will be presented.  In these,
the $p_T$ cuts used by LHCb are applied.
One must keep in mind that pair quantities are all integrated over $p_T$ from
the minimum value and thus probe, on average, higher $p_T$ and, consequently,
somewhat larger $x$ than the single meson
quantities shown here.    

\subsection{Modifications of $y_P$}

\begin{figure}[htpb]\centering
\begin{tabular}{cc}
  \includegraphics[width=0.5\columnwidth]{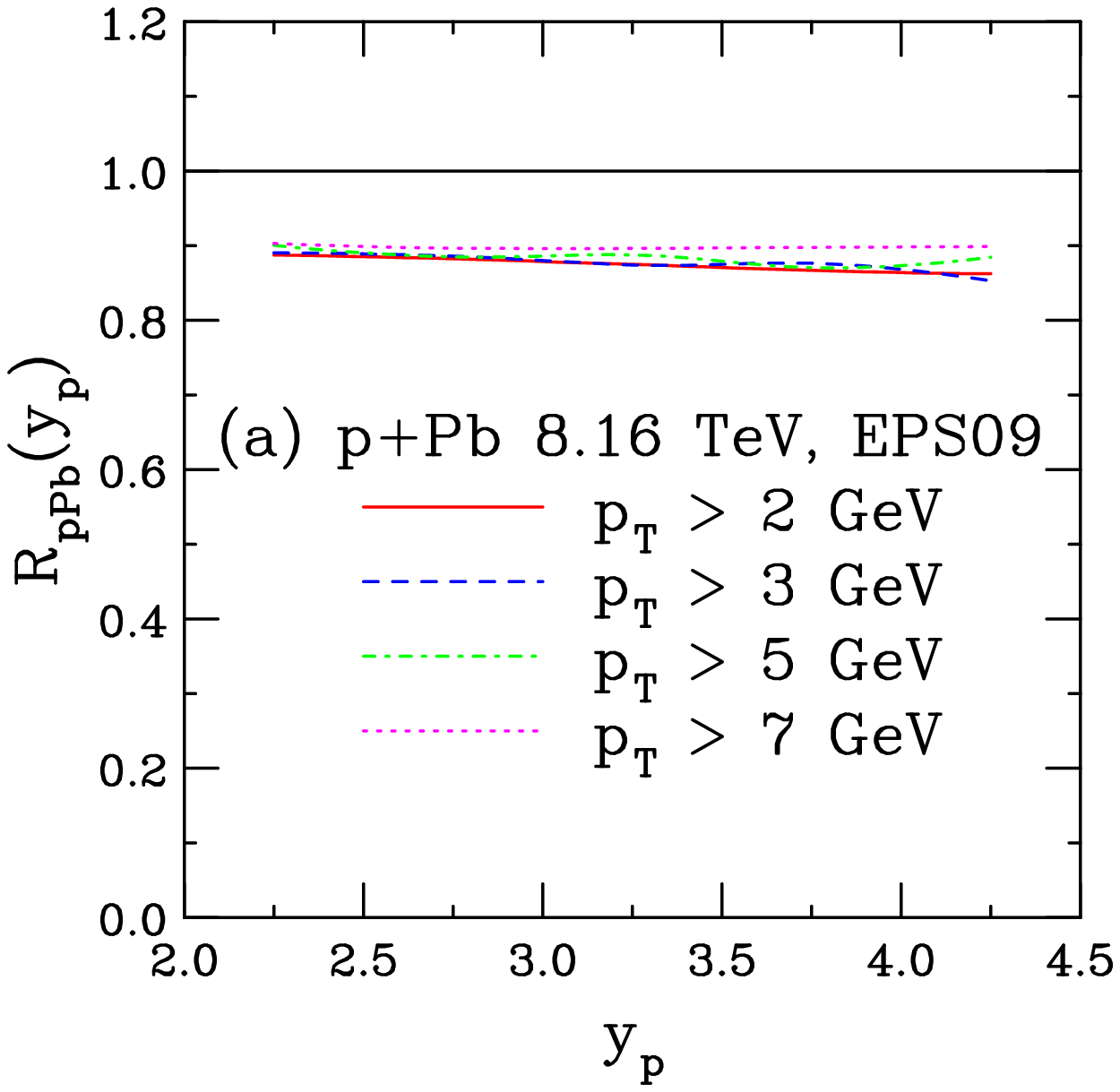} &
  \includegraphics[width=0.5\columnwidth]{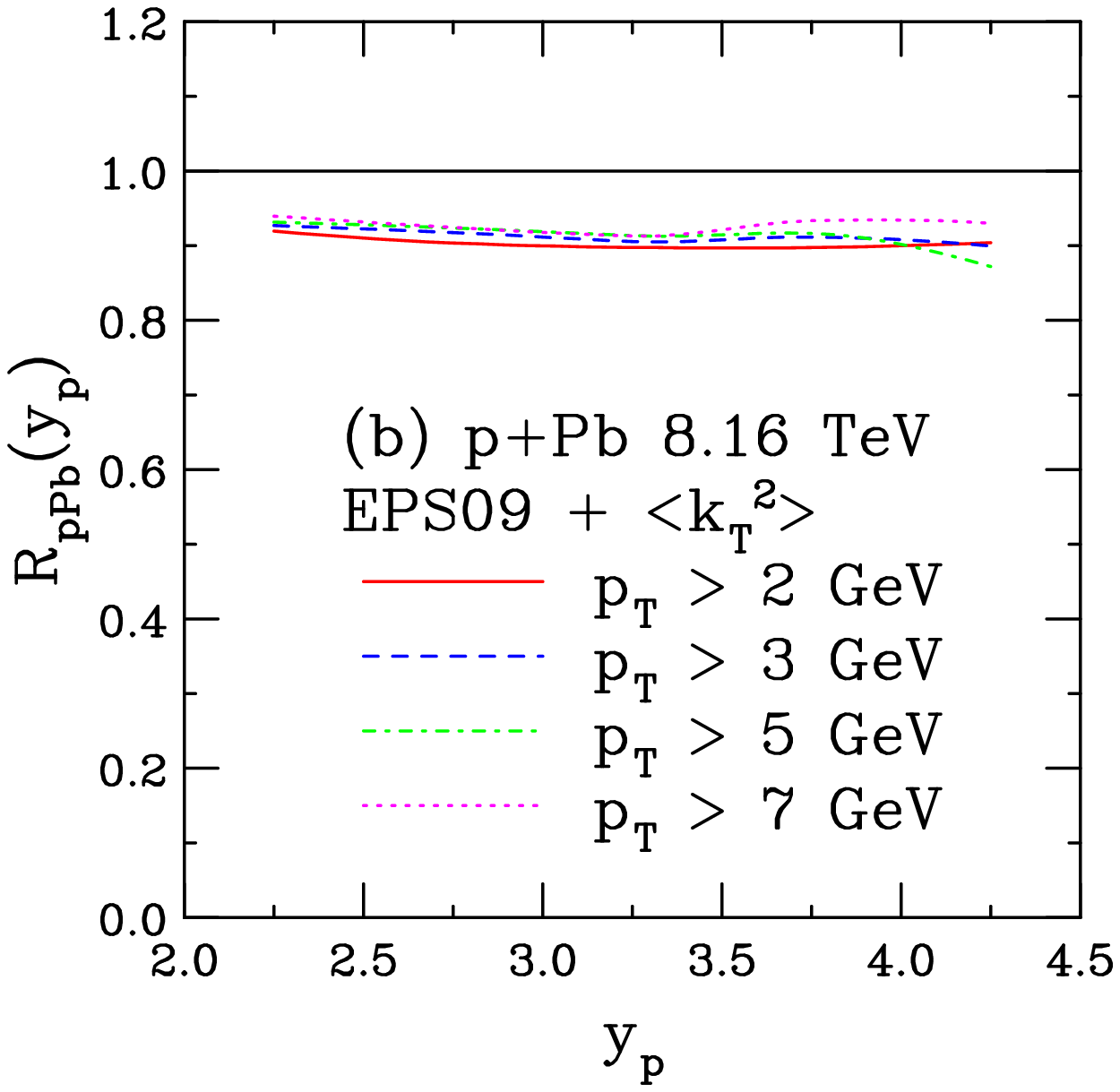} \\
  \includegraphics[width=0.5\columnwidth]{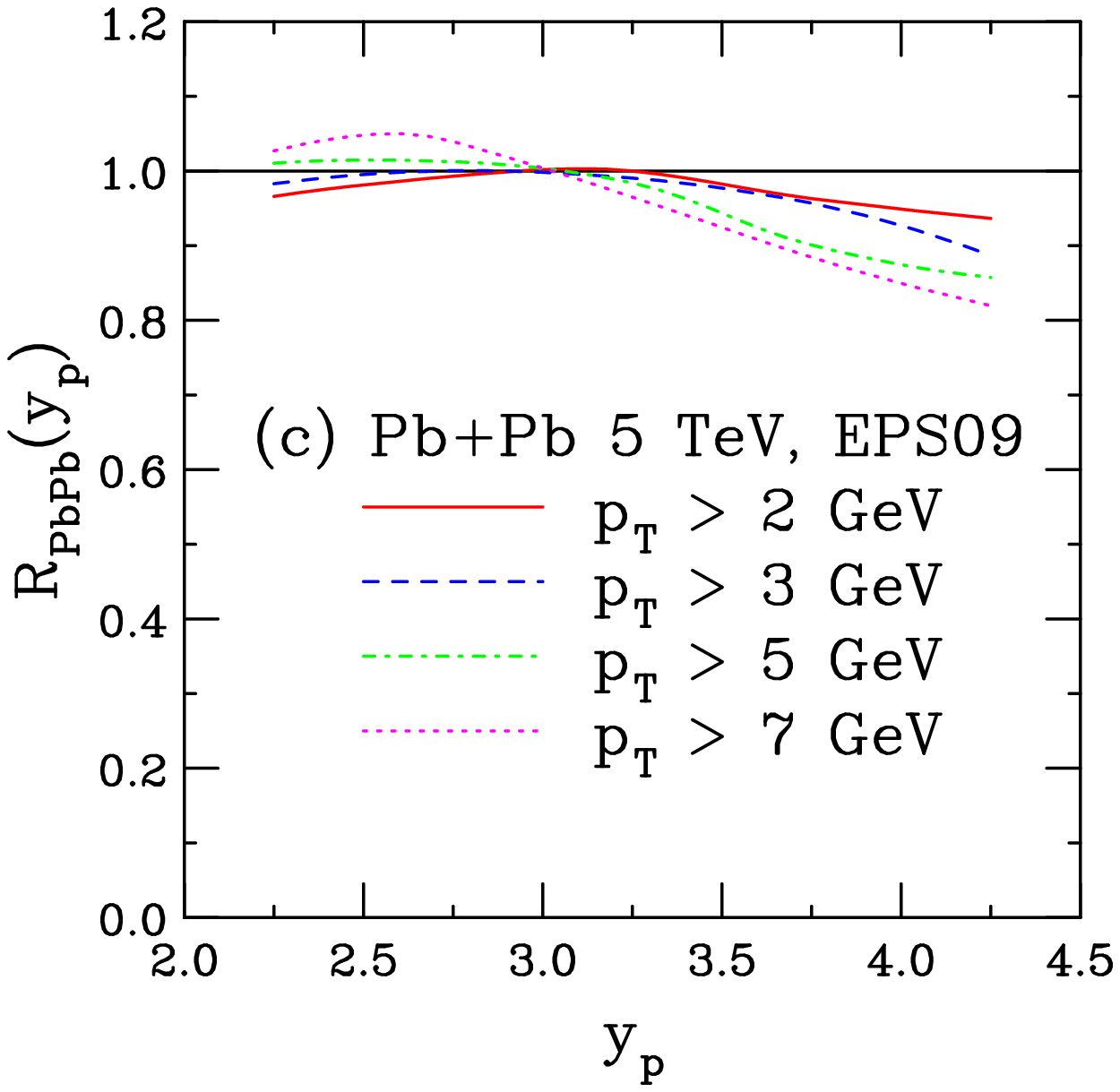} &
  \includegraphics[width=0.5\columnwidth]{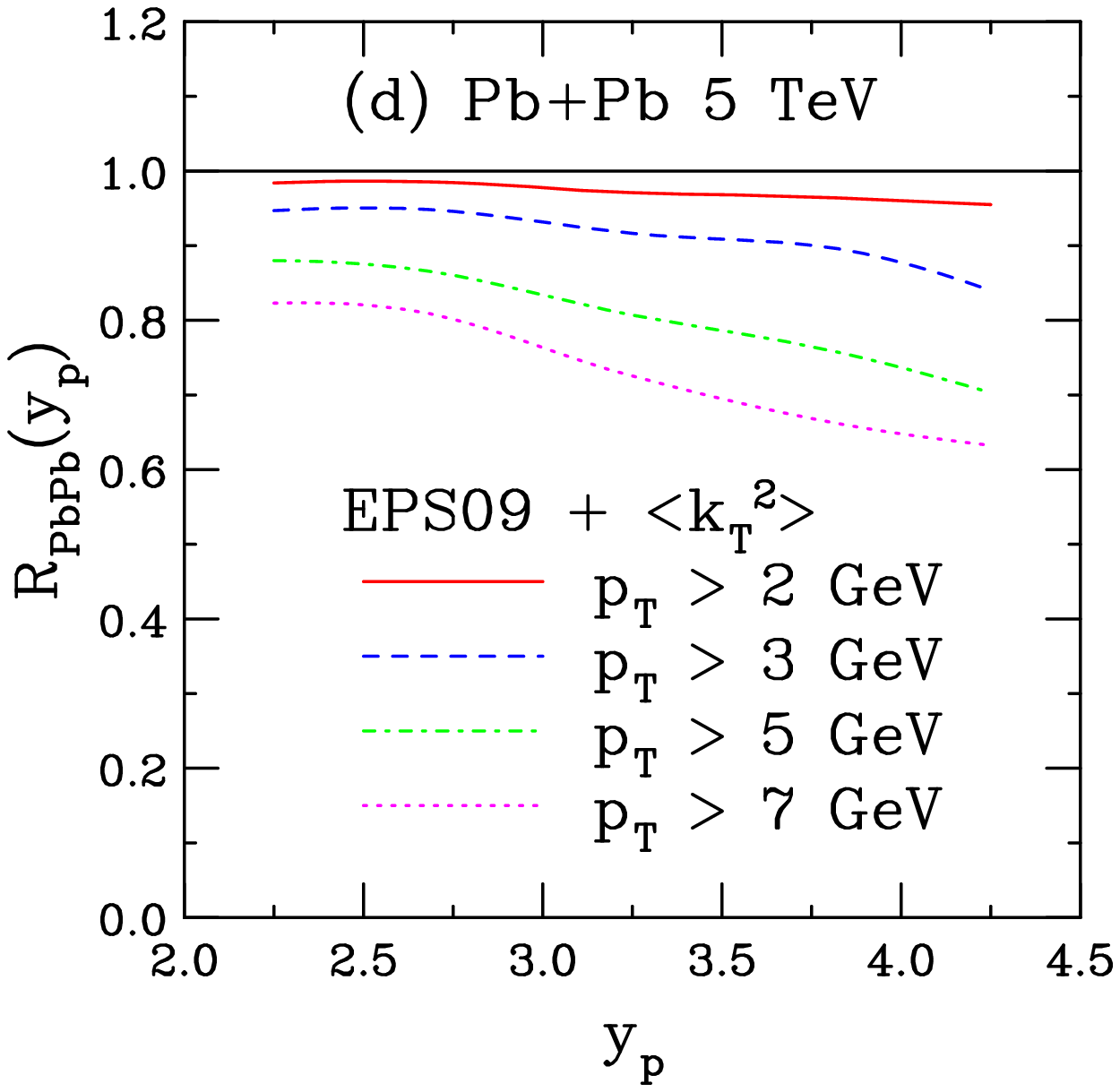} \\
 \end{tabular}
\caption[]{(Color online) Cold nuclear matter effects at forward rapidity
  ($2 < y < 4.5$) on the $b \overline b$
  pair rapidity for $p_T > 2$ (solid red), 3 (dashed blue),
  5 (dot-dashed green), and 7~GeV (dotted magenta) for (a) $p+$Pb collisions
  at 8.16~TeV with central EPS09 and the same $k_T$ kick as in $p+p$;
  (b) $R_{p{\rm Pb}}$ at 8.16 TeV with EPS09
  and additional $k_T$ broadening in Pb; (c) Pb+Pb collisions at 5~TeV with
  central EPS09 with the same $k_T$ kick in $p+p$ and $p+$Pb; and (d)
  $R_{AA}$ at 5~TeV with EPS09, additional $k_T$ broadening in the Pb nuclei,
  and a modified fragmentation function in Pb.
  }
  \label{fig_cnm_yqq}
\end{figure}

\begin{figure}[htpb]\centering
\begin{tabular}{cc}
  \includegraphics[width=0.5\columnwidth]{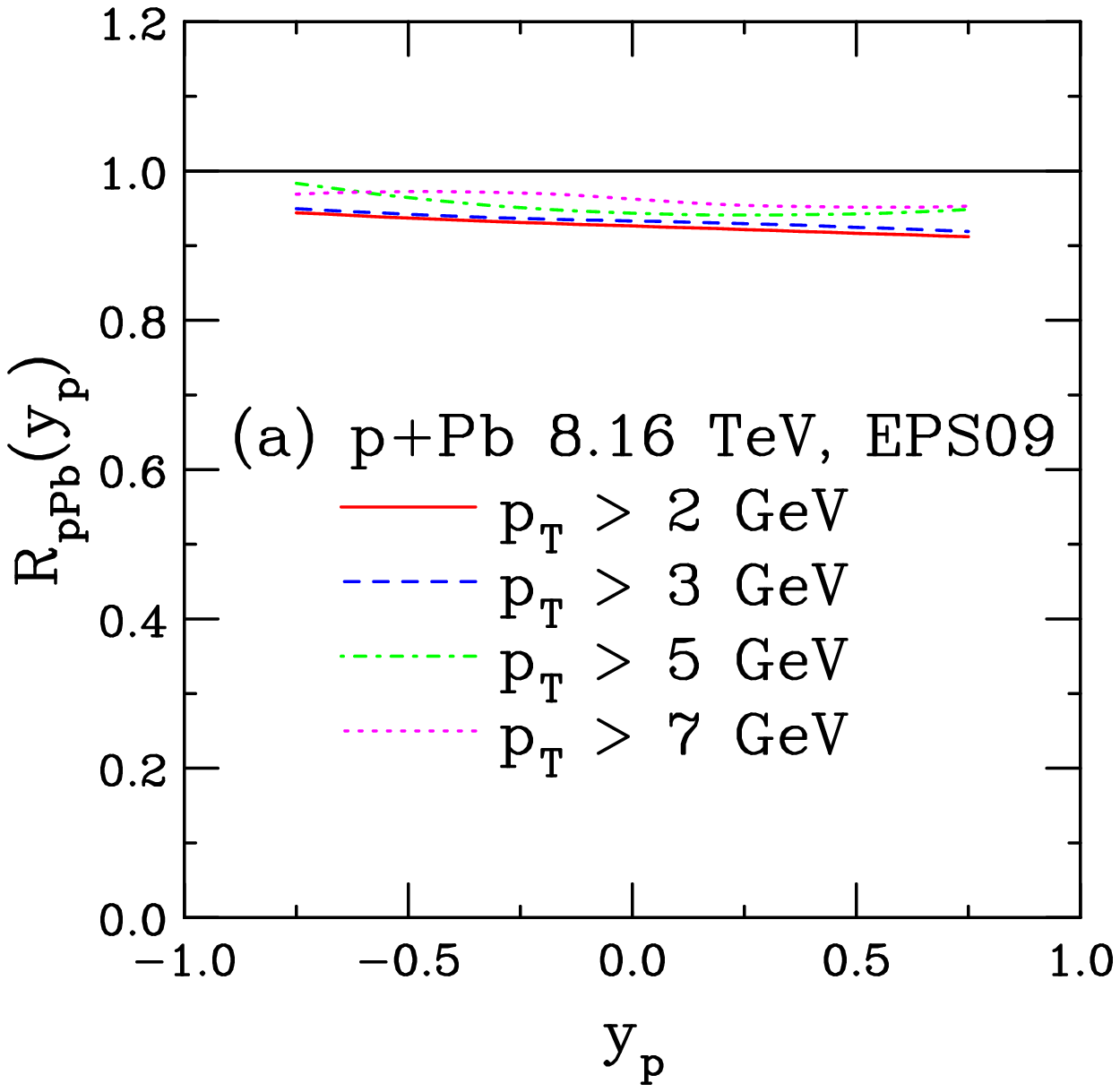} &
  \includegraphics[width=0.5\columnwidth]{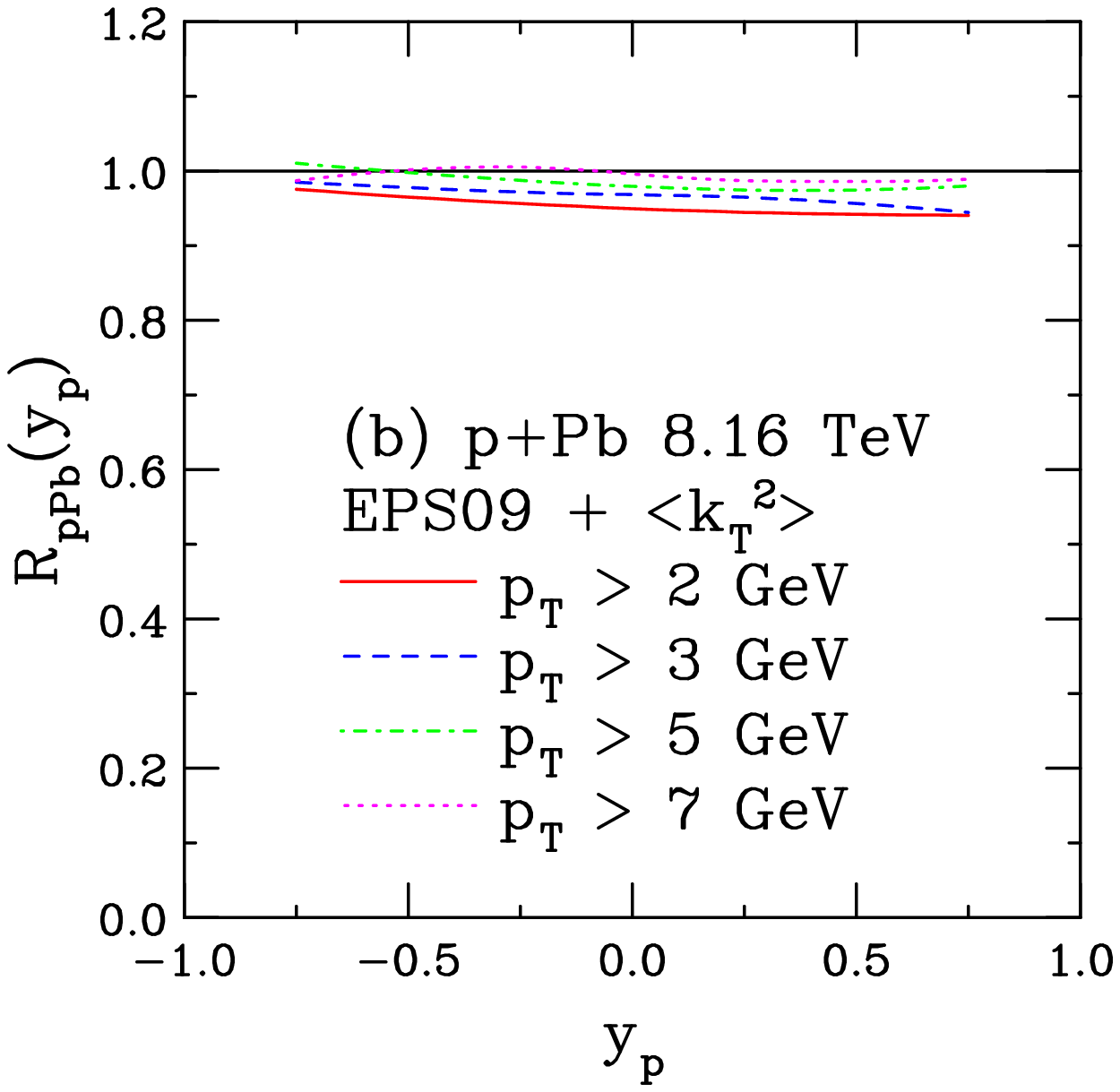} \\
  \includegraphics[width=0.5\columnwidth]{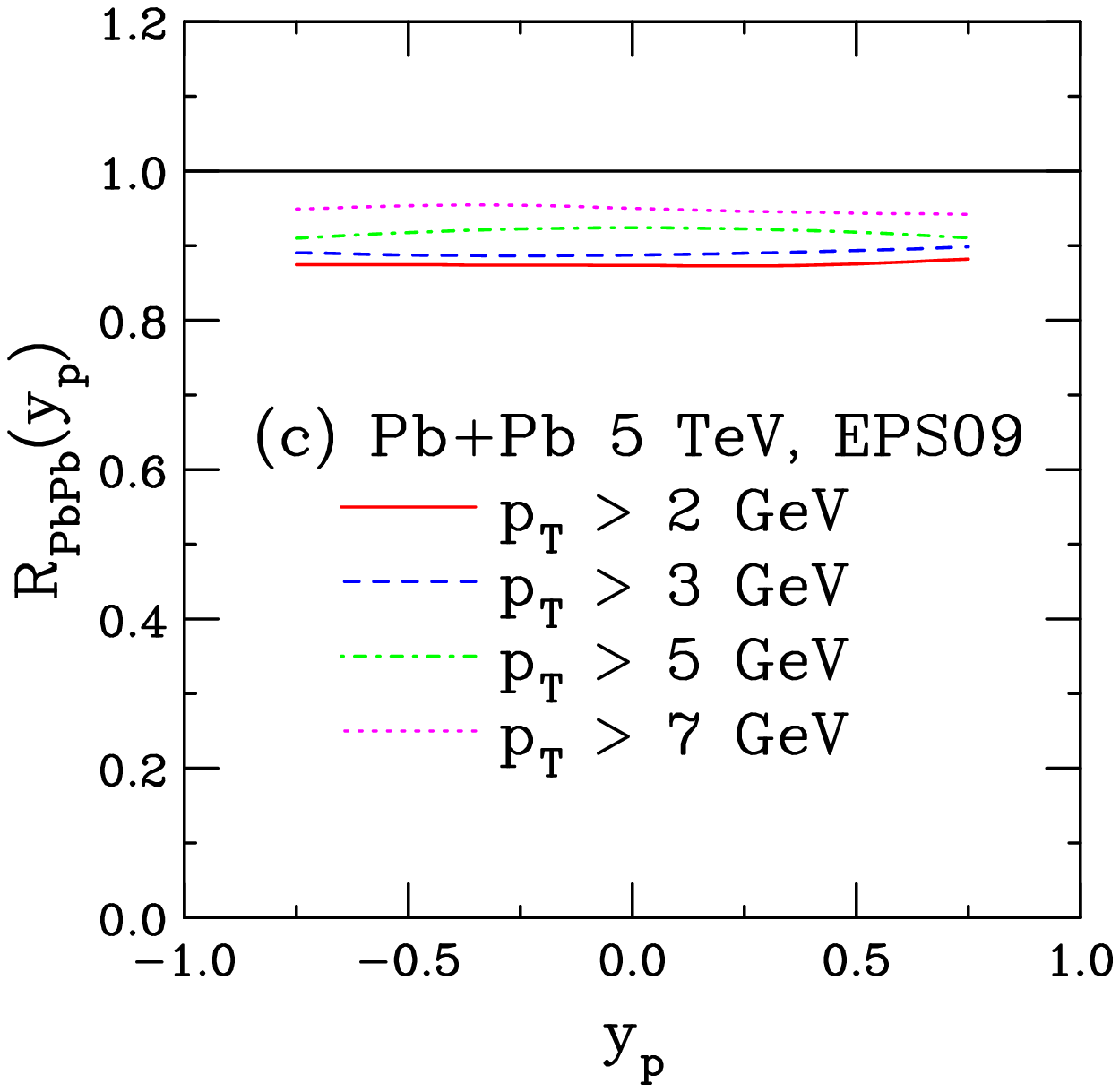} &
  \includegraphics[width=0.5\columnwidth]{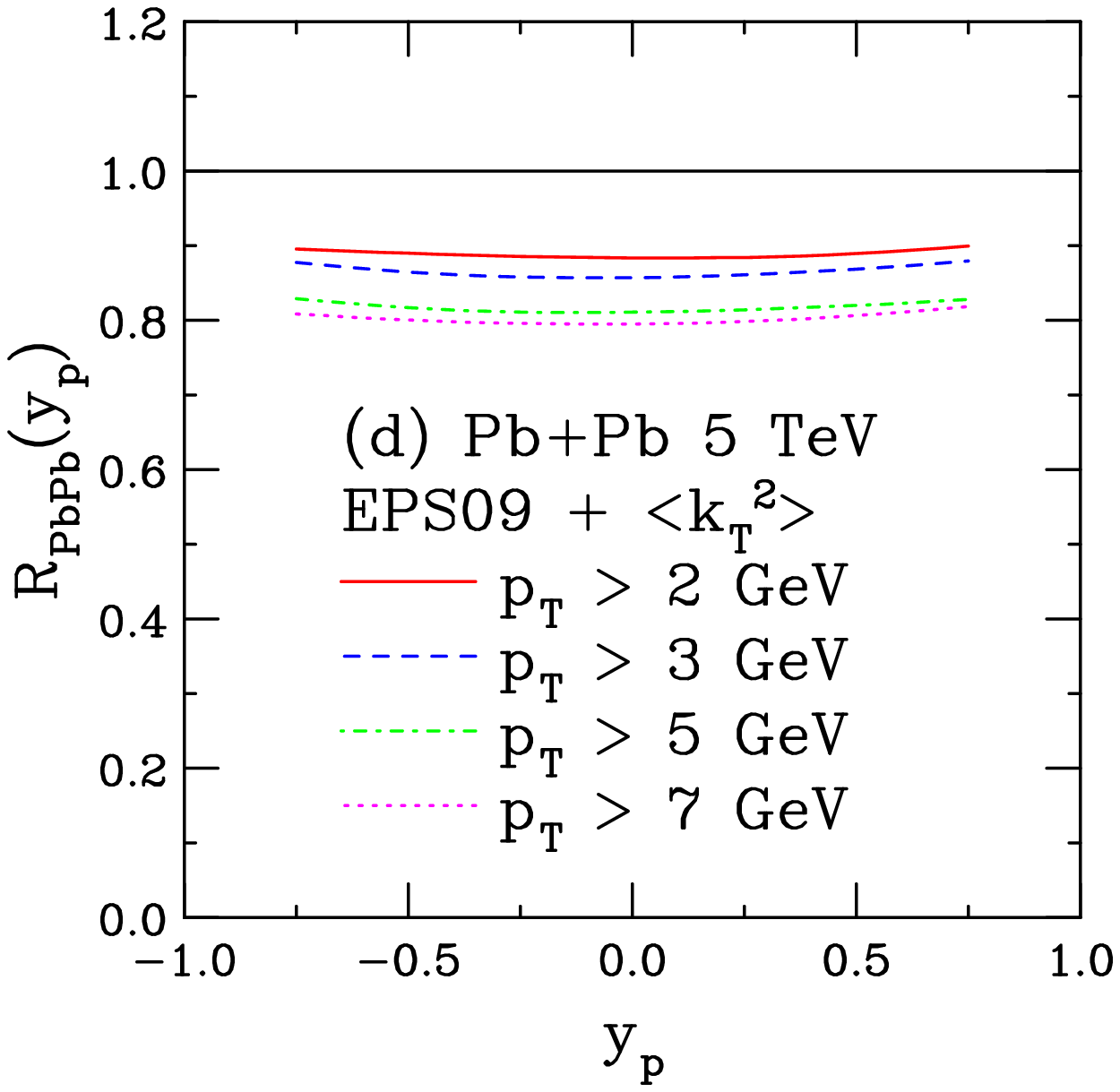} \\
 \end{tabular}
\caption[]{(Color online) Cold nuclear matter effects at central rapidity
  ($|y| \leq 0.8$) on the $b \overline b$
  pair rapidity for $p_T > 2$ (solid red), 3 (dashed blue),
  5 (dot-dashed green), and 7~GeV (dotted magenta) for (a) $p+$Pb collisions
  at 8.16~TeV with central EPS09 and the same $k_T$ kick as in $p+p$;
  (b) $R_{p{\rm Pb}}$ at 8.16 TeV with EPS09
  and additional $k_T$ broadening in Pb; (c) Pb+Pb collisions at 5~TeV with
  central EPS09 with the same $k_T$ kick in $p+p$ and $p+$Pb; and (d)
  $R_{AA}$ at 5~TeV with EPS09, additional $k_T$ broadening in the Pb nuclei,
  and a modified fragmentation function in Pb.
  }
  \label{fig_cnm_yqqC}
\end{figure}

The modifications of the pair rapidity are shown in Figs.~\ref{fig_cnm_yqq}
and \ref{fig_cnm_yqqC} for forward and central rapidities respectively.  In
$p+$Pb collisions at 8.16~TeV, for an average pair rapidity of 3 in the LHCb
acceptance, $x \approx 10^{-4}$ for $b$ quark production for the minimum $p_T$ of
2~GeV. The minimum $x$ remains of this order for all values of the minimum
$p_T$. Thus $R_{p {\rm Pb}} < 1$ for all minimum $p_T$ values.

Given the $x$ range, it is not surprising that $R_{p {\rm Pb}}$ is nearly
independent of $y_p$ since the EPS09 NLO gluon nPDF ratio is approximately
flat for $x < 0.001$ \cite{EPS09}.  The factorization scale
is also important for the nPDF ratio because the QCD scale evolution reduces the
shadowing effect at higher $p_T$ as well.  In addition, the average pair mass,
which should be considered when calculating $x$ instead of the transverse mass
of a single $b$ quark, is $\approx 15$~GeV for $p_T > 2$~GeV and $\approx 23$~GeV for
$p_T > 7$~GeV.  Thus one sees a mild tendency for $R_{p{\rm Pb}}(y_p)$ to
increase slightly as the minimum $p_T$ increases, an effect more visible
at central rapidity since the $b$ meson $R_{p{\rm Pb}}(p_T)$ rises faster with
$p_T$ at central than at forward rapidity.

When the average $k_T$ broadening is effectively doubled, as in
Figs.~\ref{fig_cnm_yqq}(b) and \ref{fig_cnm_yqqC}(b), the ratios are still
relatively independent of $y_p$
but the values of $R_{p {\rm Pb}}$ increase by a few percent relative to
calculations with shadowing alone.  A large effect is not expected because,
even for a doubling of the $k_T$ kick, $\langle k_T^2 \rangle = 5$~GeV$^2$ in
this case, $m_T$ is still larger than $\langle k_T^2 \rangle^{1/2}$ and, as seen
in Ref.~\cite{QQazi}, changing the $k_T$ kick does
not have a large effect on the shape of the single $b$ meson
$p_T$ distributions.  The change in $R_{p{\rm Pb}}(p_T)$
is also minimal, see Fig.~\ref{fig_cnm_bptdist}.
A much larger effect was seen on
the $c$ and $b$ quark $p_T$ distributions by modifying the
fragmentation function \cite{QQazi}.

Figure~\ref{fig_cnm_yqq}(c) shows the effect of shadowing alone on results
at forward rapidity from
Pb+Pb collisions at 5~TeV.  As noted earlier, the $x$ range probed is slightly
higher even though the factorization scale remains the same, due to the lower
energy.  Now, however, the ratio $R_{\rm PbPb}$ is no
longer independent of $y_p$ but shows some structure due to the combination of
nuclear effects from both nuclei.  This is because one of
the lead nuclei is now also probing the nPDFs at higher $x$, $x \geq 0.01$, and
moves through the antishadowing region as $y_p$ increases.

The increase in curvature with minimum $p_T$ is primarily due to the
antishadowing contribution from backward rapidity.  As $p_T$ and thus the input
scale of the nPDF increases, the antishadowing peak both reduces its maximum and
moves closer to midrapidity.  At the same time, the modification in the
shadowing region at forward rapidity is reduced but remains relatively
independent of rapidity.  Thus the results for shadowing only shows more change
with $y_p$ for $p_T > 7$~GeV than 2~GeV.

Perhaps the most intruiging result is seen in Fig.~\ref{fig_cnm_yqq}(d) where
the average $k_T$ kick is again doubled over that employed in $p+$Pb
collisions, to $\langle k_T^2 \rangle \approx 8.4$~GeV$^2$.  (There are, of course,
some small variations in $\langle k_T^2 \rangle$ between 5 and 8.16~TeV due to
the energy dependence assumed for $\langle k_T^2 \rangle$, given in
Eq.~(\ref{eq:kt2}).)
If that was the only effect assumed for Pb+Pb collisions, one would have
expected $R_{\rm PbPb}$ to be similar to the $p+$Pb resultsin
Fig.~\ref{fig_cnm_yqq}(b).

However, here the ratio with the highest minimum
$p_T$ is now lowest at the largest
$y_p$.  This is because, in addition to the $k_T$ broadening, an effective
energy loss has been introduced by changing the Peterson fragmentation function
parameter from the value determined in Ref.~\cite{QQazi}, to agree with the
FONLL $b$ meson $p_T$ distribution, to the $e^+ e^-$ default value,
$\epsilon_P = 0.006$.  As stated in Sec.~\ref{sec:single_b},
this is an effective
reduction in the average fraction of momentum transfered from the quark to the
meson, from $\approx 93$\% with $\epsilon_P = 0.0004$ to $\approx 83$\% for
$\epsilon_P = 0.006$ \cite{QQazi}.  As seen in Fig.~\ref{fig_cnm_bptdist}(c),
integration starting from $p_T > 2$~GeV includes the peak of the shifted $p_T$
distribution where there is an enhancement while $p_T > 7$~GeV includes a region
of relative suppression compared to $p+p$, resulting in stronger modification
for the higher $p_T$ cut than the lower.  This is an inversion of normally
expected behavior for heavy flavor $R_{\rm PbPb}$.
Note that this is in no way intended to replace a real
energy loss calculation but is rather intended to illustrate the
possible effect on $R_{\rm PbPb}$ for correlated observables.

Results as a function of $y_p$ in the central rapidity region are shown in
Fig.~\ref{fig_cnm_yqqC}.  The trends are quite similar for $p+$Pb collisions
although the level of shadowing is reduced in both cases and a slightly larger
separation of the results for different minimum $p_T$ values can be seen.
However, for Pb+Pb collisions, the results are now also independent of $y_p$.
This is because, around the narrow midrapidity window, the $x$ values probed
do not change significantly and whatever change occurs is probed symmetrically
around $y_p = 0$.
The stronger modification for higher $p_T$ with the increase of $\epsilon_P$
is still evident here, albeit with less separation
between results for different minimum $p_T$.

\begin{figure}[htpb]\centering
  \begin{tabular}{cc}
  \includegraphics[width=0.5\columnwidth]{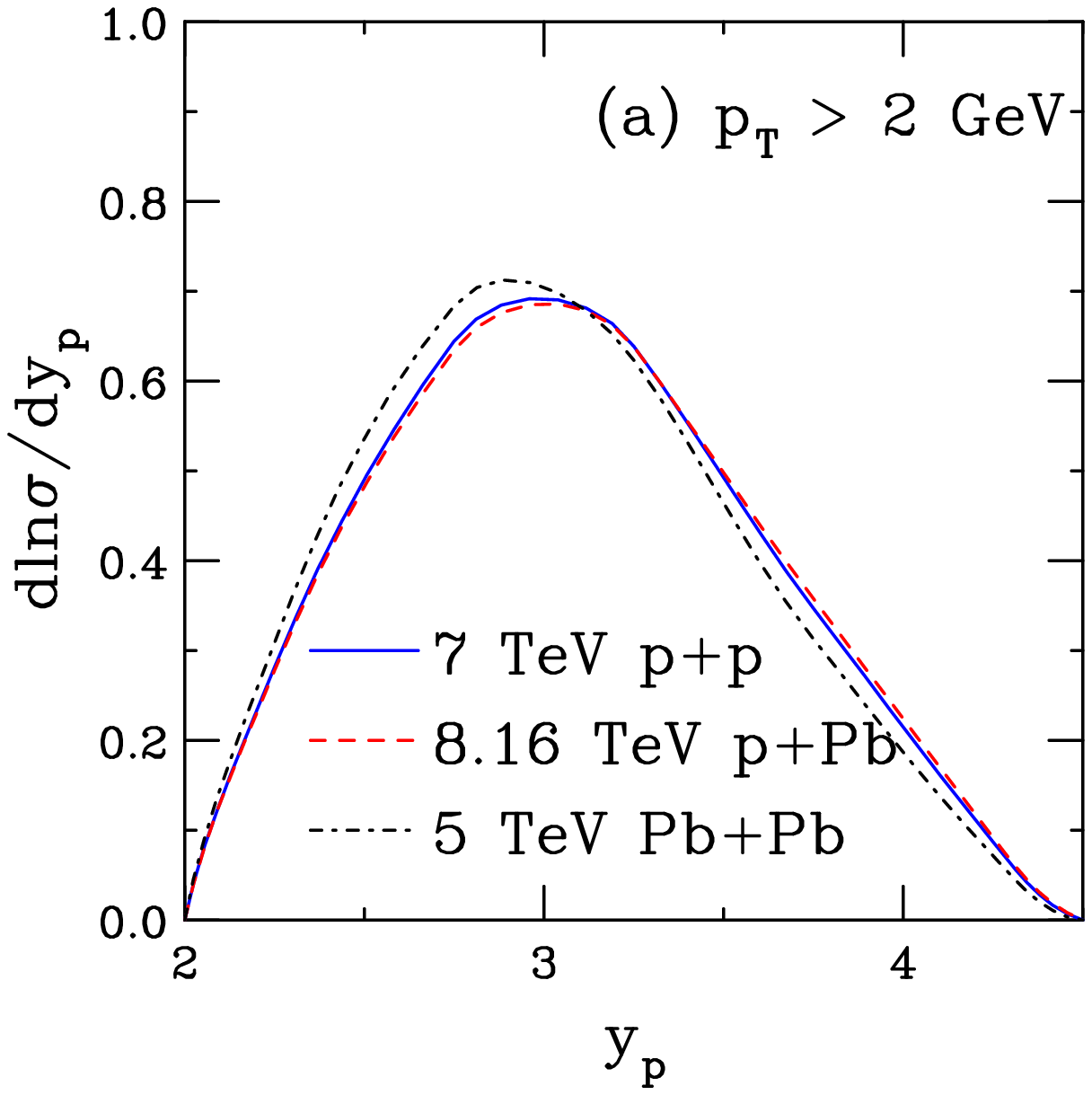} &
  \includegraphics[width=0.5\columnwidth]{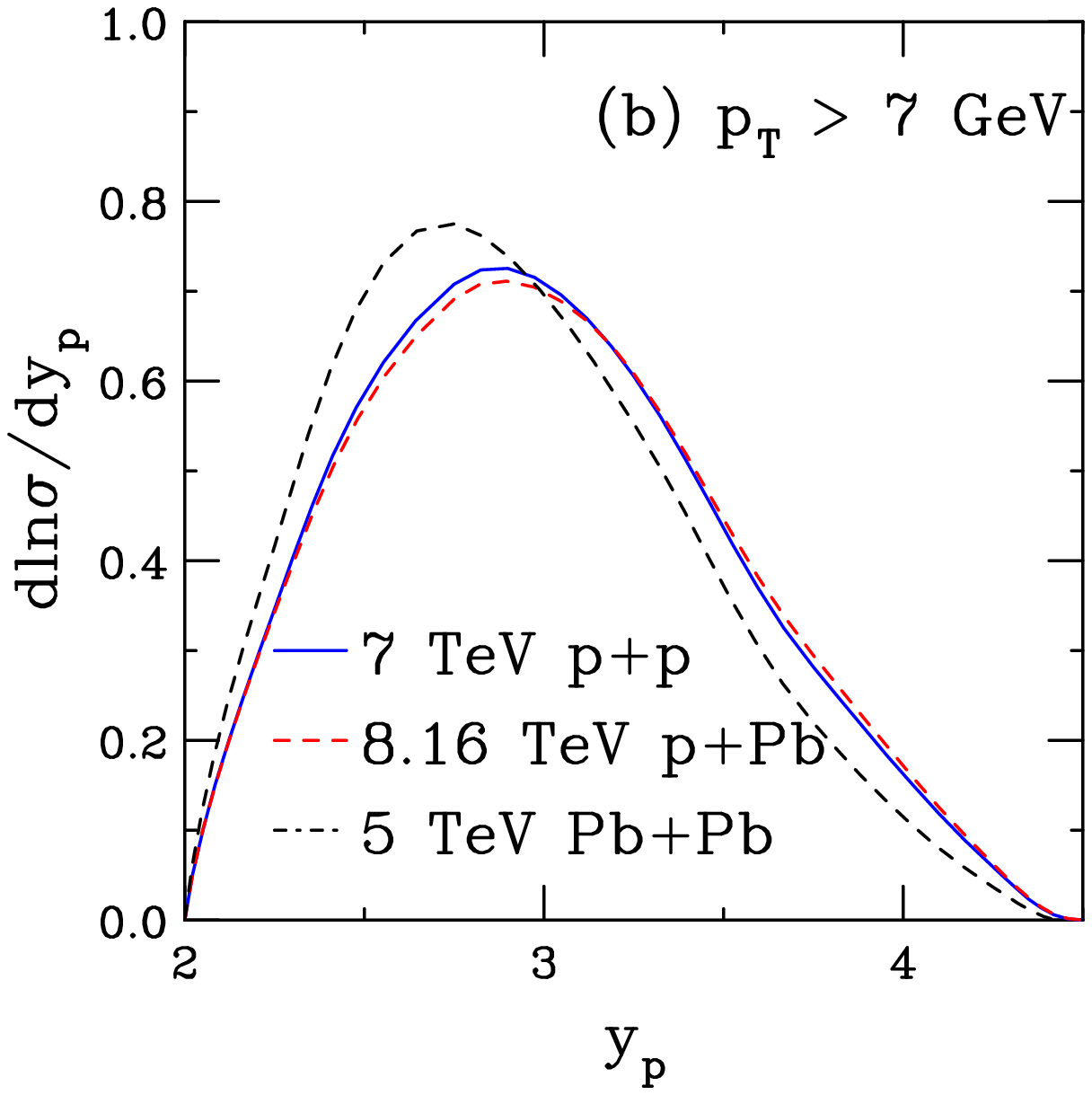} \\
  \end{tabular}
\caption[]{(Color online) The $b \overline b$
  pair rapidity in the range $2 < y_p < 4.5$
  for $p_T > 2$ (a) and 7~GeV (b) for $p+p$ collisions at
  7~TeV (solid blue), $p+$Pb collisions
  at 8.16~TeV (dashed red) and Pb+Pb collisions at 5~TeV (dot-dashed black).
  The $p+$Pb calculations include shadowing and enhanced broadening ($2\Delta$)
  while the Pb+Pb calculations include shadowing, broadening ($4\Delta$), and
  fragmentation function modification.
  }
  \label{fig_Ekt_yqq}
\end{figure}

The $y_p$ distributions are shown for $p+p$ collisions at $\sqrt{s} = 7$~TeV
(blue); $p+$Pb at $\sqrt{s_{_{NN}}} = 8.16$~TeV (red); and Pb+Pb collisions at
$\sqrt{s_{_{NN}}}= 5$~TeV (black) for $p_T > 2$ and 7~GeV in
Fig.~\ref{fig_Ekt_yqq}, including all
cold matter effects.  As one might expect from the discussion in
Sec.~\ref{sec:calcs}, the $\Delta y$ and $y_p$ distributions are unaffected by
$k_T$ broadening.  Thus the $y_p$ distributions for $p+p$ and $p+$Pb collisions
are the same shape.  However, the Pb+Pb distribution is clearly shifted
backward to lower $y_p$, a small but visible effect that increases with minimum
$p_T$.  This is due to the change in $\epsilon_P$.  The effect is stronger for
$p_T > 7$~GeV than $p_T > 2$~GeV since the larger lower limit of $p_T$
integration is more sensitive to the fragmentation function.  The steeper $p_T$
distributions with the higher value of $\epsilon_P$ mean that fewer $b$ quarks
may be found at higher rapidity, especially for higher values of the minimum
$p_T$, reducing the average $y_p$ in Pb+Pb collisions relative to the other
cases.  

\subsection{Modifications of $|\Delta \phi|$}

\begin{figure}[htpb]\centering
\begin{tabular}{cc}
  \includegraphics[width=0.5\columnwidth]{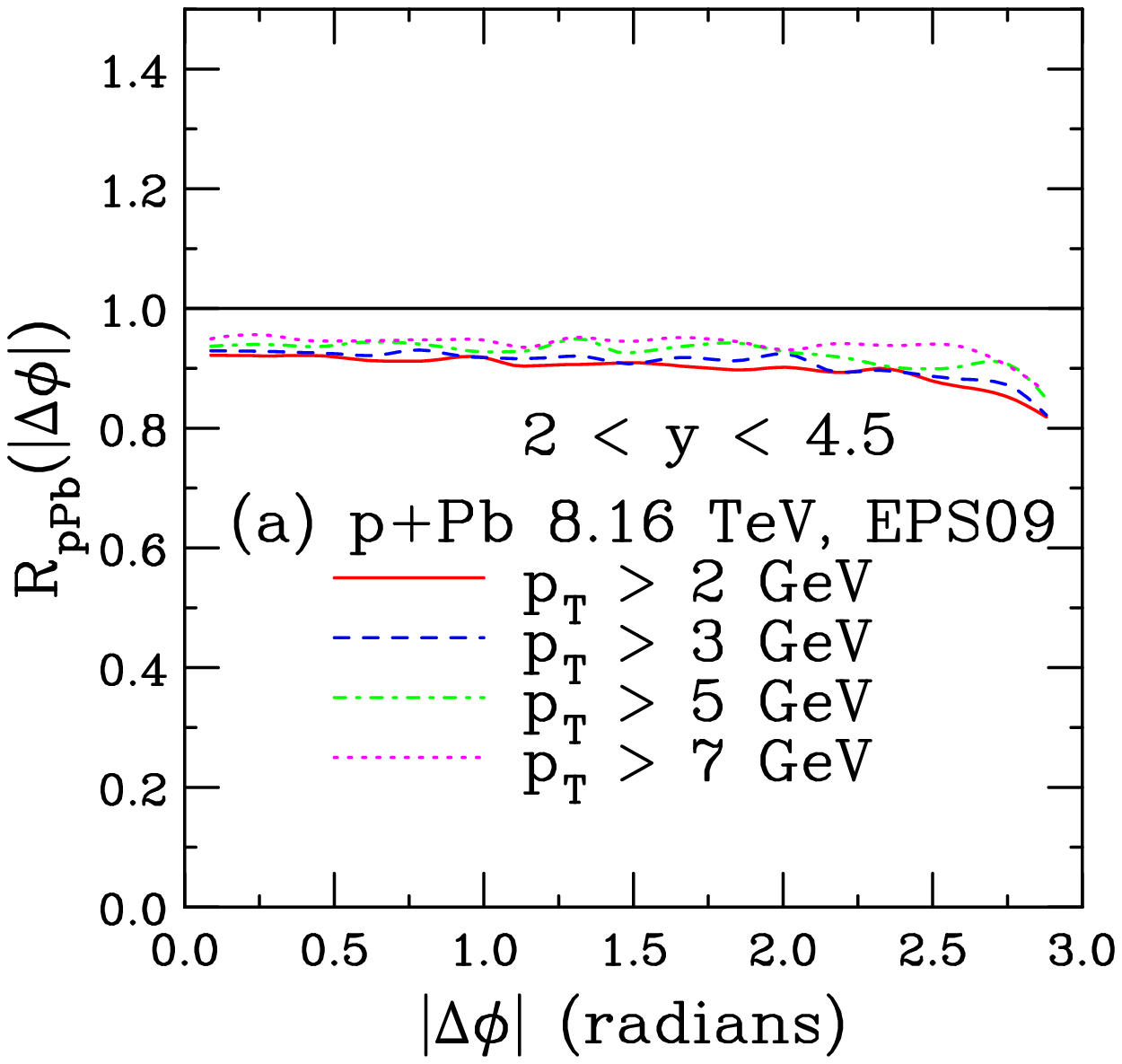} &
  \includegraphics[width=0.5\columnwidth]{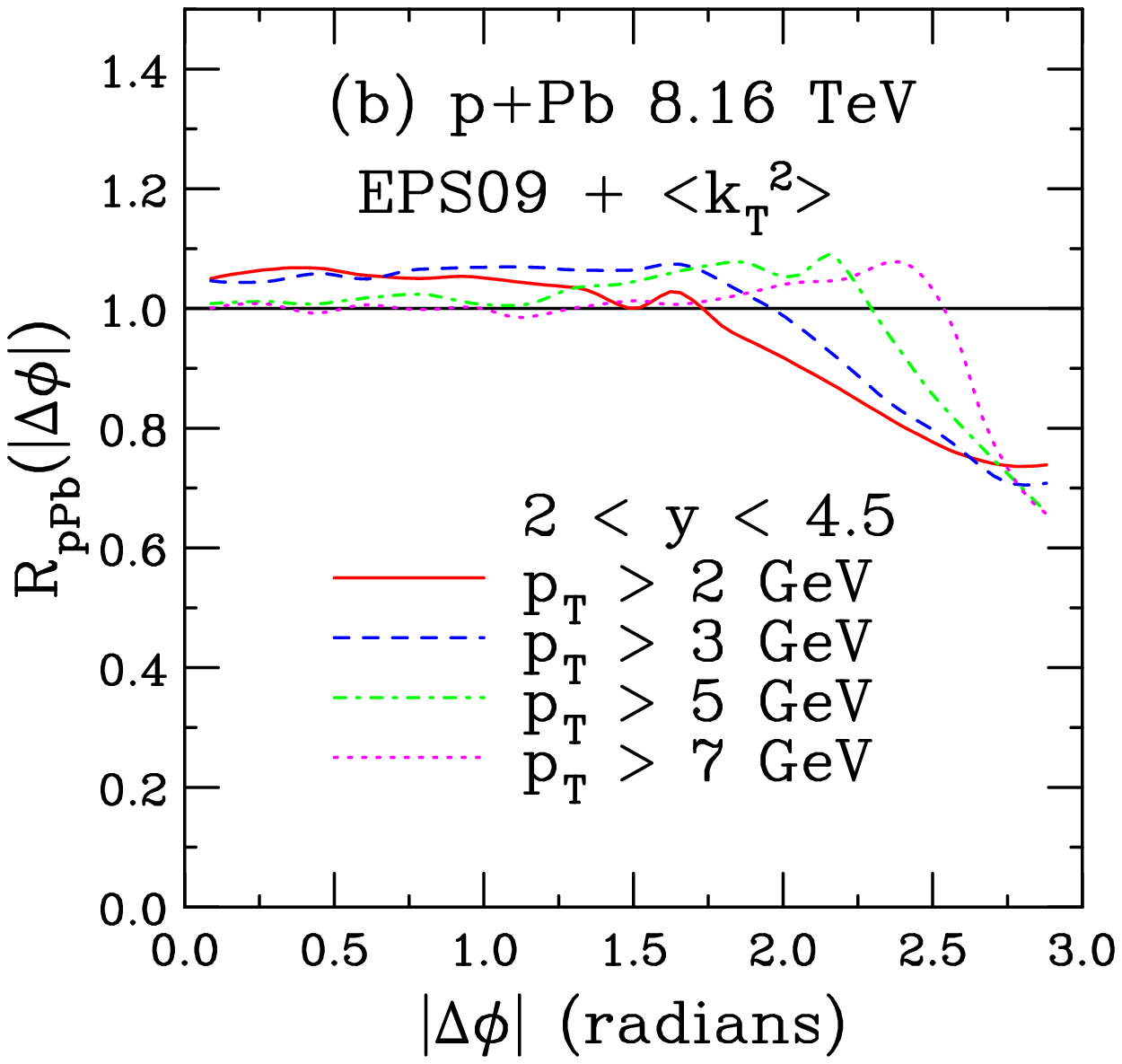} \\
  \includegraphics[width=0.5\columnwidth]{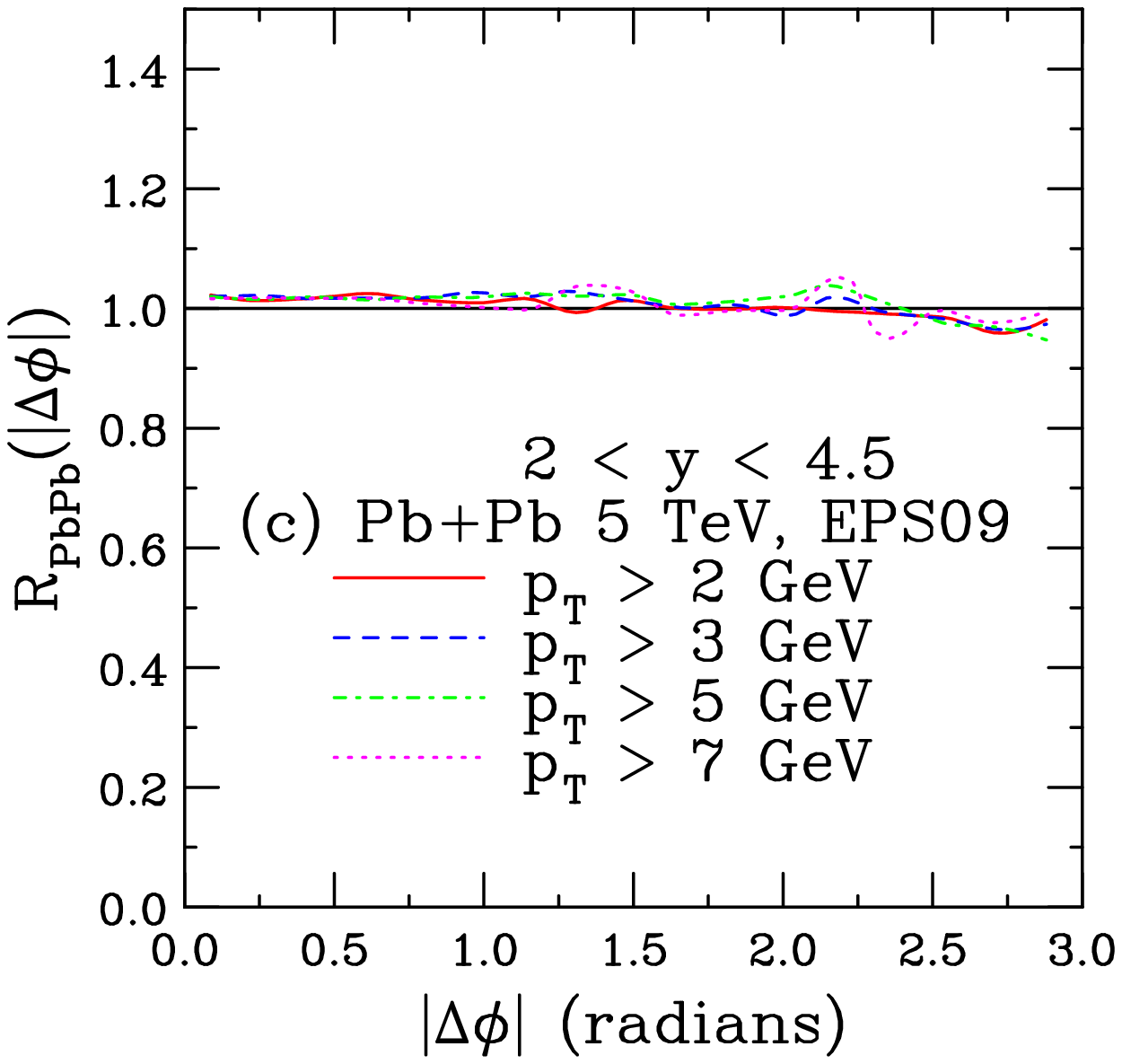} &
  \includegraphics[width=0.5\columnwidth]{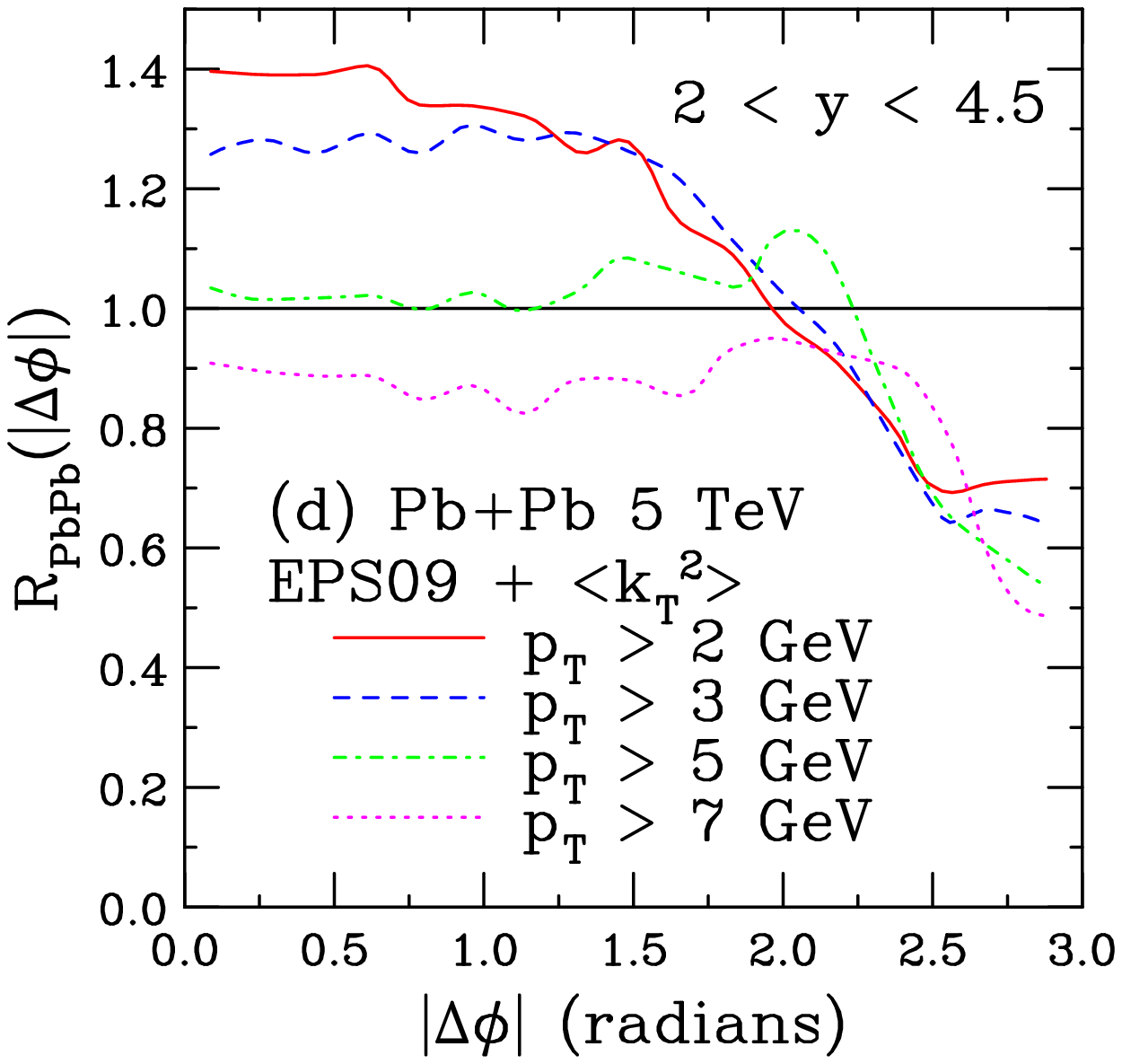} \\
 \end{tabular}
\caption[]{(Color online) Cold nuclear matter effects at forward rapidity
  ($2 < y < 4.5$) on the $b \overline b$
  azimuthal angle difference for $p_T > 2$ (solid red), 3 (dashed blue),
  5 (dot-dashed green), and 7~GeV (dotted magenta) for (a) $p+$Pb collisions
  at 8.16~TeV with central EPS09 and the same $k_T$ kick as in $p+p$;
  (b) $R_{p{\rm Pb}}$ at 8.16 TeV with EPS09
  and additional $k_T$ broadening in Pb; (c) Pb+Pb collisions at 5~TeV with
  central EPS09 with the same $k_T$ kick in $p+p$ and $p+$Pb; and (d)
  $R_{AA}$ at 5~TeV with EPS09, additional $k_T$ broadening in the Pb nuclei
  and a modified fragmentation function in Pb.
  }
  \label{fig_cnm_azi}
\end{figure}

\begin{figure}[htpb]\centering
\begin{tabular}{cc}
  \includegraphics[width=0.5\columnwidth]{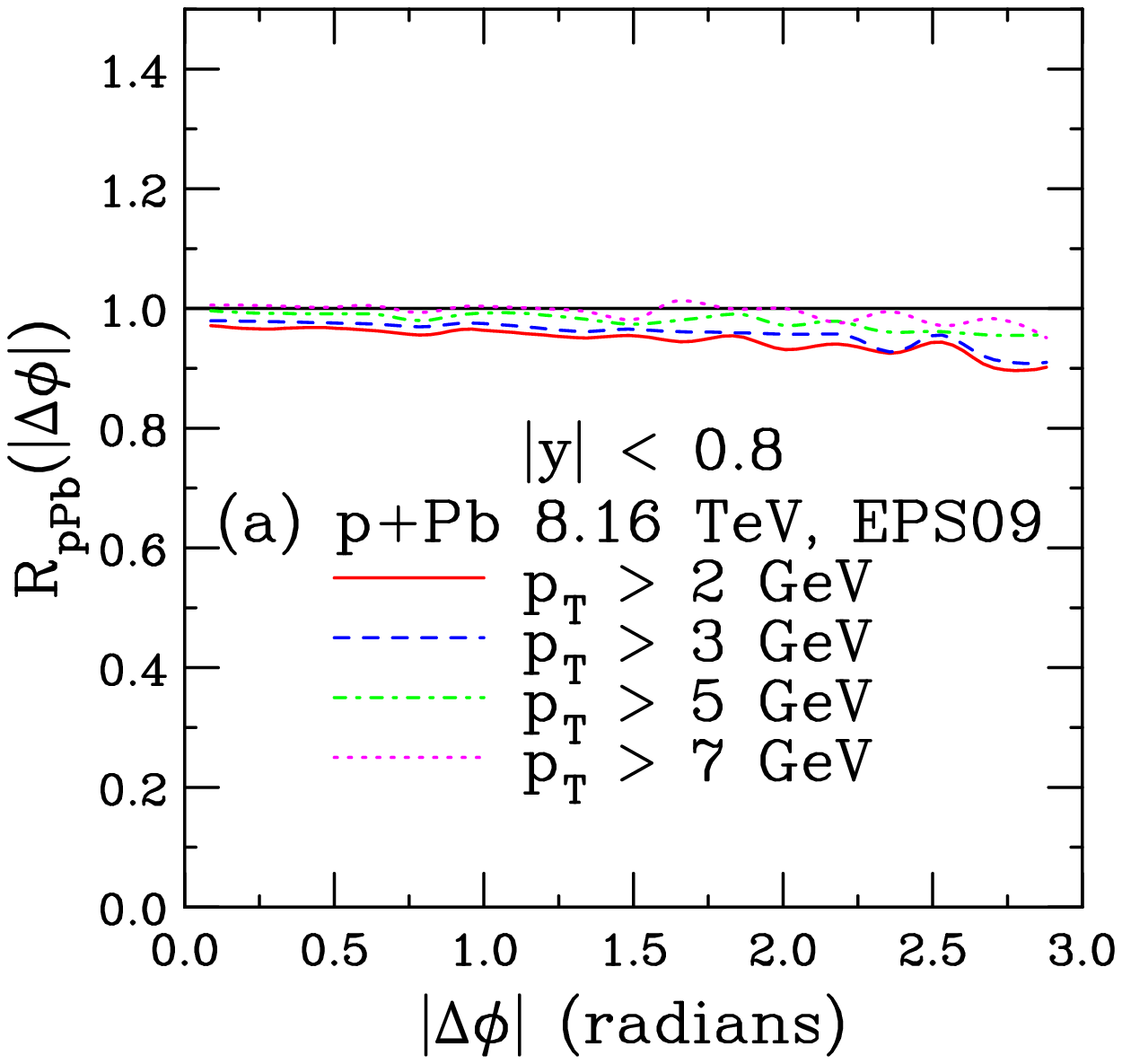} &
  \includegraphics[width=0.5\columnwidth]{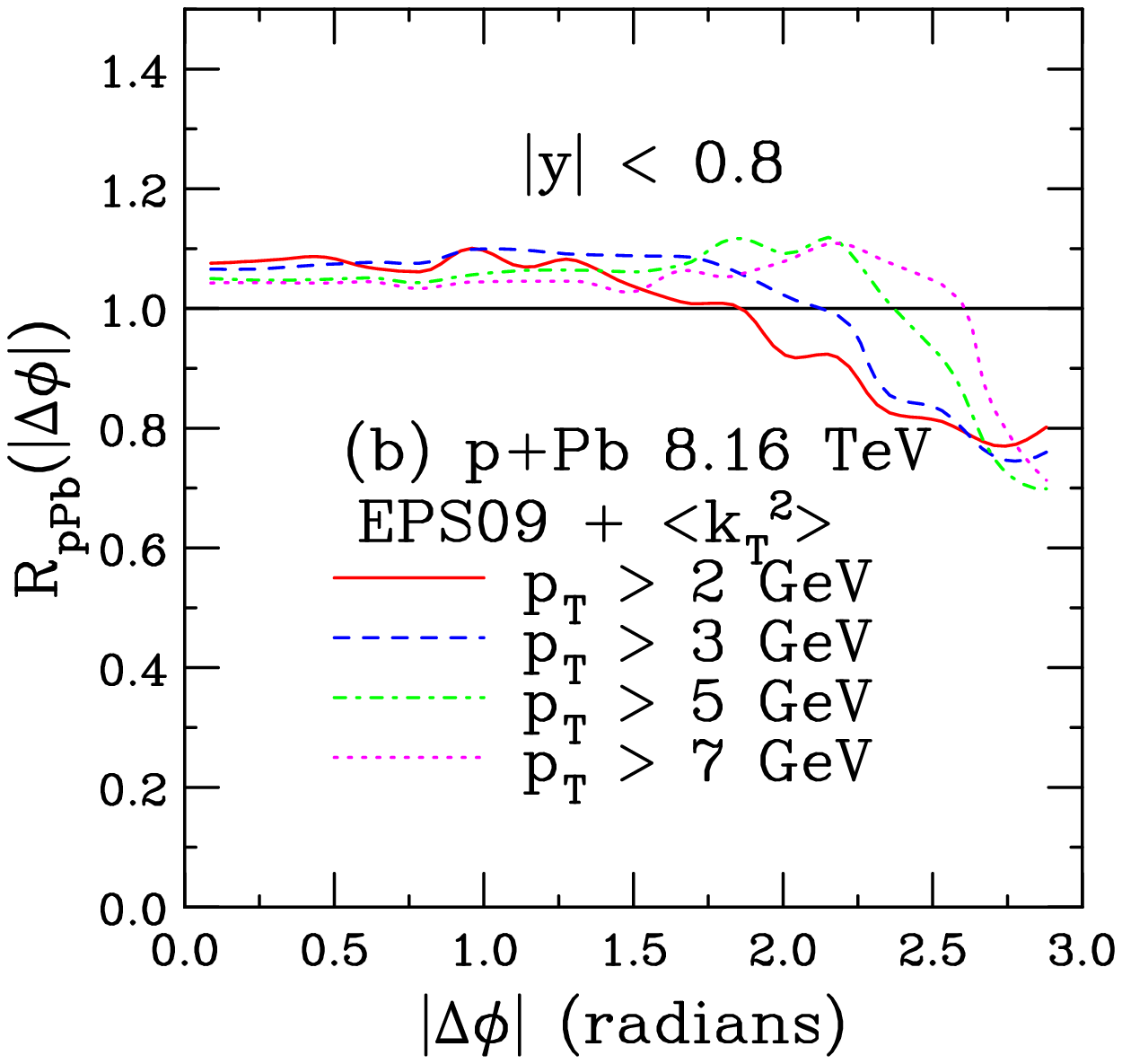} \\
  \includegraphics[width=0.5\columnwidth]{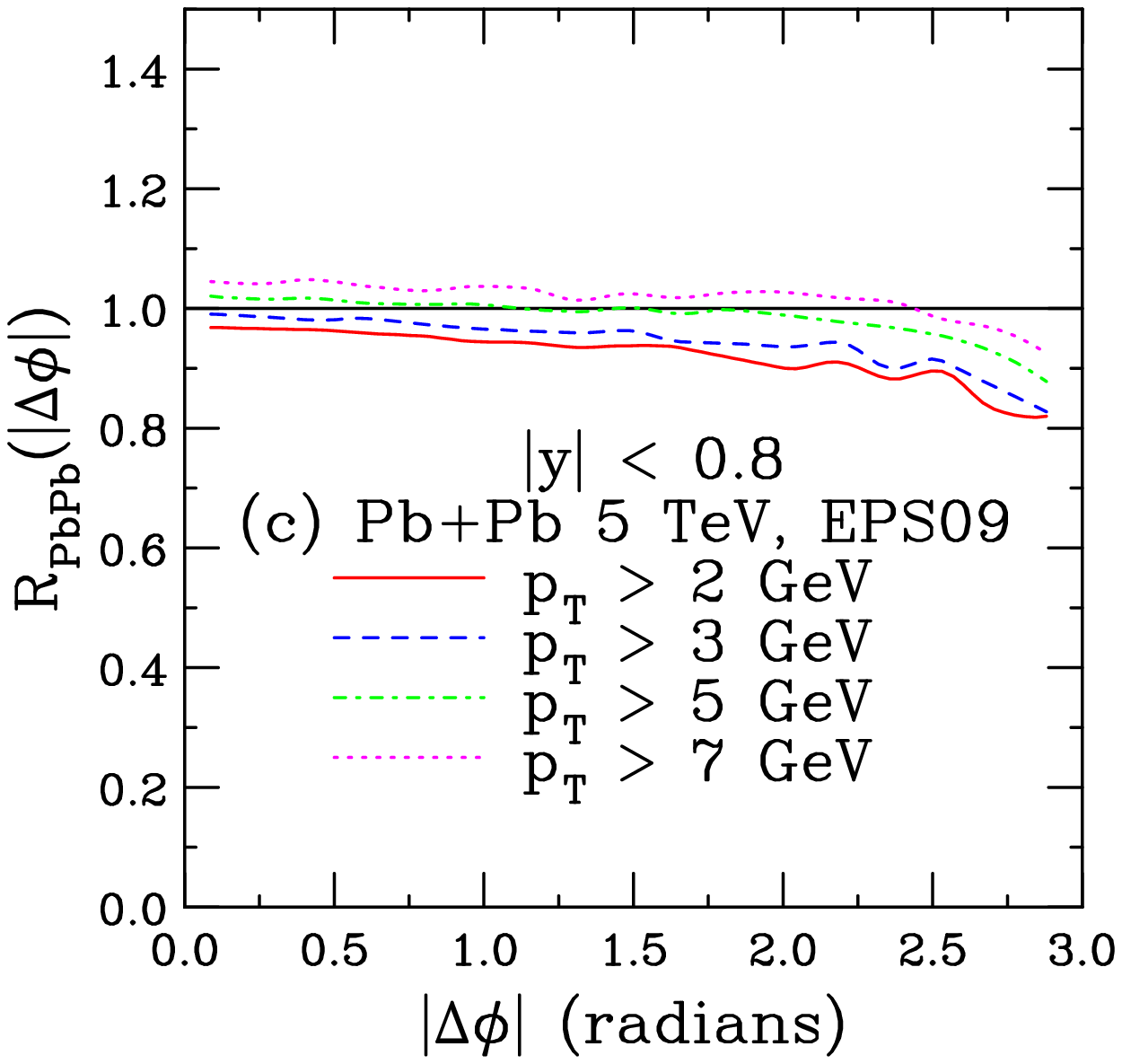} &
  \includegraphics[width=0.5\columnwidth]{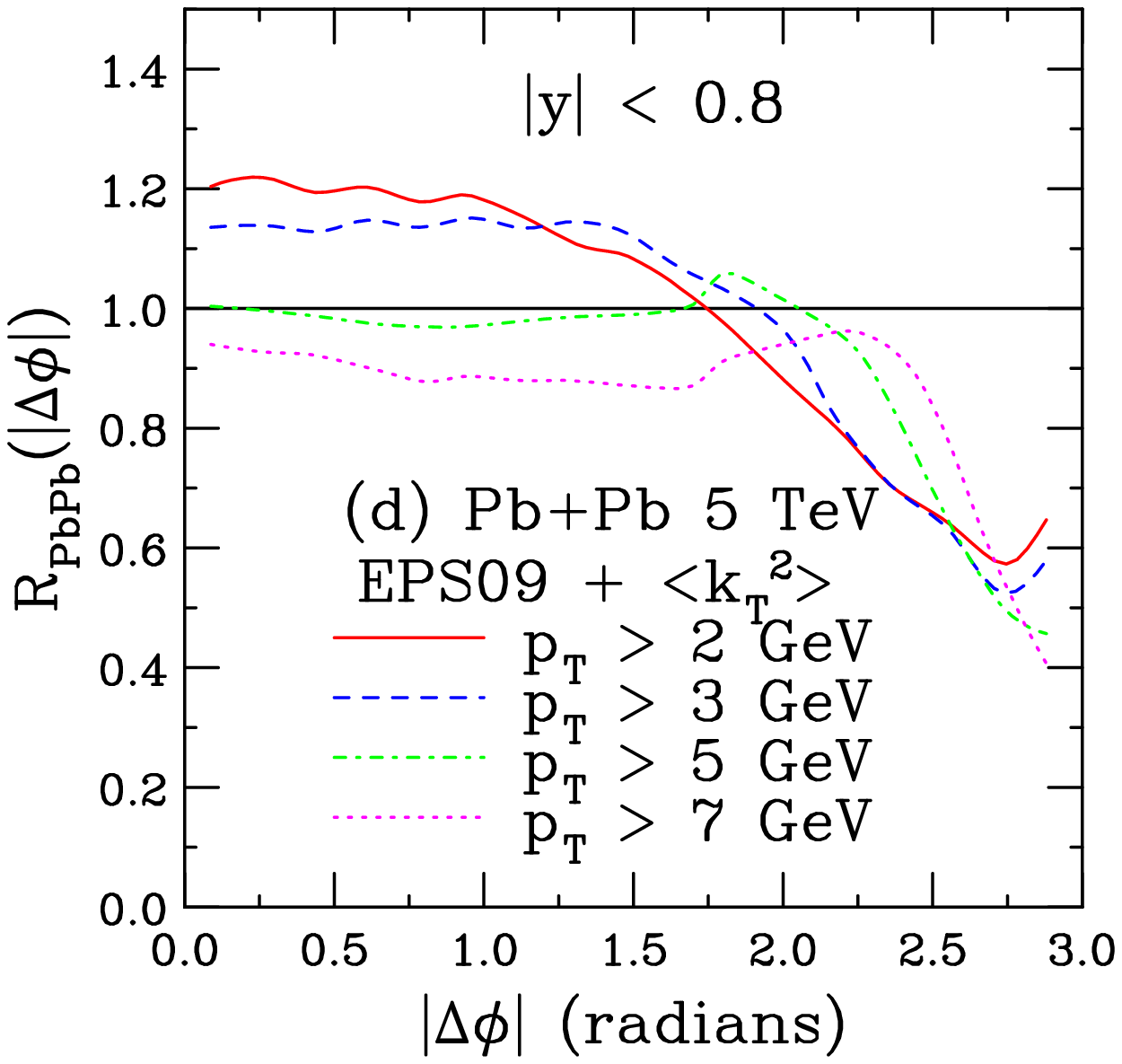} \\
 \end{tabular}
\caption[]{(Color online) Cold nuclear matter effects at central rapidity
  ($|y|<0.8$) on the $b \overline b$
  azimuthal angle difference for $p_T > 2$ (solid red), 3 (dashed blue),
  5 (dot-dashed green), and 7~GeV (dotted magenta) for (a) $p+$Pb collisions
  at 8.16~TeV with central EPS09 and the same $k_T$ kick as in $p+p$;
  (b) $R_{p{\rm Pb}}$ at 8.16 TeV with EPS09
  and additional $k_T$ broadening in Pb; (c) Pb+Pb collisions at 5~TeV with
  central EPS09 with the same $k_T$ kick in $p+p$ and $p+$Pb; and (d)
  $R_{AA}$ at 5~TeV with EPS09, additional $k_T$ broadening in the Pb nuclei
  and a modified fragmentation function in Pb.
  }
  \label{fig_cnm_aziC}
\end{figure}

Figures~\ref{fig_cnm_azi} and \ref{fig_cnm_aziC}
show $R_{p {\rm Pb}}$ and $R_{\rm PbPb}$ as a function of
the azimuthal separation between the $b$ and $\overline b$, at forward and
central rapidities respectively.  Note that for
shadowing only at forward rapidity,
the modification factors are rather independent of
$|\Delta \phi|$, with a mild decrease in the ratio
as $|\Delta \phi| \rightarrow \pi$.
At central rapidity, the shadowing only results show a somewhat stronger
decrease in the modification factor as $|\Delta \phi|$ increases.
A similar result was obtained for $c \overline c$ production at 5~TeV in
Ref.~\cite{QQazi} where the $p_T$- and rapidity-integrated $R_{p {\rm Pb}}$
shadowing ratios were independent of $|\Delta \phi|$
with $\langle k_T^2 \rangle = 0$ but showed a slight decrease with increasing
$|\Delta \phi|$ with $\langle k_T^2 \rangle \neq 0$.

Note that in Figs.~\ref{fig_cnm_azi}(a) and \ref{fig_cnm_aziC}(a), (c),
the modification factor decreases rather gradually with $|\Delta \phi|$ over
most of $|\Delta \phi|$ with an increase in the slope as $|\Delta \phi|$
approaches $\pi$.  (All the ratios are compatible with unity for Pb+Pb
collisions with shadowing alone in Fig.~\ref{fig_cnm_azi}(c), as might be
expected from the result in Fig.~\ref{fig_cnm_bptdist}(c).)
When the minimum $p_T$ is increased, the ratios are independent of
$|\Delta \phi|$ until they begin to decrease at larger $|\Delta \phi|$.
This can be attributed to the narrowing and sharpening of the peak
in the $|\Delta \phi|$ distribution with increasing $p_T$, seen in
Fig.~\ref{fig_azi}, while the enhancement
at $|\Delta \phi| \rightarrow 0$ is increasing more slowly.

\begin{figure}[htpb]\centering
  \begin{tabular}{cc}
  \includegraphics[width=0.5\columnwidth]{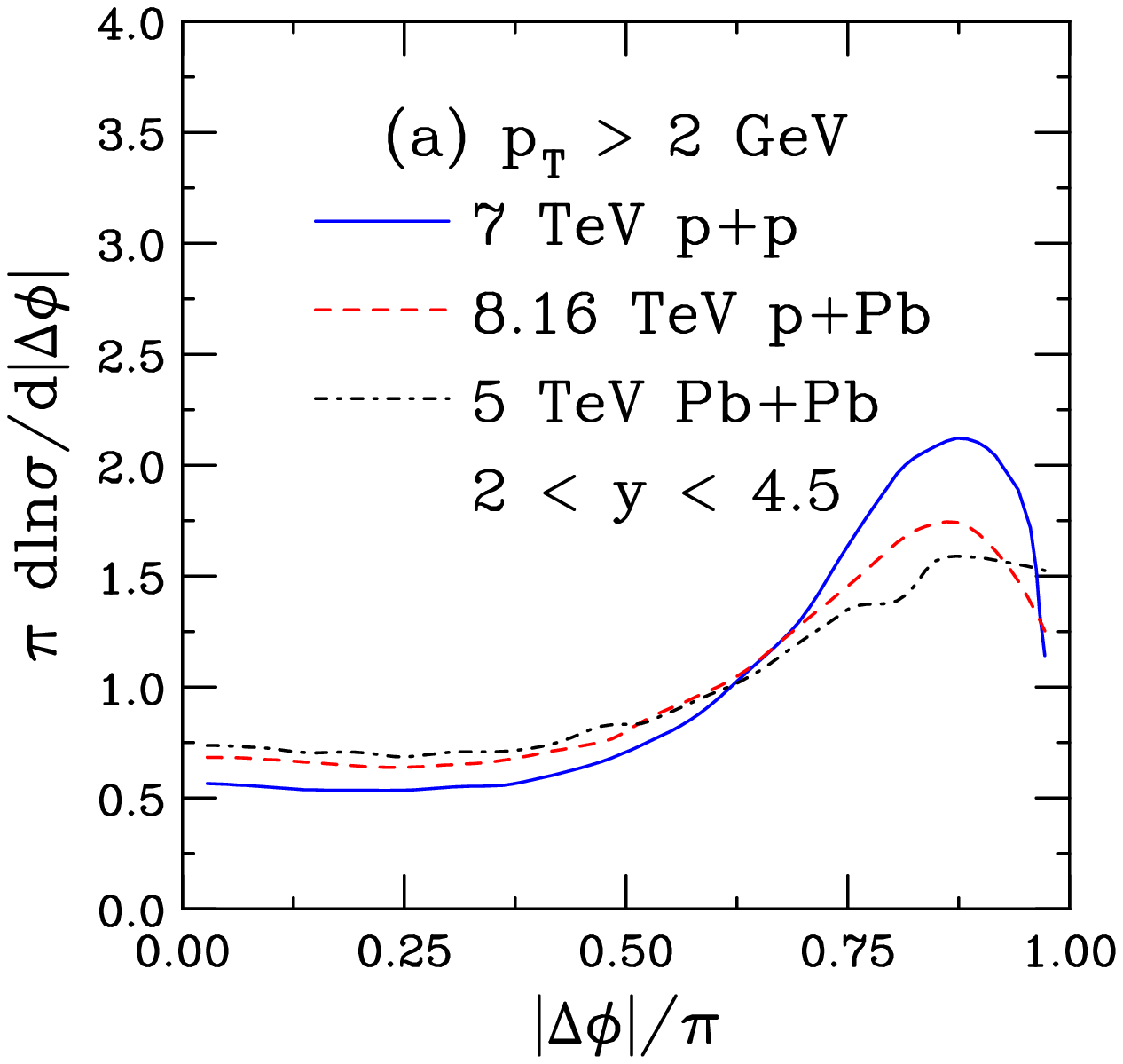} &
  \includegraphics[width=0.5\columnwidth]{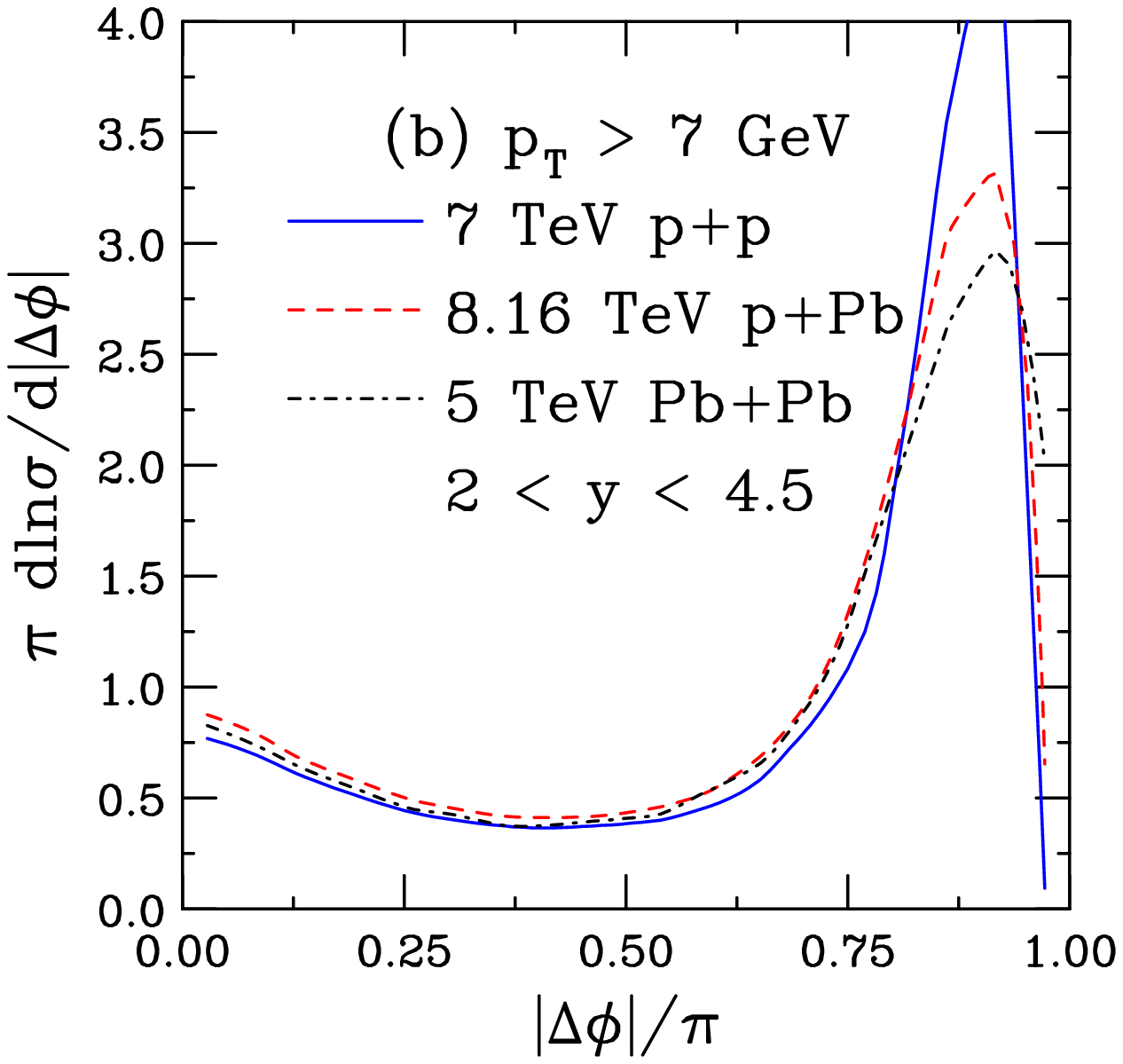} \\
  \end{tabular}
\caption[]{(Color online) The $b \overline b$
  azimuthal separation at forward rapidity ($2 < y < 4.5$)
  for $p_T > 2$ (a) and 7~GeV (b) for $p+p$ collisions at
  7~TeV (solid blue), $p+$Pb collisions
  at 8.16~TeV (dashed red) and Pb+Pb collisions at 5~TeV (dot-dashed black).
  The $p+$Pb calculations include shadowing and enhanced broadening ($2\Delta$)
  while the Pb+Pb calculations include shadowing, broadening ($4\Delta$), and
  fragmentation function modification.
  }
  \label{fig_Ekt_azi}
\end{figure}

The more striking effect is for $p+$Pb and, in particular, Pb+Pb collisions with
enhanced $k_T$ broadening.  As shown in Fig.~\ref{fig_Ekt_azi}, the effect of
broadening on the azimuthal distributions in $p+$Pb and Pb+Pb collisions
reduces and broadens the peak at $|\Delta \phi| \approx \pi$ and enhances the
distribution at $|\Delta \phi| \approx 0$.
Recall that these distributions are the same shape in $p+p$ collisions so that
the differences seen in the figure arise primarily from enhanced broadening.
Only the
results for the lowest and highest minimum $p_T$ values are again
shown to illustrate the effect.

There is an interesting change of
behavior at $|\Delta \phi| \approx 0$ in $p+$Pb relative to Pb+Pb collisions
for the two different $p_T$ cuts.  At lower $p_T$, where a change in
broadening has a larger effect on the shape of the $|\Delta \phi|$ distribution:
the Pb+Pb result is slightly enhanced over that of $p+$Pb since $\Delta = 2$
for $p+$Pb and 4 for Pb+Pb \cite{QQazi}.  On the other hand, for the higher
$p_T$ cut, the enhancement due to broadening is reduced and the change in the
fragmentation function parameter suppresses the $|\Delta \phi|$ enhancement at
$|\Delta \phi| \approx 0$ in Pb+Pb relative to $p+$Pb, even though the $k_T$
broadening is larger in Pb+Pb collisions, see Ref.~\cite{QQazi}.

The $p+$Pb ratios with enhanced $k_T$ broadening in both rapidity regions
exhibit a kink that occurs at
higher $\Delta \phi$ for increasing minimum $p_T$.  This can be understood
from the ratios of increasing $\langle k_T^2 \rangle$ relative to the results
with no broadening, $\langle k_T^2 \rangle = 0$. Reference~\cite{QQazi}
studied the
turn on of the effect at $\langle k_T^2 \rangle > 0$, becoming increasingly
isotropic as $\langle k_T^2 \rangle$ increases.  As shown in Ref.~\cite{QQazi},
the $|\Delta \phi|$ distributions peak more sharply at both
$|\Delta \phi| \rightarrow \pi$ and $|\Delta \phi| \rightarrow 0$.  The
effect at $|\Delta \phi| = 0$ is reduced in $b \overline b$ production relative
to $c \overline c$ since it requires a much harder gluon to balance a more
massive $b \overline b$ pair than the lighter $c \overline c$ pair.  This
change in
relative height of the peak for fixed $\langle k_T^2 \rangle$ and increasing
minimum $p_T$ causes the location of the kink in the ratio to increase from
$|\Delta \phi| \approx 1.7$ to 2.5 radians as the minimum $p_T$ increases from 2 to
7~GeV.  

The heirarchy is more clearly reversed for Pb+Pb collisions, shown in
Figs.~\ref{fig_cnm_azi}(d) and \ref{fig_cnm_aziC}(d).
The fragmentation function parameter
$\epsilon_P$ has almost no
effect on the shape of the $\Delta \phi$ distribution, as also shown in
Ref.~\cite{QQazi} when integrated over all $p_T$.  However, it will change the
number of $b \overline b$ pairs with both quarks in the rapidity acceptance,
as illustrated in Fig.~\ref{fig_Ekt_yqq}, producing the inverted hierarchy of
ratios seen here.  Note that the larger $k_T$ kick assumed for Pb+Pb collisions
also result in the kink in $R_{p {\rm Pb}}$ seen in Figs.~\ref{fig_cnm_azi}(b)
and \ref{fig_cnm_aziC}(b),
moving to lower $\Delta \phi$, now between 1.5 to 2.4 radians in
Figs.~\ref{fig_cnm_azi}(d) and \ref{fig_cnm_aziC}(d).

\section{Summary}
\label{sec:summary}

The $b \overline b \rightarrow J/\psi J/\psi$ pair observables measured by
LHCb in $p+p$ collisions
were studied in detail in an exclusive NLO calculation with fragmentation
and $k_T$ broadening, as first described in Ref.~\cite{QQazi}.  The calculations
reproduced the data very well in all cases and for all $p_T$ cuts.  The
sensitivity of the results to the $k_T$ broadening is shown and, while
the direct $b \overline b$ observables are indeed sensitive to the $k_T$
broadening, the resulting $J/\psi$ pairs are not since the decays
produce a decorrelation of the $J/\psi$s relative to the parent $b$ hadrons.
The mass and scale dependence has also been studied
and shown not to be large, as expected for $b \overline b$ production.  The
dependence of the results on rapidity were also shown.

Finally, the nuclear modification factors for enhanced
$k_T$ broadening and fragmentation function modification in cold nuclear
matter were presented.  The potential cold nuclear matter effects calculated
here for $p+$Pb and Pb+Pb
collisions are not intended to be definitive but illustrative only.
The calculations have demonstrated how effects like broadening and energy loss
could be disentangled by specific correlated observables more sensitive to each.
Although both observables discussed are affected by the two effects, the pair
rapidity is more senstive to fragmentation while the azimuthal correlation
depends most strongly on the $k_T$ broadening.
While the effects were modeled in the context of
cold nuclear matter, similar decorrelation, as produced by
enhanced $k_T$ broadening, could be due to hot matter effects, as produced in
the quark-gluon plasma \cite{gossiaux}.  A thermal medium also results in heavy
quark energy loss, as modeled by the modified $\epsilon_P$.  These calculations
thus suggest that
additional correlated observables are required to better quantify
such effects, regardless of the medium.\\

{\bf Acknowledgments:}
I would like to thank A.~Mischke, T.~Dahms, 
and M.~Winn for discussions.
This work was performed under the auspices of the 
U.S. Department of Energy by Lawrence Livermore National Laboratory under 
Contract DE-AC52-07NA27344 and supported by the U.S. Department of Energy, 
Office of Science, Office of Nuclear Physics (Nuclear Theory) under contract 
number DE-SC-0004014.


\begin{thebibliography}{}
%
%

\bibitem{UA1} C. Albajar {\it et al.} (UA1 Collaboration), 
Measurement of $b \overline b$ correlations at the CERN $p \overline p$
collider, 
Z. Phys. C {\bf 61} (1994) 41.

\bibitem{D0} B. Abbot {\it et al.} (D0 Collaboration),
The $b \overline b$ production cross section and angular correlations
in $p \overline p$ collisions at $\sqrt{s} = 1.8$~TeV,
Phys. Lett. B {\bf 487} (2000) 264.

\bibitem{CDF180}
D. Acosta {\it et al.} (CDF Collaboration), 
Measurements of $b \overline b$ azimuthal
production correlations in $p \overline p$ collisions at $\sqrt{s} = 1.8$~TeV,
Phys. Rev. D {\bf 71} (2005) 092001.

\bibitem{CDF196} 
T. Aaltonen {\it et al.} (CDF Collaboration),
Measurement of correlated $b \overline b$ production in $p \overline p$
colliosions at $\sqrt{s} = 1960$~TeV,
Phys. Rev. D {\bf 77} (2008) 072004.

\bibitem{ATLAS}
  M. Aaboud {\it et al.} (ATLAS Collaboration), 
Measurement of $b$-hadron pair production
with the ATLAS detector in proton-proton collisions at $\sqrt{s} = 8$~TeV,  
JHEP {\bf 1711} (2017) 062.

\bibitem{PHENIX}
  C. Aidala {\it et al.} (PHENIX Collaboration),
  Correlations of $\mu \mu$, $e \mu$, and $ee$ pairs in $p+p$ collisions at
  $\sqrt{s} = 200$~GeV and implications for $c \overline c$ and $b \overline b$
  production mechanisms,
  arXiv:1805.04075.

\bibitem{ALICE7}
  S.~Acharya {\it et al.} [ALICE Collaboration],
  Dielectron production in proton-proton collisions at $ \sqrt{s}=7 $ TeV,
  JHEP {\bf 1809}, 064 (2018).

\bibitem{ALICE13}
  S.~Acharya {\it et al.} [ALICE Collaboration],
  Dielectron and heavy-quark production in inelastic and high-multiplicity
  proton–proton collisions at $\sqrt {s_{NN}}= 13$~TeV,
  Phys.\ Lett.\ B {\bf 788}, 505 (2019).
  
\bibitem{ISAJET}
  F.~E.~Paige and S.~D.~Protopopescu,
  Isajet 5.20: A Monte Carlo Event Generator for $p p$ and $\bar{p} p$
  Interactions,
  Conf.\ Proc.\ C {\bf 860115}, 213 (1986).
  
\bibitem{HVQJET}
  Cited in Ref.~\cite{D0} as: M. Baarmand and F. Paige. HVQJET Monte Carlo Event
  Generator, private communication.
  
\bibitem{HERWIG}
 G.~Corcella {\it et al.}, HERWIG 6: An event generator for hadron emission
  reactions with interfering gluons (including supersymmetric processes),
JHEP {\bf 0101}, 010 (2001).

\bibitem{PYTHIA}
T. Sjostrand {\it et al.},
High-energy physics event generation with PYTHIA 6.1,
Comput.\ Phys.\ Commun.  {\bf 135}, 238 (2001);
arXiv:hep-ph/0308153.
  
\bibitem{MNR}  M. L. Mangano, P. Nason, and G. Ridolfi,
  Heavy quark correlations in hadron collisions at next-to-leading
  order, Nucl. Phys. B {\bf 373}, 295 (1992).

\bibitem{QQazi} R. Vogt,
Heavy Flavor Azimuthal Correlations in Cold Nuclear Matter,
Phys. Rev. C {\bf 98} (2018) 034907.
  
\bibitem{LHCb}  R. Aaij {\it et al.} (LHCb Collaboration),
  Study of $b \overline b$ correlations in high energy proton-proton collisions,
  JEHP {\bf 11} (2017) 030.

\bibitem{CDFDDpairs}
  B. Reisert {\it et al.} [CDF Collaboration], Charm Production Studies at CDF,
  Nucl. Phys. Porc. Suppl. {\bf 170}, 243 (2007).

\bibitem{LHCbDDpairs}
R.~Aaij {\it et al.} [LHCb Collaboration],
Observation of double charm production involving open charm in pp collisions
at $\sqrt{s}$ = 7 TeV,
  JHEP {\bf 1206}, 141 (2012),
  [JHEP {\bf 1403}, 108 (2014)].
  
\bibitem{CMSbbpairs}  
  V. Khachatryan {\it et al.} [CMS Collaboration], Measurement of
  $B \overline B$ Angular Correlations based on Seconary Verte Reconstruction
  at $\sqrt{s} = 7$~TeV, JHEP {\bf 1105}, 136 (2011).

\bibitem{PYTHIA6}
T. Sjostrand, S. Mrenna and P. Z. Skands,
PYTHIA 6.4 physics and manual, JHEP {\bf 05} (2006) 026.

\bibitem{PYTHIA8}
T. Sjostrand, S. Mrenna and P. Z. Skands,
A brief introduction to PYTHIA 8.1,
Comp. Phys. Comm. {\bf 178} (2008) 852.

\bibitem{POWHEG}
  S. Frixione, P. Nason, and G. Ridolfi,  A positive-weight
  next-to-leading-order Monte Carlo for heavy flavour hadroproduction,
  JHEP {\bf 0709}, 126 (2007);
  arXiv:0707.3081 [hep-ph].

\bibitem{LHCb1}
  R.~Aaij {\it et al.} [LHCb Collaboration],
  Measurement of $J/\psi$ production in $pp$ collisions at
  $\sqrt{s}=7~\rm{TeV}$,
  Eur.\ Phys.\ J.\ C {\bf 71}, 1645 (2011).

\bibitem{LHCb2}
  R.~Aaij {\it et al.} [LHCb Collaboration],
  Production of $J/\psi$ and $\Upsilon$ mesons in $pp$ collisions at
  $\sqrt{s} = 8$~TeV,
  JHEP {\bf 1306}, 064 (2013).

\bibitem{NorrbinSjo}
  E. Norrbin and T. Sj\"{o}strand, Production and Hadronization of Heavy Quarks,
  Eur. Phys. J C {\bf 17}, 137 (2000).

\bibitem{Bedjidian:2004gd} 
  M.~Bedjidian {\it et al.},
  Hard probes in heavy ion collisions at the LHC: Heavy flavor physics,
  arXiv:hep-ph/0311048.

\bibitem{Peterson}
    C. Peterson, D. Schlatter, I. Schmitt, and P. Zerwas, Scaling Violations
  in Inclusive $e^+ e^-$ Annihilation Spectra, Phys. Rev. D {\bf 27} (1983) 105.

\bibitem{FONLL}
  M. Cacciari, M. Greco and P. Nason, The $p_T$ spectrum in heavy flavor
  hadroproduction, JHEP {\bf 05}, 007 (1998).

\bibitem{GMVFN} B. A. Kniehl, G. Kramer, I. Schienbein and H. Spiesberger,
  Inclusive $D^{*+}$ production in $p \overline p$ collisions with massive
  charm quarks,
  Phys. Rev. D {\bf 71}, 014018 (2005).

\bibitem{Helenius:2018uuf} 
  I.~Helenius and H.~Paukkunen,
  Revisiting the D meson hadroproduction in general-mass variable flavour number
  scheme,
  JHEP {\bf 1805}, 196 (2018).
  
\bibitem{CYLO}
  C. Y. Lo and J. D. Sullivan, Transverse Momentum Distributions in Drell-Yan
  Processes, Phys. Lett. B {\bf 86}, 327 (1979).
  
\bibitem{MLM1} 
  S. Frixione, M. L. Mangano, P. Nason and G. Ridolfi, Charm and bottom
  production: theoretical results versus experimental data, Nucl. Phys. B
  {\bf 431}, 453 (1994).

\bibitem{APP} G. Altarelli, G. Parisi and R. Petronzio, Transverse Momentum
  in Drell-Yan Processes, Phys. Lett. B {\bf 76}, 351 (1978).

\bibitem{CGreco}
  P. Chiappeta and M. Greco, Transverse Momentum Distributions for Drell-Yan
  Pairs in QCD, Phys. Lett. B {\bf 106}, 219 (1981).

\bibitem{BtoJpsi_decay}
  M.~Tanabashi {\it et al.} [Particle Data Group],
  Review of Particle Physics,
  Phys.\ Rev.\ D {\bf 98}, 030001 (2018).

\bibitem{NVF}
  R. E. Nelson, R. Vogt and A. D. Frawley,
  Narrowing the uncertainty on the total charm cross section and its effect on 
  the $J/\psi$ cross section,
  Phys. Rev. C {\bf 87}, 014908 (2013).

\bibitem{NVFinprep}
  R. E. Nelson, R. Vogt and A. D. Frawley, in preparation.

\bibitem{lhc_ppb}
  R.~Vogt,
  Shadowing effects on $J/\psi$ and $\Upsilon$ production at energies available
  at the CERN Large Hadron Collider,
  Phys.\ Rev.\ C {\bf 92}, 034909 (2015).

\bibitem{gossiaux}
  M.~Nahrgang, J.~Aichelin, P.~B.~Gossiaux and K.~Werner,
  Azimuthal correlations of heavy quarks in Pb + Pb collisions at
  $\sqrt{s}=2.76$ TeV at the CERN Large Hadron Collider,
  Phys.\ Rev.\ C {\bf 90}, 024907 (2014).

\bibitem{younus1}
  M.~Younus, U.~Jamil and D.~K.~Srivastava,
  Correlations of Heavy Quarks Produced at Large Hadron Collider,
  J. Phys. G {\bf 39}, 025001 (2012).
  
\bibitem{younus2}
  M.~Younus and D.~K.~Srivastava,
  Effect of Energy Loss on Azimuthal Correlations of charm and correlated charm
  decay in collision of lead nuclei at $\sqrt{s} = 2.76$ $A$TeV,
  J. Phys. G {\bf 40}, 065004 (2013).

\bibitem{younus3}
  M.~Younus, S.~K.~Tripathy, P.~K.~Sahu and Z.~Niak,
  Azimuthal correlations of $D$-mesons in $p+p$ and $p+$Pb collisions at LHC
  energies, Eur. Phys. J. A {\bf 53}, 112 (2017).
  
\bibitem{EPS09} K. J. Eskola, H. Paukkunen and C. A. Salgado, 
  EPS09: A New Generation of NLO and LO Nuclear Parton Distribution Functions, 
  JHEP {\bf 0904}, 065 (2009). 

\bibitem{EPPS16}
  K.~J.~Eskola, P.~Paakkinen, H.~Paukkunen and C.~A.~Salgado,
  EPPS16: Nuclear parton distributions with LHC data,
  Eur.\ Phys.\ J.\ C {\bf 77}, 163 (2017).

\bibitem{816TeV_pred}
  J.~L.~Albacete {\it et al.},
  Predictions for Cold Nuclear Matter Effects in $p+$Pb Collisions at
  $\sqrt{s_{_{NN}}} = 8.16$~TeV,
  Nucl.\ Phys.\ A {\bf 972}, 18 (2018).

\bibitem{RVcent}
  R.~Vogt,
  Relation of hard and total cross-sections to centrality,
  Acta Phys.\ Hung.\ A {\bf 9}, 339 (1999).

\bibitem{RVdilep2}
R.~Vogt, B.~V.~Jacak, P.~L.~McGaughey and P.~V.~Ruuskanen,
  Rapidity distribution of dileptons from a hadronizing quark-gluon plasma,
  Phys.\ Rev.\ D {\bf 49}, 3345 (1994)
  
\bibitem{Andre}
  A.~Mischke, A new correlation method to identify and separate charm and
  bottom production processes at RHIC, Phys.\ Lett.\  B {\bf 671}, 361 (2009).

\bibitem{KPfins}
  I.~Helenius and H.~Paukkunen, Double $D$-meson production in proton-proton and
  proton-lead collisions at the LHC, arXiv:1906.06971.

\bibitem{LHCb_NPpsi}
  R.~Aaij {\it et al.} [LHCb Collaboration],
  Prompt and nonprompt J/$\psi$ production and nuclear modification in $p$Pb
  collisions at $\sqrt{s_{\text{NN}}}= 8.16$ TeV,
 Phys.\ Lett.\ B {\bf 774}, 159 (2017).
 
\bibitem{LHCb_Bp}
  R.~Aaij {\it et al.} [LHCb Collaboration],
  Measurement of $B^+$, $B^0$ and $\Lambda_b^0$ production in
  $p\mkern 1mu\mathrm{Pb}$ collisions at
  $\sqrt{s_\mathrm{NN}}=8.16\,{\rm TeV}$,
  Phys.\ Rev.\ D {\bf 99}, 052011 (2019).

\bibitem{LandsbergShao}
  J.~P.~Lansberg and H.~S.~Shao,
  Towards an automated tool to evaluate the impact of the nuclear modification
  of the gluon density on quarkonium, D and B meson production in proton–nucleus
  collisions, Eur.\ Phys.\ J. C {\bf 77}, 1 (2017).
  
\bibitem{Vitev}
  I.~Vitev, T.~Goldman, M.~B.~Johnson and J.~W.~Qiu,
  Phys.\ Rev.\ D {\bf 74}, 054010 (2006).
  
\bibitem{Andronic}
  A.~Andronic {\it et al.},
  Heavy-flavour and quarkonium production in the LHC era: from proton–proton
  to heavy-ion collisions,
  Eur.\ Phys.\ J.\ C {\bf 76}, 107 (2016).

\end{thebibliography}
\end{document}